\documentclass[a4paper,fleqn,usenatbib]{mnras}

\usepackage{newtxtext,newtxmath}
\usepackage[T1]{fontenc}
\usepackage{ae,aecompl}

\usepackage{graphicx}	% Including figure files
\usepackage{amsmath}	% Advanced maths commands
\usepackage{amssymb}	% Extra maths symbols
\usepackage{lscape}
\usepackage{geometry}
\usepackage{comment}
\usepackage{gensymb}
\usepackage{float}

\newcommand{\Cl}{Cl~J1449+0856}
\newcommand{\Htwo}{H$_{2}$}
\newcommand{\kms}{kms$^{-1}$}

\title[Merger driven star-formation activity]{Merger driven star-formation activity in \Cl\ at z=1.99 as seen by ALMA and JVLA}

\author[R. T. Coogan et al.]{R. T. Coogan$^{1,2}$\thanks{E-mail: r.coogan@sussex.ac.uk}, E. Daddi$^{2}$, M. T. Sargent$^{1}$, V. Strazzullo$^{3}$, F. Valentino$^{4}$, R. Gobat$^{5}$,  \newauthor
 G. Magdis$^{4,6}$, M. Bethermin$^{7, 8}$, M. Pannella$^{3}$, M. Onodera$^{9,10}$, D. Liu$^{11}$, A. Cimatti$^{12}$, \newauthor H. Dannerbauer$^{13,14}$, M. Carollo$^{15}$, A. Renzini$^{16}$, E. Tremou$^{2,17}$
\\
$^{1}$Astronomy Centre, Department of Physics and Astronomy, University of Sussex, Brighton BN1 9QH, UK\\
$^{2}$Irfu/Service d'Astrophysique, CEA Saclay, Orme des Merisiers, F-
91191 Gif sur Yvette, France\\
$^{3}$Department of Physics, Ludwig-Maximilians-Universitat, Scheinerstr.
1, 81679 Munchen, Germany\\
$^{4}$Dark Cosmology Centre, Niels Bohr Institute, University of Copenhagen, Juliane Mariesvej 30, 2100, Copenhagen, Denmark\\
$^{5}$Instituto de F\'{i}sica, Pontificia Universidad Cat\'{o}lica de Valpara\'{i}so, Casilla 4059, Valpara\'{i}so, Chile\\
$^{6}$Institute for Astronomy, Astrophysics, Space Applications and Remote Sensing, National Observatory of Athens, 15236, Athens, Greece\\
$^{7}$Aix Marseille Univ, CNRS, LAM, Laboratoire d'Astrophysique de Marseille, Marseille, France\\
$^{8}$European Southern Observatory, Karl-Schwarzschild-Str. 2, 85748 Garching, Germany\\
$^{9}$Subaru Telescope, National Astronomical Observatory of Japan, National
Institutes of Natural Sciences (NINS), 650 North A'ohoku Place,
Hilo, HI, 96720, USA\\
$^{10}$Department of Astronomical Science, SOKENDAI (The Graduate
University for Advanced Studies), 650 North A'ohoku Place, Hilo, HI, 96720, USA\\
$^{11}$Max Planck Institute for Astronomy, Konigstuhl 17, D-69117 Heidelberg,
Germany\\
$^{12}$Dipartimento di Fisica e Astronomia, Universit\'{a} di Bologna, Via Gobetti 93/2, I-40129, Bologna, Italy\\
$^{13}$Instituto de Astrof\'{i}sica de Canarias (IAC), E-38205 La Laguna, Tenerife, Spain\\
$^{14}$Universidad de La Laguna, Dpto. Astrof\'{i}sica, E-38206 La Laguna, Tenerife, Spain\\
$^{15}$Institute for Astronomy, ETH Zurich, CH-8093 Zurich,
Switzerland\\
$^{16}$INAF - Osservatorio Astronomico di Padova, Vicolo
dell'Osservatorio 5, I-35122 Padova, Italy\\
$^{17}$AIM/CEA Paris-Saclay, Universit\'{e} Paris Diderot, CNRS, F-91191 Gif-sur-Yvette, France\\
}

\date{Accepted 2018 May 22. Received 2018 May 14; in original form 2017 October 19}

\pubyear{2018}

\begin{document}
\label{firstpage}
\pagerange{\pageref{firstpage}--\pageref{lastpage}}
\maketitle

\begin{abstract}

We use ALMA and JVLA observations of the galaxy cluster \Cl\ at z=1.99, in order to study how dust-obscured star-formation, ISM content and AGN activity are linked to environment and galaxy interactions during the crucial phase of high-z cluster assembly. We present detections of multiple transitions of $^{12}$CO, as well as dust continuum emission detections from 11 galaxies in the core of \Cl{}. We measure the gas excitation properties, star-formation rates, gas consumption timescales and gas-to-stellar mass ratios for the galaxies.

We find evidence for a large fraction of galaxies with highly-excited molecular gas, contributing $>$50\% to the total SFR in the cluster core. We compare these results with expectations for field galaxies, and conclude that environmental influences have strongly enhanced the fraction of excited galaxies in this cluster. We find a dearth of molecular gas in the galaxies' gas reservoirs, implying a high star-formation efficiency (SFE) in the cluster core, and find short gas depletion timescales $\tau$<0.1-0.4~Gyrs for all galaxies. Interestingly, we do not see evidence for increased specific star-formation rates (sSFRs) in the cluster galaxies, despite their high SFEs and gas excitations. We find evidence for a large number of mergers in the cluster core, contributing a large fraction of the core's total star-formation compared with expectations in the field. We conclude that the environmental impact on the galaxy excitations is linked to the high rate of galaxy mergers, interactions and active galactic nuclei in the cluster core.

\end{abstract}


%%%%%%%%%%%%%%%%%%%%%%%%%%%%%%%%%%%%%%%%%%%%%%%%%%

%%%%%%%%%%%%%%%%% BODY OF PAPER %%%%%%%%%%%%%%%%%%

\section{Introduction}
The evolutionary path of galaxies between the peak epoch of mass assembly (z$\sim$2) and the local Universe is complex, with many different processes influencing galaxy properties throughout cosmic time. The physical mechanisms that lead to the cessation of star-formation, so-called galaxy `quenching', are still highly debated, and are among of the most important unsolved issues in galaxy evolution to date. Having well established that a population of early-type galaxies dominate the cores of galaxy clusters in the local Universe, the environmental effect on the evolution of galaxies has become a central aspect of current galaxy evolution studies. In order to better understand the origins and transformations of local cluster galaxies, we must trace these galaxies back in time, to the epoch of both peak galaxy and cluster assembly.

Several studies have suggested a reversal in the SFR-density relation at z$\sim$1 (e.g. \citealt{ref:D.Elbaz2007, ref:M.Cooper2008, ref:TonnesenCen2014}), where we see the star-formation rate (SFR) of galaxies start to increase with local galaxy density, contrary to what is observed in the Universe at z=0. High-z galaxy clusters are therefore a perfect laboratory for studying the environmental effects on the evolution of the galaxies that will form the most massive quenched galaxies in the local Universe. Direct observation of cluster galaxies at z$\sim$2 gives us insight into the relevance of (and link between) star-formation, active galactic nuclei (AGN) activity and mergers in over-dense regions. An increase in SFRs in clusters at z$\geq$1.5 with respect to local clusters has been observed, building up the bulk of the stellar mass in these huge structures (e.g. \citealt{ref:M.Hilton2010}, \mbox{\citealt{ref:M.Hayashi2011}}, \mbox{\citealt{ref:M.Brodwin2013}}, \mbox{\citealt{ref:K.Tran2015}}). It has also been shown that galaxy mergers, interactions and AGN activity are enhanced in high redshift (proto-)clusters (e.g. \mbox{\citealt{ref:J.Lotz2013}}, \mbox{\citealt{ref:C.Krishnan2017}}), and whether or not this causes a suppression or enhancement of the overall star-formation activity in these structures with respect to the coeval field is still a matter of debate.

Many processes have been invoked to explain the evolution and quenching of cluster galaxies. If galaxies in dense environments have an increased star-formation efficiency with respect to the coeval field, synonymous with rapid gas depletion, this could give rise to rapid galaxy evolution and quenching in cluster environments. Cluster-specific interactions with the intra-cluster medium (ICM) such as ram-pressure stripping, strangulation and harassment are effective at removing the atomic Hydrogen gas surrounding galaxies in clusters at z=0, and could potentially give rise to the population of quenched galaxies in the cores of local clusters if these processes are still efficient at z$>$0 (e.g. \citealt{ref:J.Gunn1972}, \citealt{ref:L.Aguilar1985}, \citealt{ref:Y.Peng2015}). However, the direct environmental effect on the fuel for star-formation, molecular gas (\Htwo), is not well understood, and obtaining observations of the star-forming gas in these systems becomes increasingly difficult at higher redshifts.

This study investigates the environmental effect on the cold gas reservoirs of galaxies in one of the highest redshift spectroscopically-confirmed, X-ray detected clusters discovered to date, \Cl\ at z=1.99 \mbox{\citep{ref:R.Gobat2011}}. This is an important time for both galaxy mass assembly and cluster evolution, as it is the last epoch before star-formation must rapidly quench to form the massive, passive galaxies that dominate the cores of clusters seen at lower redshift (e.g. \mbox{\citealt{ref:R.Gobat2008}}, \mbox{\citealt{ref:C.L.Mancone2010}}, \mbox{\citealt{ref:A.Vanderwel2010}}). Unlike the less evolved proto-clusters more commonly found at z$>$2, \Cl\ shows extended X-ray emission originating from hot plasma in the ICM, and already hosts a population of red, quiescent galaxies in its core alongside a number of highly star-forming galaxies. These passive galaxies are in fact starting to form a red sequence, a defining signature of dense environments, as observed in local galaxy clusters (e.g. \mbox{\citealt{ref:L.Spitler2012}}, \mbox{\citealt{ref:V.Strazzullo2016}}). \Cl\ also contains a diverse star-forming galaxy population, with a number of very low-metallicity, highly star-forming galaxies identified towards the edges of the cluster, possibly due to large-scale gas inflow \citep{ref:F.Valentino2015}. There are also two active galactic nuclei confirmed in the cluster, thought to be powering a vast Lyman-$\alpha$ bubble in the core \citep{ref:F.Valentino2016}. \Cl\ is a typical progenitor of present-day galaxy clusters with a mass of $\sim$5$\times$10$^{13}$M$_{\odot}$ \citep{ref:R.Gobat2011}, and to date has tens of spectroscopically confirmed members \mbox{\citep{ref:R.Gobat2013, ref:V.Strazzullo2013}}. It is therefore a perfect candidate for studying the effects of a maturing cluster environment on the star-formation and evolution of galaxies at high-z.

Thanks to recent developments in the capabilities of sub-mm and radio observatories, increasing numbers of studies have started to emerge focussing on the molecular gas content of field galaxies at high redshift (e.g.  \mbox{\citealt{ref:A.Saintonge2011}}, \citealt{ref:E.Schinnerer2016}, \mbox{\citealt{ref:M.Huynh2017}}, \mbox{\citealt{ref:L.Tacconi2017}}). Use of sub-mm and radio data allows us to measure both the dust and cold molecular gas contents of galaxies, essential to understanding the baryonic processes shaping the galaxies. The H$_{2}$ gas content of the interstellar medium (ISM) is measured indirectly, based on the tracer molecule $^{12}$CO (CO), via the CO-to-\Htwo\ conversion factor $\alpha_{\rm CO}$, (see e.g. \citealt{ref:A.Bolatto2013}). Measurement of the extended cold gas reservoir in galaxies is estimated from the J=1-0 transition of CO, where J is the rotational quantum number of the electron state within the CO molecule. The J=1-0 transition is important in the context of galaxy evolution as it gives an indication of the total amount of fuel that is available to form stars, and therefore the time until the galaxy will quench (assuming that star-formation continues at its current rate and gas reservoirs are not replenished through accretion). Different populations of star-forming galaxies have been shown to have different relative abundances of molecular gas in different excitation states, with local Ultra Luminous Infra-Red Galaxies (ULIRGS) and starburst (SB) galaxies having relatively more dense gas driving their high levels of star-formation, possibly due to mergers and interactions compacting the gas.

For the majority of field galaxies, the increase of SFR with increasing look-back time comes hand-in-hand with an increase in molecular gas fraction (and, to a lesser extent, star-formation efficiency). The interplay between these has been encapsulated in so-called `scaling relations' in several recent studies referring to large, statistical samples (e.g. \citealt{ref:P.Santini2014, ref:M.Sargent2014, ref:R.Genzel2015, ref:N.Scoville2017, ref:L.Tacconi2017}). Similarly, cosmological simulations and semi-analytical models have started to predict the relative importance of molecular gas as a function of redshift, and find that the role of H$_{2}$ in galaxies becomes increasingly dominant at higher redshifts (e.g. \citealt{ref:C.Lagos2011, ref:C.dPLagos2015}). However, similarly complete studies have not yet been achieved at z$>$1.5 in cluster environments. There is a real need for convincing detections of molecular gas in high redshift clusters, that will allow us to better characterise cluster galaxies and identify the impact of environment on the star-formation processes at these early times.

Previous studies of galaxies in clusters tend to be split between mature cluster environments at z$\leq$1.6, and very gas-rich environments at earlier epochs. \citet{ref:M.Lee2017} recently measured the gas fractions and environmental trends in a proto-cluster at z=2.49, finding the molecular gas masses and fractions to be on average the same as in similar field galaxies, and that the gas fraction increases with decreasing galaxy number density per unit area. \mbox{\citet{ref:T.Wang2016}} also study the environmental effects at high redshift, in the most distant X-ray detected cluster to date at z$\sim$2.5. This cluster is dominated by gas-rich, highly star-forming galaxies with very few quiescent galaxies yet evolved in the core. They find short gas depletion timescales ($\tau_{dep}$) of order $\sim$200~Myrs, driven by a high fraction of starburst galaxies amongst the numerous star-forming galaxies. On the other hand, \citet{ref:H.Dannerbauer2017} reported the first CO[1-0] detection of a normal, star-forming disk galaxy in a proto-cluster at z=2.15, and compiling previous CO[1-0] detections found no environmental dependence of star-formation efficiency in proto-cluster galaxies. \citet{ref:A.G.Noble2017} recently studied three separate clusters at z$\sim$1.6 containing a higher fraction of quenched galaxies. They found evidence for star-forming galaxies with higher gas fractions than in the field, and with significantly longer gas depletion timescales, averaging $\tau_{dep}$=1.6~Gyrs. They find that the evolution of gas fractions in cluster galaxies mimics that of the field, and suggest that dense cluster environments may encourage the formation of molecular gas compared to the field, or somehow prevent a portion of this gas from actively forming stars. \citet{ref:G.Rudnick2017} were also able to obtain deep CO[1-0] data for a forming cluster at z$\sim$1.6, and find that the two detected galaxies have gas fractions and star-formation efficiencies consistent with field scaling relations.

Reaching down to z$\sim$1.46, two groups studying cluster XCS~J2215 both conclude that there is a suppression of the cold molecular gas in the core, having suffered environmental gas depletion that will lead to the quenching of star-formation on short timescales \citep{ref:MHayashi2017, ref:S.Stach2017}. \mbox{\citet{ref:S.Stach2017}} suggest that environmental processes may be stripping the more diffuse ISM, reducing the gas fractions in the galaxies and increasing their ratio of excited to diffuse gas.

The work presented here probes the molecular gas content of cluster galaxies, in a higher redshift environment than the majority of similar studies. \Cl\ provides a high-density environment for comparison with field galaxies during an important transitional phase of cluster evolution, allowing us to place better constraints on theoretical models of galaxy evolution in cluster progenitors (e.g. \citealt{ref:A.Saro2009}). In Section~\ref{sec:method}, we describe the available datasets and their reduction routines. In Section~\ref{sec:results}, we present our results for several excitations of CO in our cluster galaxies, and derive physical galaxy properties such as star-formation rates and dust masses. In Section~\ref{sec:discussion}, we compare our findings with similar field studies and assess the environmental impact on the molecular gas properties of the cluster. We summarise our conclusions in Section~\ref{sec:conc}. Throughout this paper, a $\Lambda$CDM cosmology is adopted, with H$_{0}$=70kms$^{-1}$Mpc$^{-1}$, $\Omega_{M}$=0.3 and $\Omega_{\Lambda}$=0.7. We use a Chabrier initial mass function (IMF, \citealt{ref:G.Chabrier2003}).

\begin{figure*}
    \centering
    \begin{minipage}{175mm}
    \includegraphics[width=0.5\textwidth]{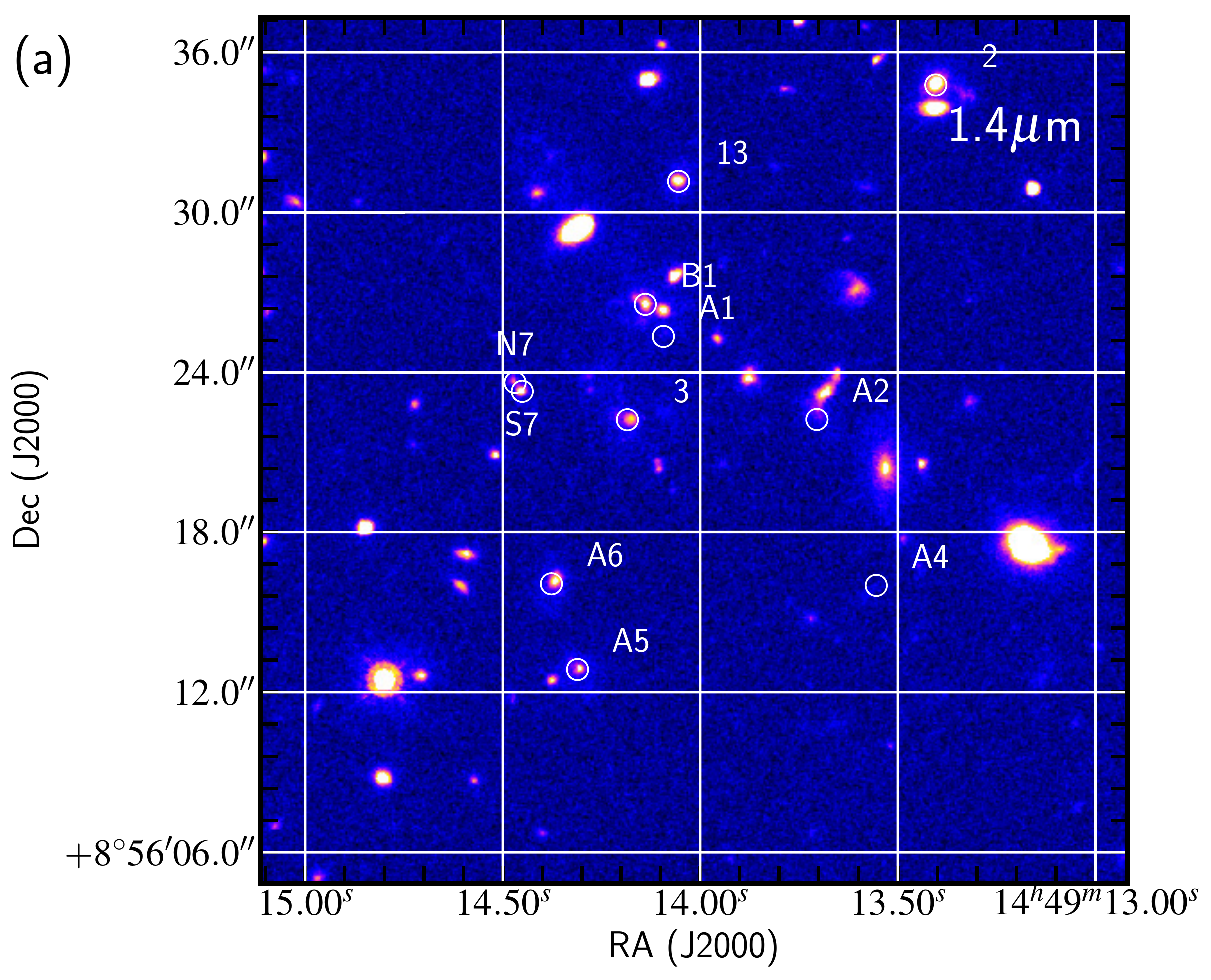}\hfill 
    \includegraphics[width=0.5\textwidth]{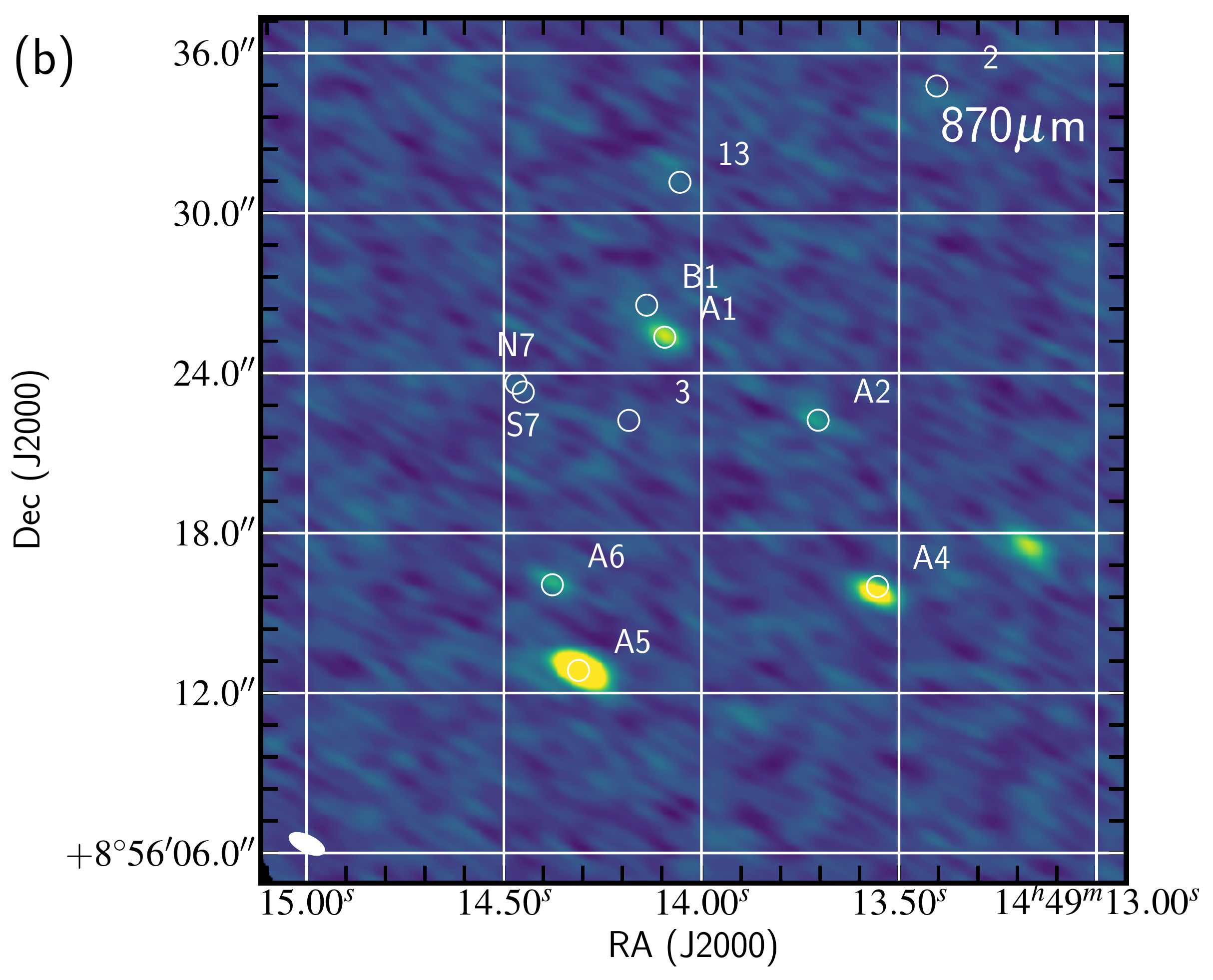}\hfill
    \end{minipage}
    \begin{minipage}{175mm}
    \centering
    \includegraphics[width=0.5\textwidth]{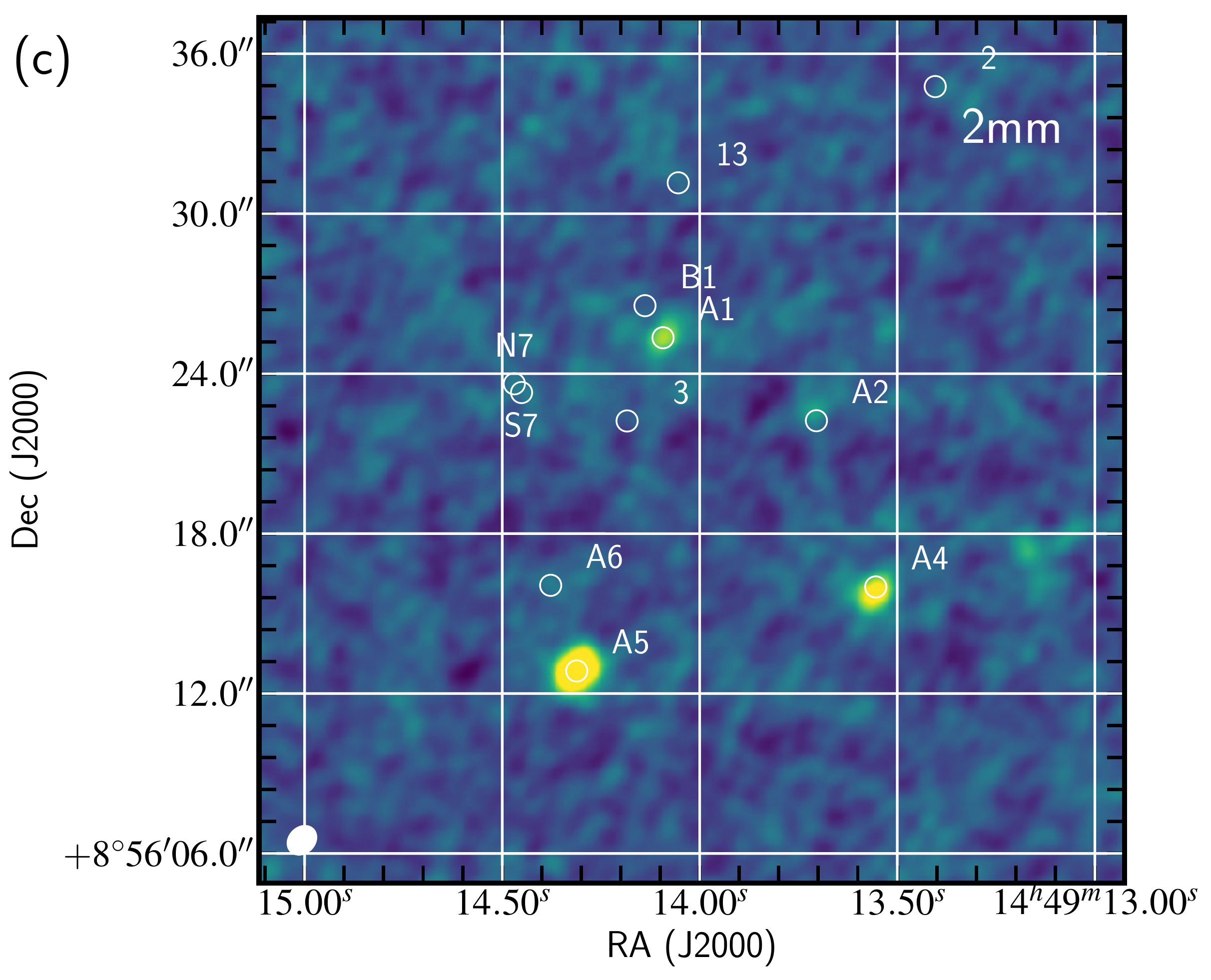}\hfill
    \includegraphics[width=0.5\textwidth]{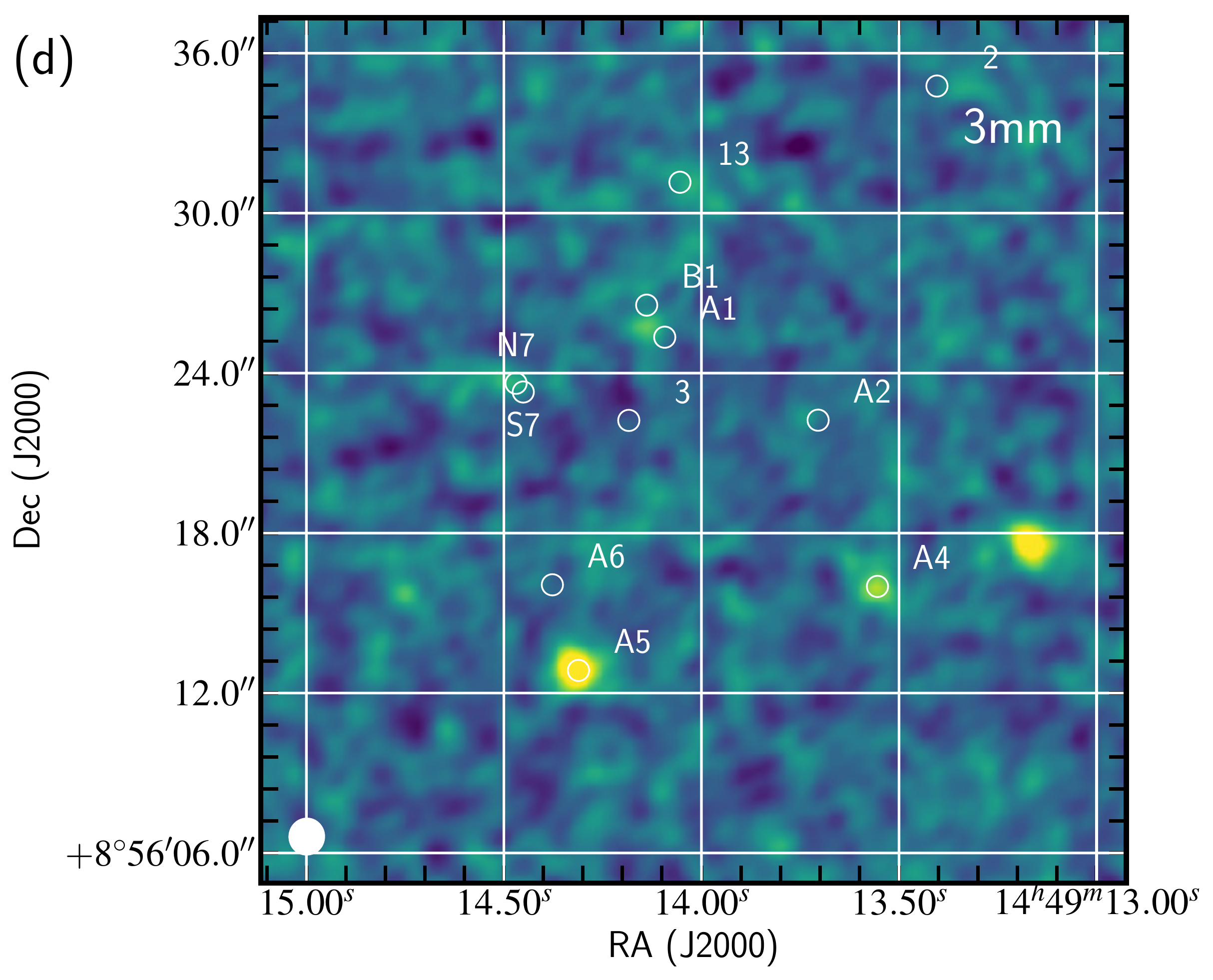}\hfill
    \end{minipage}
    \begin{minipage}{175mm}
    \centering
    \includegraphics[width=0.5\textwidth]{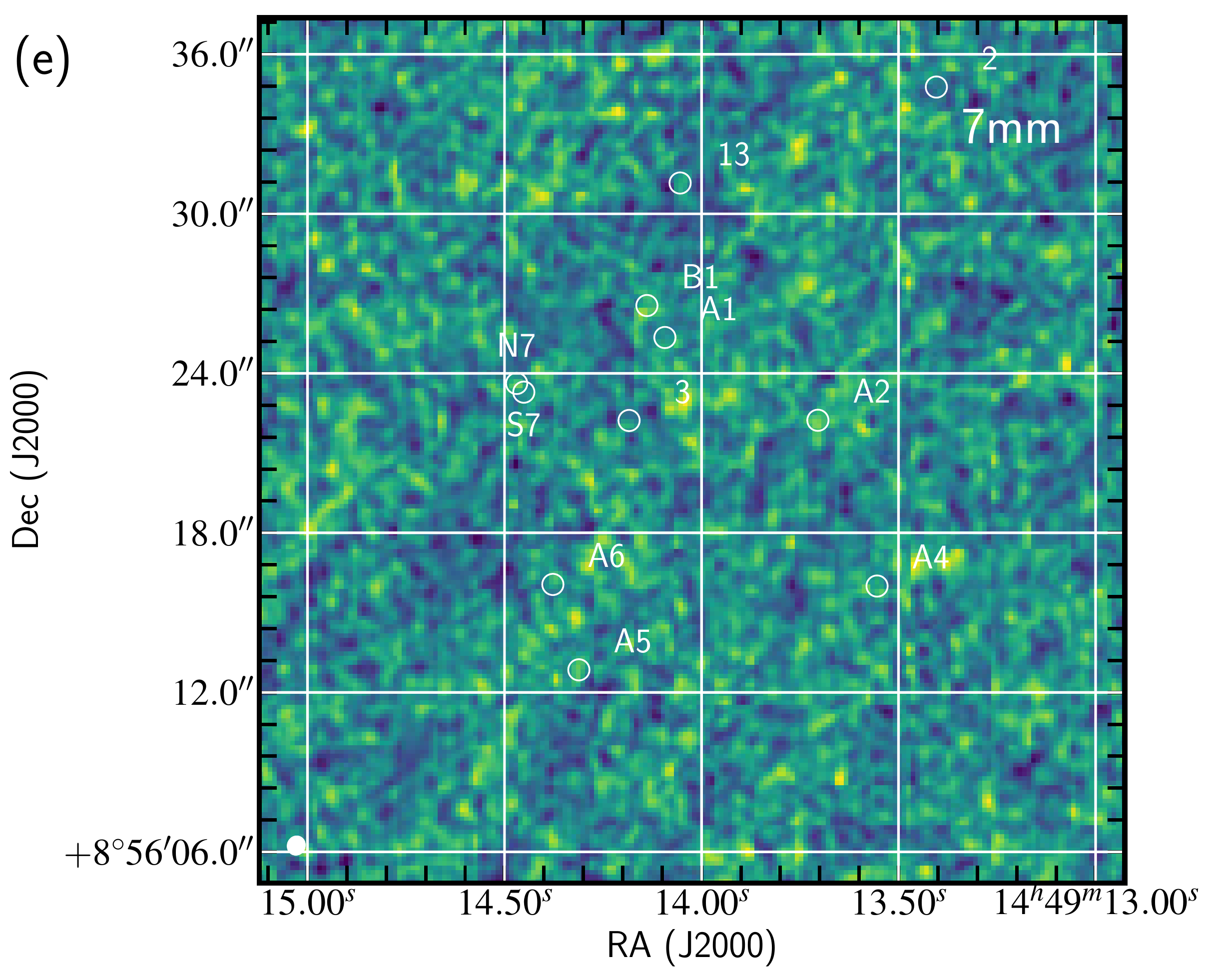}\hfill
    \includegraphics[width=0.5\textwidth]{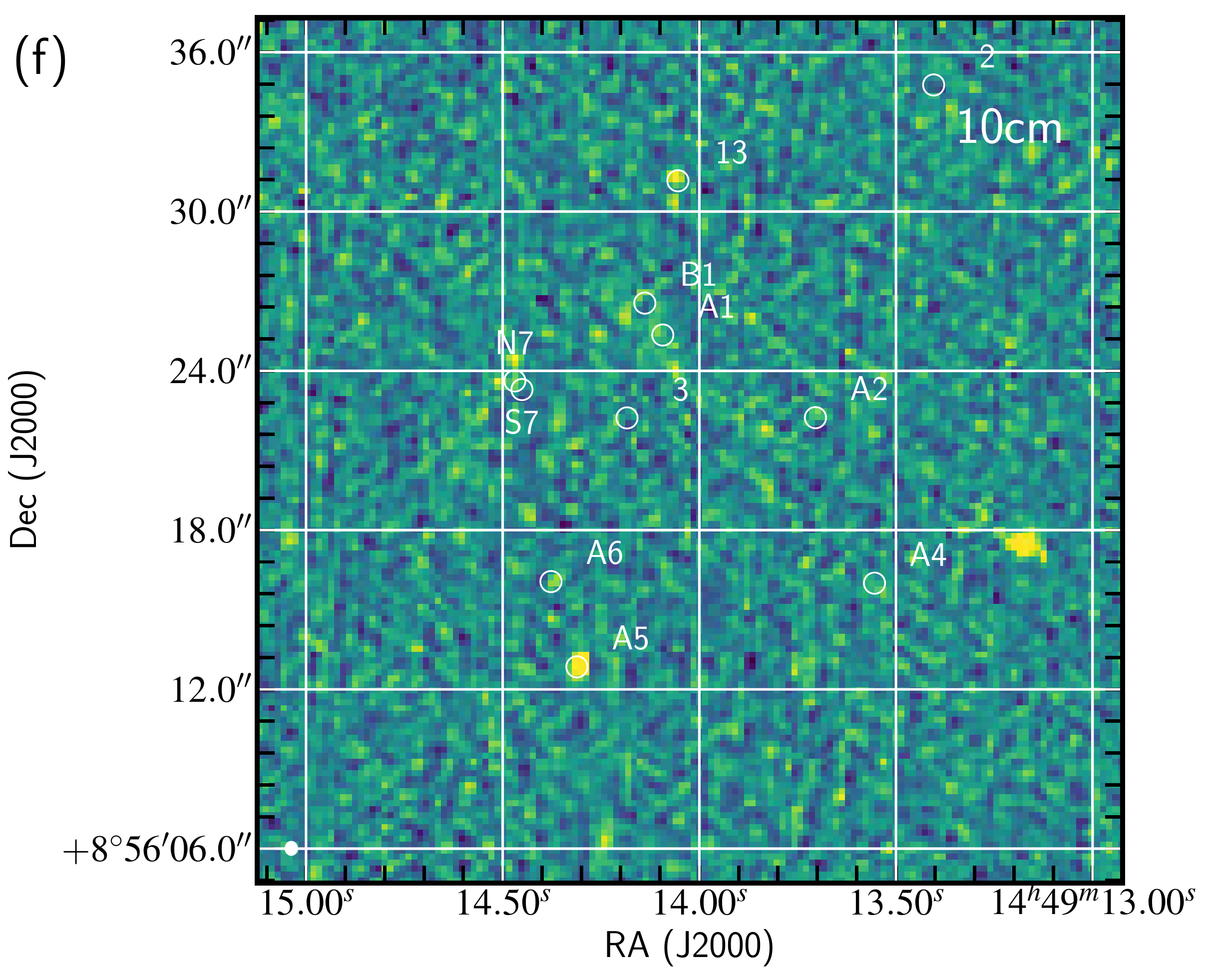}\hfill
    \caption{Continuum images of \Cl{}: (a) near infra-red 1.4$\micron$ (HST/WFC3), (b) 870$\micron$ (ALMA), (c)-(d) 2mm and 3mm (ALMA), (e)-(f) 7mm and 10cm (JVLA). Any line emission present has not been removed from the images. The signal-to-noise of the galaxies can be seen to decrease towards the longer wavelengths. In the 10cm image, only the deep, high-resolution A-configuration data is shown, without the C-configuration. The images have not been corrected for the primary beam attenuation of the instruments. The galaxies indicated show the positions of the galaxies for which we find CO[4-3] detections, with the exception of A4 and A5, which are discussed in Section~\ref{sec:A4A5}, and ID 2, which is only detected at 870$\micron$. A bright foreground galaxy can be seen to the West of A4, which has been spectroscopically confirmed at z=1.3, and will therefore not be discussed in this paper.}  
    \label{fig:ContData}
    \end{minipage}
\end{figure*}

\section{Observations and data reduction}
\label{sec:method}
We observed \Cl\ over a wide range of radio and submillimetre frequencies, presented below. This allowed us to measure the line fluxes of three separate CO transitions and the continuum underneath these lines, in order to study the molecular gas content and excitations of the galaxies. The ALMA band 4 and band 3 observations covered the expected frequency of the CO[4-3] and CO[3-2] lines respectively, and the JVLA Ka-band data targeted the CO[1-0] line.

We also study the submillimetre-radio spectral energy distributions of the galaxies, using the continuum underneath the CO lines from the above observations, as well as ALMA band~7 and JVLA 3~GHz continuum observations. These observations were used to calculate the dust contents of the galaxies.

The observing runs, data reduction and flux extraction performed on each of these datasets are outlined in the following subsections, and summarised in Table~\ref{tab:ObsInfo}.

\subsection{ALMA Band 4 Observations}
We collected Atacama Large Millimetre Array (ALMA) band 4 observations of the cluster in two single pointings during Cycle 3 (Project ID: 2015.1.01355.S, PI: V. Strazzullo). One of these pointings was a deep observation of the core of \Cl{}, and the other was a shallower observation of the low metallicity members in the outskirts of the cluster \citep{ref:F.Valentino2015}. Observations were completed in May 2016, for a total on-source time of $\sim$2h in the core, and $\sim$30~minutes in the outskirts. The band 4 observations target the CO[4-3] line.

For both pointings, the CO[4-3] line at this redshift was contained in a spectral window (SPW) centered at 153.94~GHz, with a bandwidth of 1.875~GHz and a spectral resolution of $\sim$1953.1~kHz ($\sim$3.8~\kms). The remaining three SPWs were set up for continuum observations, centered at 152.00~GHz, 140.00~GHz and 141.70~GHz respectively, each with bandwidths of 1.875~GHz and spectral resolutions of 62.5~MHz ($\sim$123.27\kms). The FWHM of the primary beam at this frequency is $\sim$40.9". For the core pointing, quasar J1550+0527 was used for flux calibration, and the synthesised beam was 1.19"$\times$0.96" at a position angle (PA) of -44.8\degree. The 1$\sigma$ noise for the data in the core pointing was 10~mJy\kms\ over 100\kms. Over the entire bandwidth, the continuum root-mean-squared (RMS) noise was 7.98$\mu$Jy/beam. From this, we were able to detect an equivalent star-forming population down to a 5$\sigma$ detection limit at z=2 of $\sim$32M$_{\odot}$yr$^{-1}$, for a Main Sequence galaxy with a CO[4-3] linewidth of 400\kms.

For the off-center pointing of the low-metallicity galaxies, Titan was used for flux calibration. The synthesised beam was 1.17"$\times$0.97" at PA = -34.1\degree, with 1$\sigma$ noise of 23.7mJy\kms\ over 100\kms. The RMS over the whole continuum bandwidth was 19.1$\mu$Jy/beam. From this, we were able to detect an equivalent star-forming population down to a 5$\sigma$ detection limit at z=2 of $\sim$75M$_{\odot}$yr$^{-1}$, for a Main Sequence galaxy with a CO[4-3] linewidth of 400\kms.

\subsection{ALMA Band 3 and 7 Observations}
ALMA band 3 observations were taken in a single pointing on the cluster core in Cycle 1 (Project ID: 2012.1.00885.S, PI: V.~Strazzullo). Observations were completed in June 2015 for a total on-source time of $\sim$2h. For band 3 observations, Titan was used for flux calibration. The CO[3-2] line was contained in a SPW centered at 114.93~GHz. The other SPWs were centered at 112.94~GHz, 102.90~GHz and 101.00~GHz respectively. As with the CO[4-3] observations, the line-free SPWs were used to measure continuum fluxes. Each SPW covered a bandwidth of 1.875~GHz and had a spectral resolution of 1953~kHz (5.09 - 6.00\kms). The FWHM of the primary beam was $\sim$54.8", with a synthesised beam FWHM of 1.36"$\times$1.31" at PA~=~-5.8\degree. The noise over 100\kms\ was 12~mJy\kms\ and the continuum RMS was 8.36$\mu$Jy/beam.

Band 7 observations were taken as part of the same project, using seven single pointings centered at 345~GHz (870$\micron$) to cover the cluster core over an 0.3arcmin$^{2}$ area. Observations were completed in December 2014 for a total on-source time of $\sim$2.3h. Quasar J1337-1257 was used for flux calibration. Band 7 observations allowed us to derive continuum SFRs for the cluster galaxies, down to a 3$\sigma$ SFR detection limit at z=2 of $\sim$21M$_{\odot}$yr$^{-1}$, for a Main Sequence galaxy. These observations also allowed us to put constraints on the normalisation of the submillimetre portion of the spectral energy distributions (SEDs). All four SPWs were set up for continuum observations, covering a bandwidth of 7.5~GHz between $\sim$338-340~GHz and $\sim$350-352~GHz. The spectral resolution for these observations was lower, at 31.25~MHz (26.6 - 27.7\kms). The RMS noise was 67.5$\mu$Jy/beam, with a FWHM resolution of the synthesised beam of 1.41"$\times$0.62" at PA~=~64.5\degree.

\subsection{JVLA Ka-band Observations}
Deep high frequency data of \Cl\ was taken using the Karl G. Jansky Very Large Array (JVLA), in order to measure the CO[1-0] lines in the cluster galaxies (project code: 12A-188, PI: V. Strazzullo). These observations were completed in March 2012, for a total on-source time of $\sim$15.5h in configuration C. Quasar J1331+3030 was used for flux calibration. A total bandwidth of 2.048~GHz centered at 38.15~GHz was observed over 16 SPWs in a single pointing, with a spectral resolution of 4000~kHz (29.2 - 32.3\kms). The CO[1-0] line falls in either SPW 10 or 11 or is split across both depending on the redshift, with the rest of the SPWs being used for continuum observations. The FWHM of the primary beam was $\sim$79.1", with a synthesised beam of 0.70"$\times$0.64" at PA~=~-20.2\degree. The 1$\sigma$ spectral noise over a 100\kms\ width was 4~mJy\kms\, with a continuum RMS of 3.30$\mu$Jy/beam. The depth of these observations also allowed for the 5$\sigma$ detection of CO[1-0] emission from Main Sequence star-forming galaxies with SFR$\sim$33M$_{\odot}$yr$^{-1}$ assuming a CO[1-0] linewidth of 400\kms.

\subsection{JVLA S-band Observations}
Continuum observations centered at 3~GHz were also obtained for the cluster using the JVLA (project code: 12A-188, PI: V. Strazzullo). These observations were carried out between February and November 2012, for a total on-source time of $\sim$1.1h in configuration C and $\sim$11.4h in configuration A. Quasar J1331+3030 was used for flux calibration. These observations were used to model the radio frequencies of the galaxies' SEDs, and calculate SFRs from the radio luminosities. The primary beam FWHM of the 3~GHz observations was 15', and the FWHM of the synthesised beam at this wavelength was $\sim$0.50"$\times$0.44" at PA~=~-9.61\degree\ for the A-configuration data, with an RMS noise of 1.6$\mu$Jy/beam. In the C-configuration, the synthesised beam was $\sim$7.0"$\times$6.1" at PA~=~-16.47\degree, with an RMS noise of 8.8$\mu$Jy/beam.\newline

The ALMA band 4, band 3, band 7, and JVLA Ka-band data were calibrated using the standard reduction pipeline in Common Astronomy Software Applications (CASA, \citealt{ref:J.P.McMullin2007}). For the band 3 and 7 observations, CASA version 4.2.2 was used for reduction. For band 4, version 4.5.3 was used. For the JVLA K-band data, version 4.4.0 was used. For the JVLA S-band data, the standard reduction pipeline was used to calibrate the A-configuration data, and manual calibration using CASA 4.6.0 was performed for the shorter configuration-C scheduling blocks.

\begin{table*}
\begin{tabular}{cccccccc}
\hline
Telescope & Pointing & Date & Ave. Freq. & Bandwidth & On-source time & FWHM$_{syn. beam}$ & PA$_{syn. beam}$\\
 & & Completed & (GHz) & (GHz) & (h) & (") & ($\degree$)\\
\hline
ALMA & Core, mosaic & December 2014 & 345.0 & 7.5 & 2.3 & 1.41$\times$0.62 & 64.5\\
ALMA & Core, single & May 2016 & 146.9 & 7.5 & 2.0 & 1.19$\times$0.96 & -44.8\\
ALMA & Outskirts, single & May 2016 & 146.9 & 7.5 & 0.4 & 1.17$\times$0.97 & -34.1\\
ALMA & Core, single & June 2015 & 107.9 & 7.5 & 2.0 & 1.36$\times$1.31 & -5.8\\
JVLA & Core, single & March 2012 & 38.2 & 2.0 & 15.5 & 0.70$\times$0.64 & -20.2\\
JVLA & Core, single & November 2012 & 3.0 & 2.0 & 12.5 & 0.50$\times$0.44 & -9.61\\
\end{tabular}
\caption{Summary of the observations discussed in this paper. The telescope, pointing target (\Cl\ core/outskirts) and mode (pointing/mosaic), date, average frequency, bandwidth, on-source time, FWHM and position angle of the synthesised beam are shown. For the 3~GHz data, the synthesised beam given is for the A-configuration data. The on-source time given is the addition of the A- and C-configuration data.}
\label{tab:ObsInfo}
\end{table*}

\begin{figure*}
    \centering
    \begin{minipage}{175mm}
    \includegraphics[width=0.25\textwidth]{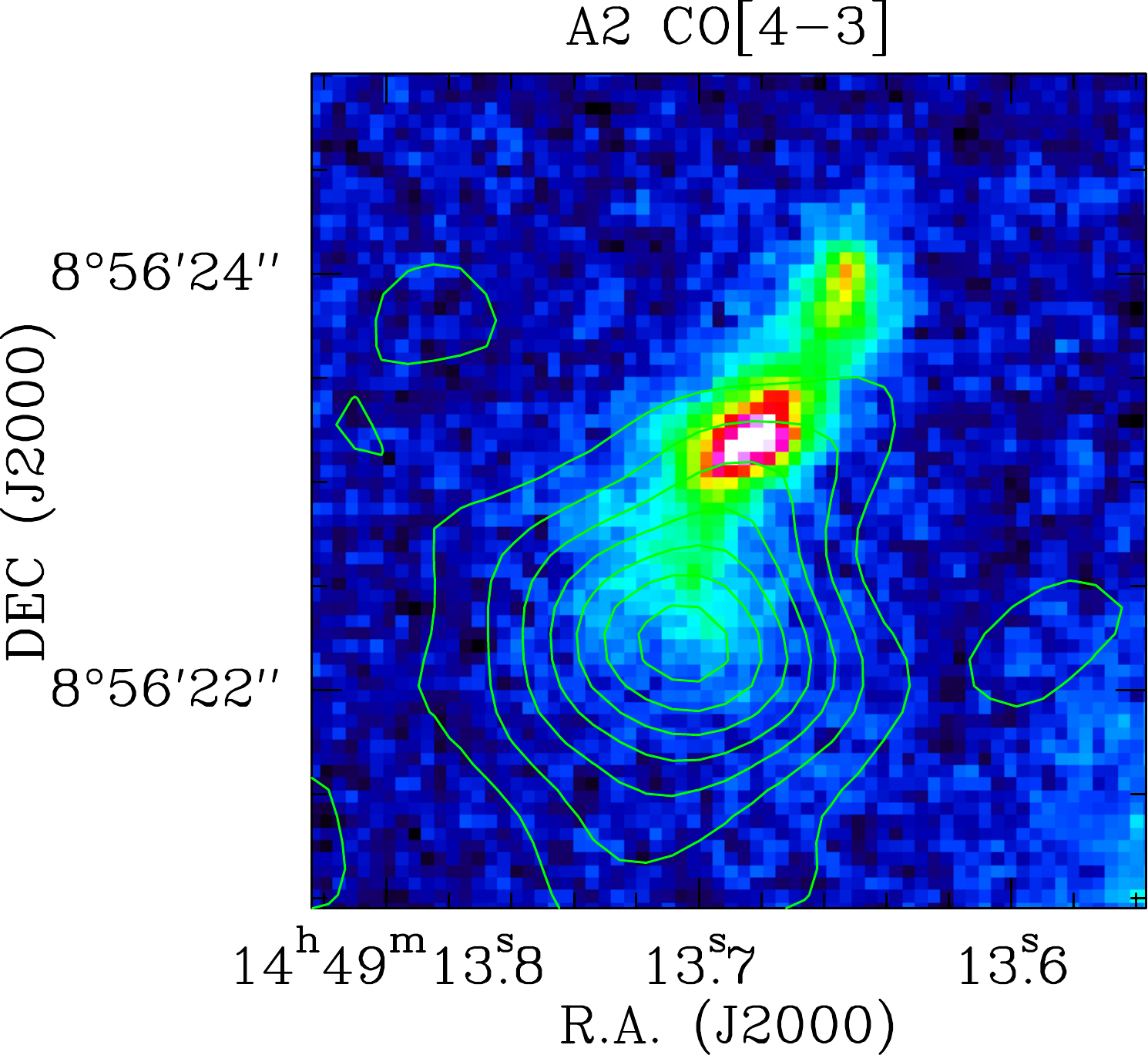}\hfill
    \includegraphics[width=0.25\textwidth]{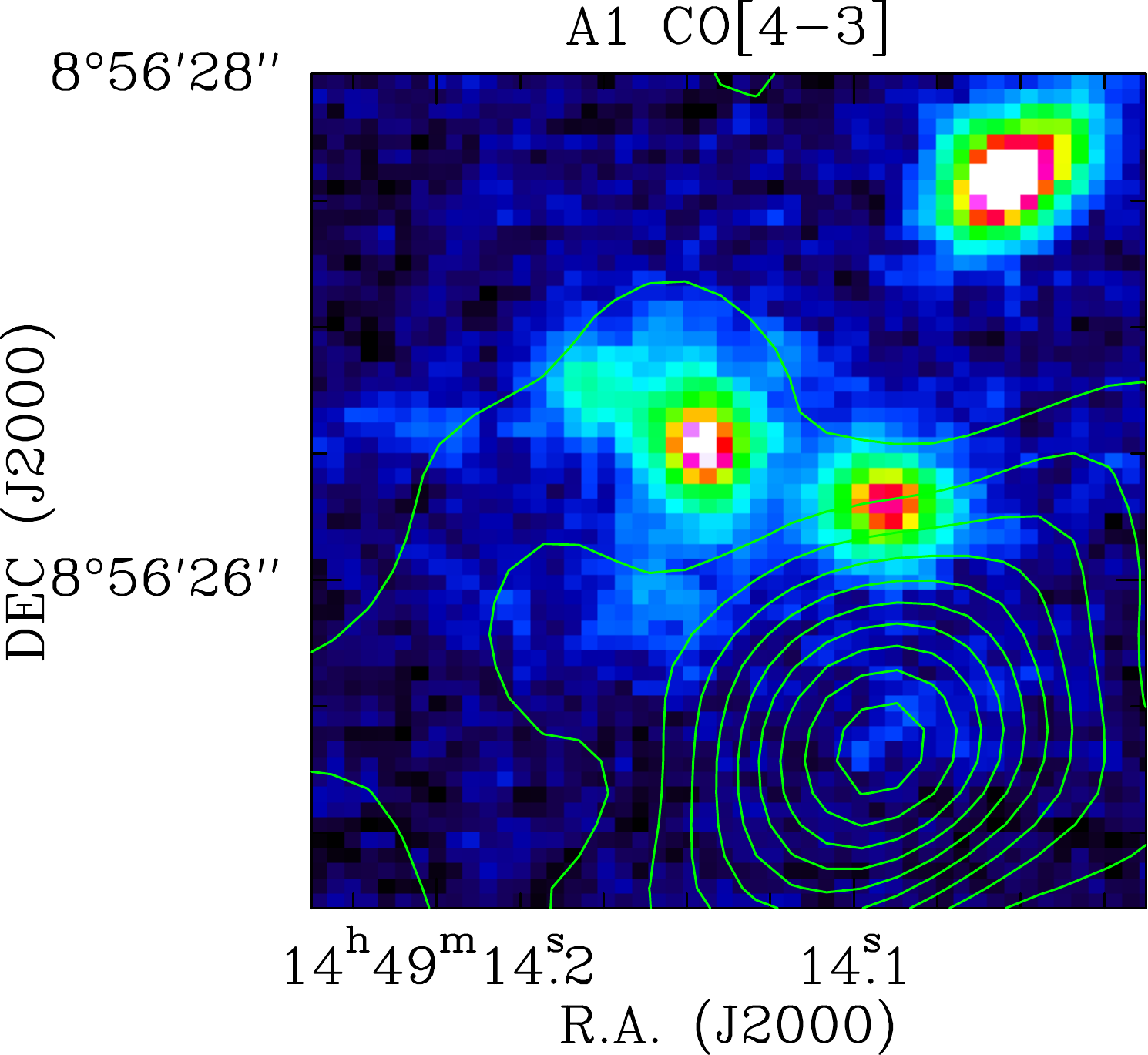}\hfill
    \includegraphics[width=0.25\textwidth]{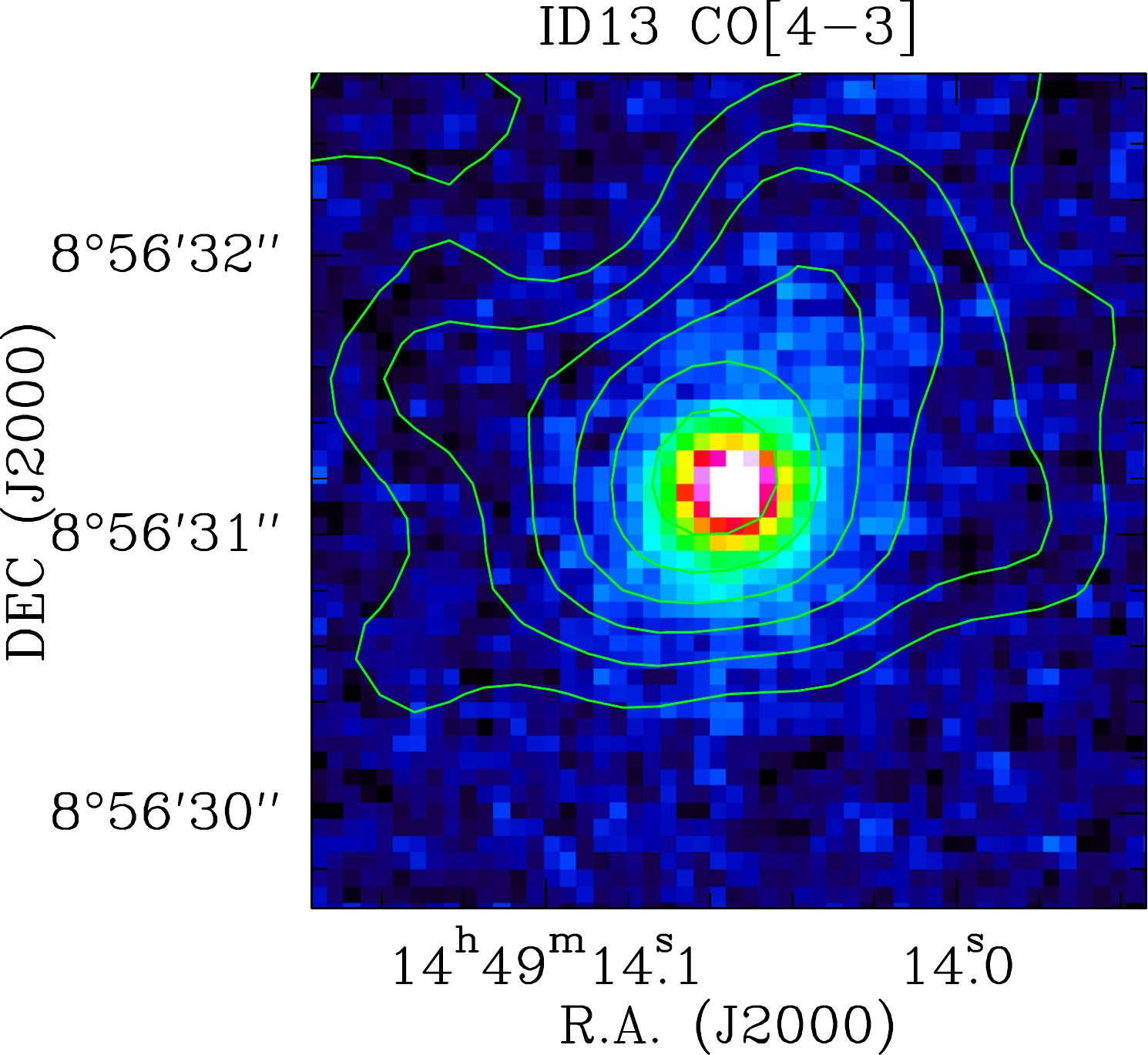}\hfill
    \includegraphics[width=0.25\textwidth]{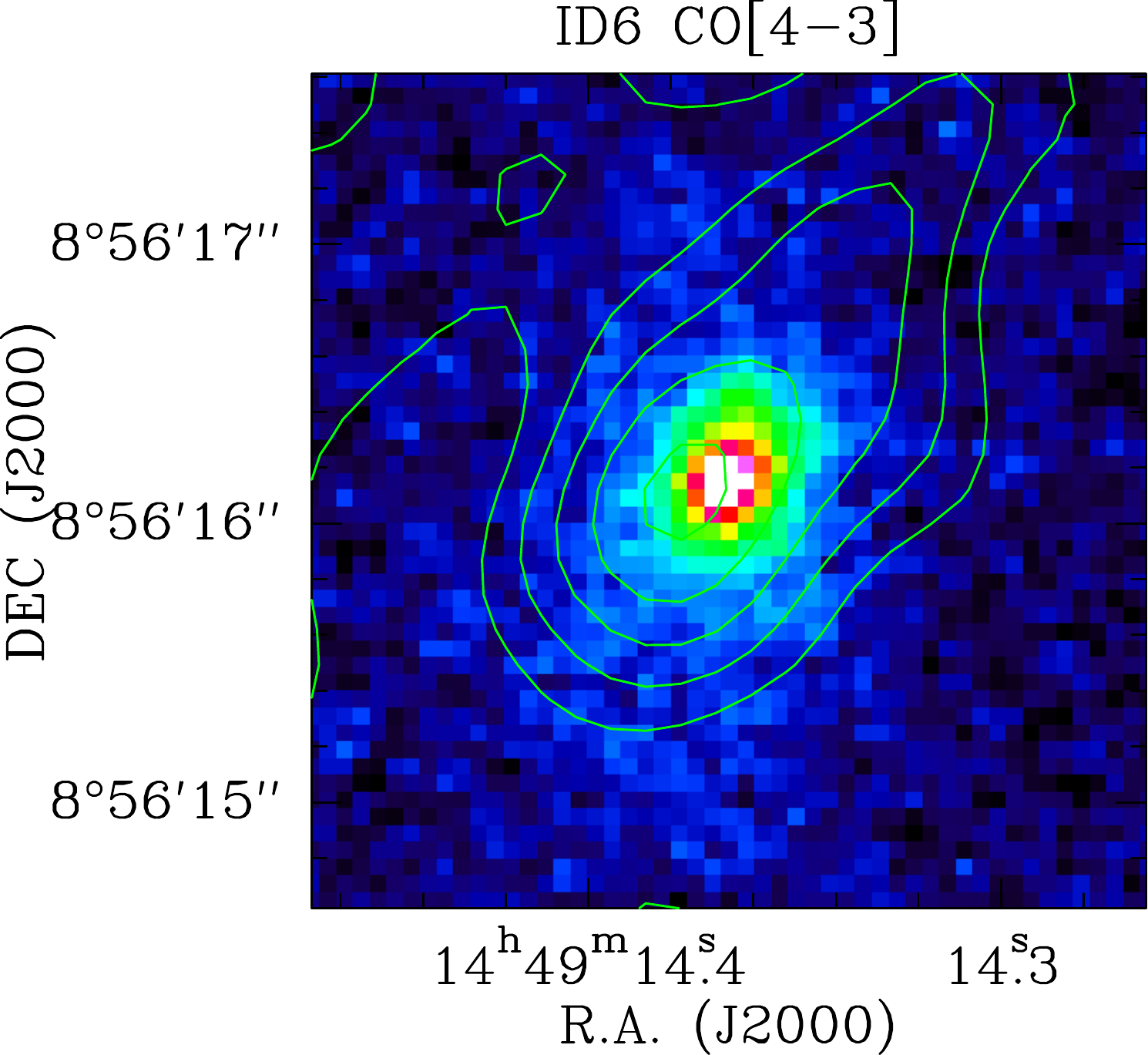}\hfill
    \end{minipage}
    \begin{minipage}{175mm}
    \centering
    \includegraphics[width=0.25\textwidth]{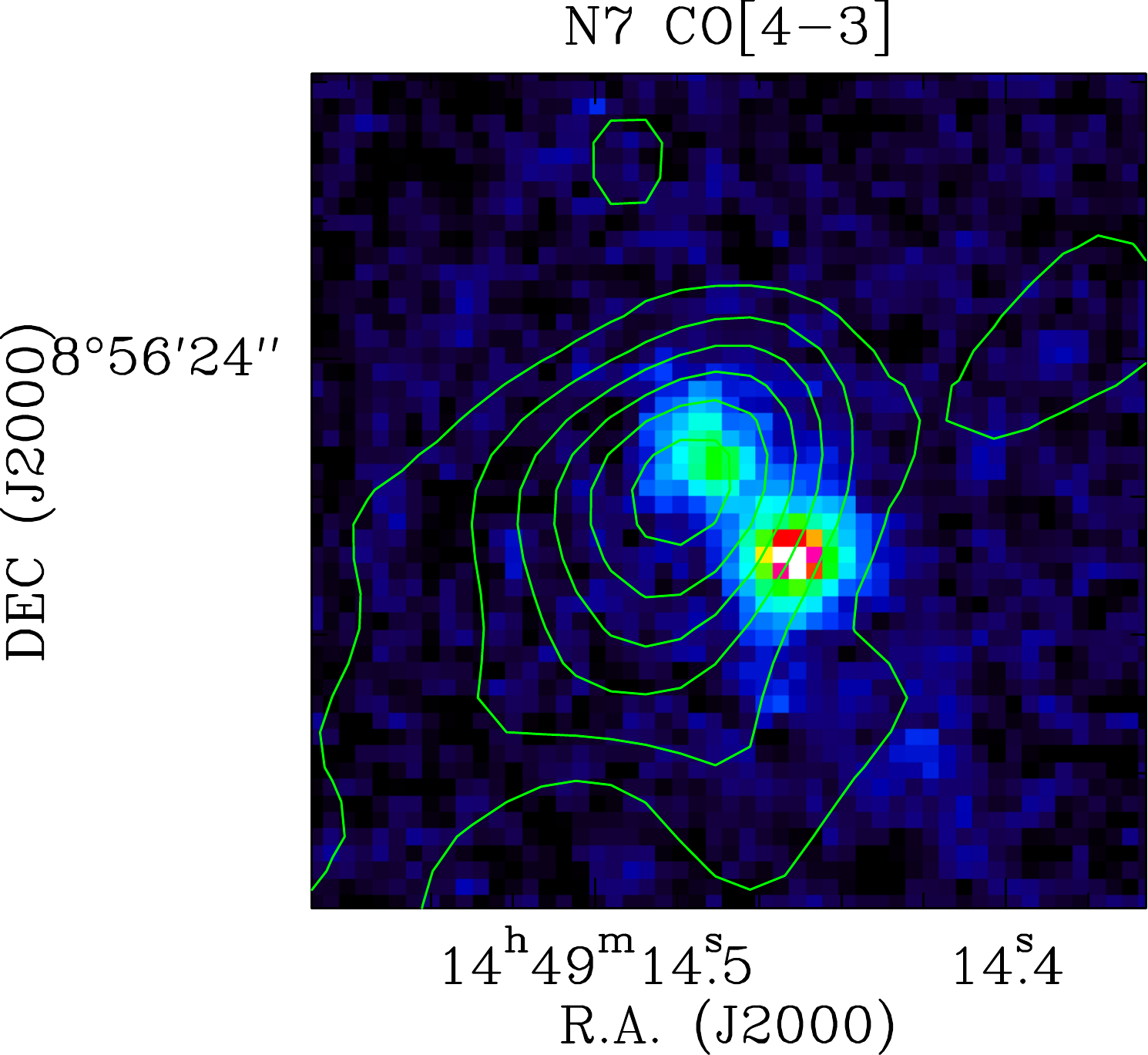}\hfill
    \includegraphics[width=0.25\textwidth]{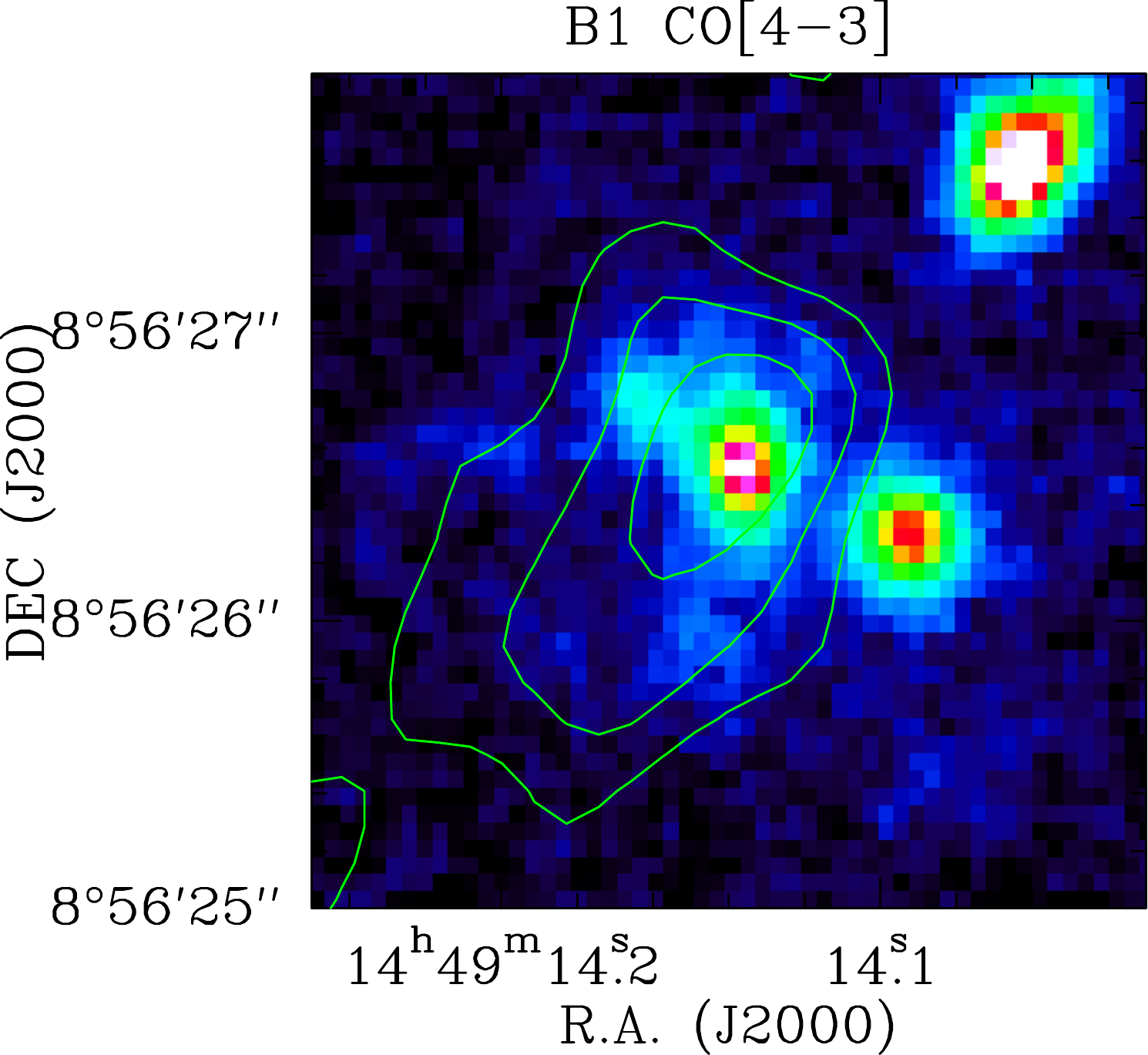}\hfill
    \includegraphics[width=0.25\textwidth]{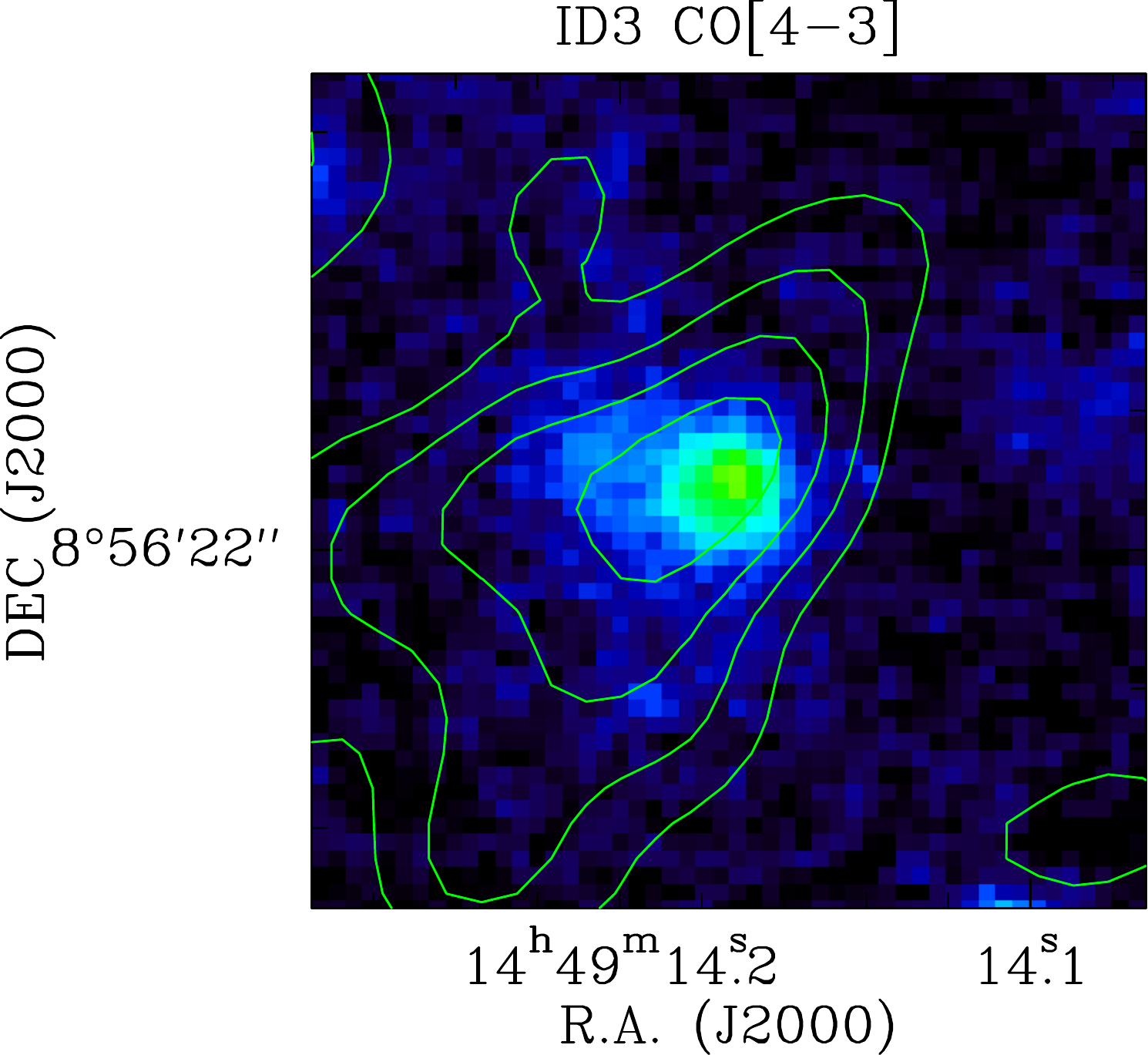}\hfill
    \includegraphics[width=0.25\textwidth]{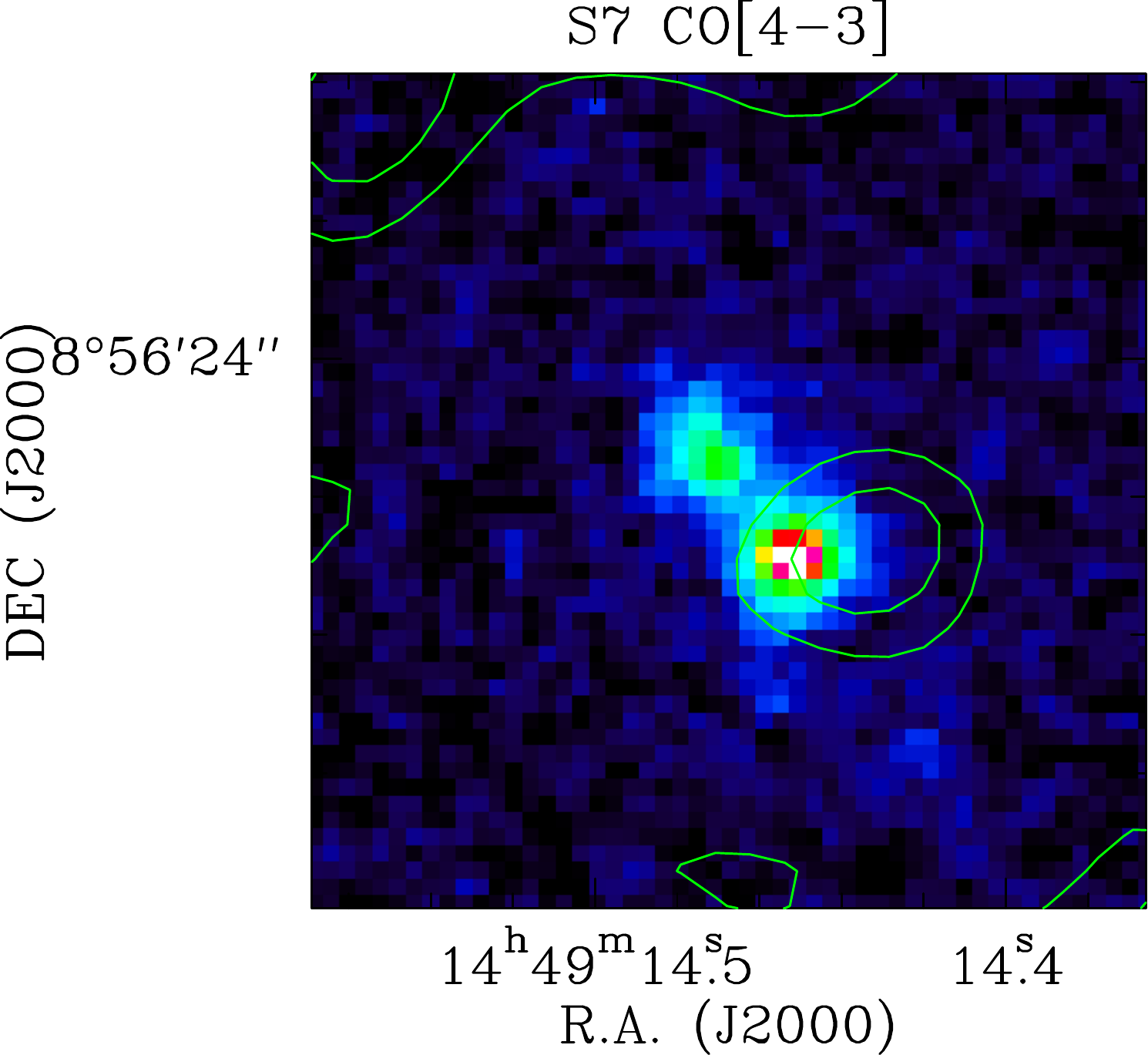}\hfill
    \caption{Rest-frame optical HST/WFC3 images of the galaxies in \Cl\ with CO[4-3] detections, overlaid with the CO[4-3] line contours in increments of 1$\sigma$. All contours start at 1$\sigma$. A large offset can be seen between bright optical counterparts and the dense CO[4-3] gas in A1 and A2.}
    \label{fig:A126Colour}
    \end{minipage}
\end{figure*} 

\subsection{Spectral flux extraction}
\label{sec:method:extrac}
The spectral line fluxes for all three CO transitions (CO[4-3], CO[3-2] and CO[1-0]) were extracted using the Gildas\footnote{http://www.iram.fr/IRAMFR/GILDAS} tool \texttt{uvfit} directly on the uv-visibility data. The line-free continuum fluxes underneath the CO lines were also extracted from the uv-visibilities.

We see in Figs~\ref{fig:ContData} and \ref{fig:A126Colour} (see also galaxy IDs) that for spectroscopically confirmed cluster members with HST/WFC3 counterparts, the signals in the submillimetre maps are consistent with the rest-frame optical positions, and no systematic offset is suggested by the data. Spectral extractions initially searching for the CO[4-3] line were therefore made at fixed rest-frame optical position of spectroscopically confirmed cluster members, and also at the positions of ALMA continuum detections (SNR$>$3). Once these spectra had been extracted, a line-searching algorithm was run over each 1D spectrum. The algorithm returned both the signal-to-noise (S/N, SNR) of the brightest line found in the spectrum, and the optimum velocity range to be considered as the linewidth, in order to maximise the S/N.

Having identified the galaxies with detected CO[4-3] lines in this way, the size and flux of the CO[4-3] emission in each galaxy was measured using the collapsed observational data across the optimised CO[4-3] linewidth. As stated, the flux extraction over this collapsed data was performed using \texttt{uvfit} directly on the uv-visibilities. In all cases except A1 and A2, the CO[4-3] line fluxes of the galaxies were measured at fixed rest-frame optical position in the collapsed line data. In the case of strong band 4 continuum detections with faint optical counterparts (A1 and A2, see Figs~\ref{fig:ContData} and \ref{fig:A126Colour}), the position of the emission was free to be optimised by the uv-fitting on the collapsed data. Sources that were not resolved in CO[4-3] were extracted as point-spread-functions (PSFs). Similarly, if the emission could be significantly resolved by \texttt{uvfit} (SNR$>$3 for the measured size in uv-space), the shape of the emission was fit as a 2D Gaussian with freely varying amplitude, shape and position angle. \texttt{uvfit} measures the sizes of sources by inspecting the relation between measured flux and uv-distance.

Given the crowded nature of the field, it is important to ensure that all fluxes are extracted without contamination from neighbouring sources. The \texttt{uvfit} tool allows simultaneous modelling of up to four sources at any time. We therefore first extracted the spectra of the brightest sources in the field of view simultaneously, and then subtracted these bright sources from the data. We then modelled the next brightest sources simultaneously in the data, with the first four sources already removed. This process was repeated iteratively until all of the sources had been modelled and then removed from the data. In this way, we ensured that flux from bright sources did not contaminate the extracted fluxes of fainter galaxies. We also verify that the order in which the sources were modelled does not significantly affect our results, as  the power contained in the side-lobes of the known ALMA beam contain only a fraction of the power of the beam core. As our detections are not extremely bright, the side-lobes of the PSF do not carry significant amounts of flux. Additionally, the small redshift variations between the cluster galaxies result in the spectral lines being offset in frequency. However, in the case of A1 and B1, line fluxes were extracted for both sources simultaneously in each collapsed dataset, to prevent any contamination between the two, due to their very small redshift separation. No other line or continuum sources were identified in our maps other than those modelled in this way. 

In order to remove the continuum contributions to the measured line fluxes, the line-free spectral channels were collapsed and averaged, and an average continuum flux was measured by \texttt{uvfit} at the same position and size as was done for the line flux. The frequency of the measured continuum was therefore close to the frequency mid-point of the total bandwidth. In order to find the continuum contribution under the line flux, we used a two-component model, $y=a\nu^{b}$, to model the continuum, where y is continuum flux and $\nu$ is frequency. We calculated the normalisation factor, $a$, from our average continuum flux and frequency, and extrapolated to the frequency under the line using a fixed spectral slope $b=3.5$. This was then subtracted from the measured line flux.

Once the flux, position and size of the line emission had been measured from the collapsed data in this way, the full 1D spectra were re-extracted at fixed position and size in each spectral channel, binned into velocity slices of 30\kms. These 1D spectra were used to model the shapes of the line profiles, but the line fluxes used for analysis are from the collapsed data only.

For the CO[3-2] and CO[1-0] lines, line fluxes were extracted over the same linewidth as was used for the CO[4-3] lines, centered at the expected frequency of the relevant transition based on the CO redshift derived from the CO[4-3] line. This was again done in the collapsed data by \texttt{uvfit}. The fluxes were extracted at the same positions as the CO[4-3] lines, and over the same size. The continuum contributions to the line fluxes were treated as for the CO[4-3] line. All measured fluxes were then corrected for primary beam attenuation. In this way, all line fluxes were extracted consistently. If the different CO excitations were to have different underlying linewidths, they would be physically distinct components. By studying the emission properties over the same sizes and velocity ranges, we are ensuring a self-consistent, physically meaningful extraction of the CO spectral line energy distributions (SLEDs).

It is useful to note here that previous studies claim evidence for extended CO[1-0] reservoirs in certain galaxies, compared to their CO[4-3] emission (e.g. \citealt{ref:S.Stach2017}). If the sources were meaningfully resolved by our smaller beams (e.g. the CO[1-0] dataset) but not resolved in our CO[4-3] dataset where we determine emission sizes, this could introduce a small source of bias. However, both the measured sizes of the cold CO[1-0] gas for our resolved galaxies, as well as the cold dust at 870$\micron$, are consistent with the galaxy CO[4-3] sizes, verifying that our method does not exclude significant CO[1-0] emission. A comparison of these sizes can be seen in Table~\ref{tab:sizeComp}, where only the emission that was significantly resolved (SNR$_{size}$$>$3) with flux emission SNR$>$5 is presented. In order to measure sizes in CO[3-2], CO[1-0], 2mm and 3mm for this comparison, we fixed the position of the source at the position of the CO[4-3] emission, and left the shape of the emission to be fit as a circular Gaussian whose size was left free to vary. This was done in the uv-plane. For the 870$\micron$ sizes, those sources that were fit over an extended shape in the flux extraction, with free position, were also left free to fit as a circular Gaussian for comparison here, again using GALFIT. It is important to note that these circular Gaussian sizes are presented for comparison only, they do not give the flux extraction sizes. As can be seen in Table~\ref{tab:sizeComp}, all of the emission sizes are consistent within 2$\sigma$.

Overlays of the HST, 870$\micron$ data and CO[4-3] data are also shown in figure 2 of the companion paper to this study, \citet{ref:V.Strazzullo2017}, suggesting that there is no evidence of substantial cold gas reservoirs at scales much larger than the beams that we are using. This is reasonable, as we are observing with resolutions that are larger than the typical size of our CO detections, and those sizes determined by HST imaging \citep{ref:R.Gobat2013, ref:V.Strazzullo2013, ref:V.Strazzullo2016}. By measuring the emission over the same size for each transition we are taking measurements on the same physical scales. If we were in fact missing components in CO[1-0] at wider scales, our measurements and the ratios between the different CO transitions would still be self-consistent. Although we do not see any evidence for this in our galaxies, if the CO[1-0] was more extended than the CO[4-3] gas, much of the gas in the outskirts would be at broader velocities than the CO[4-3] transition, and to use this high velocity CO[1-0] flux for our excitation analysis would result in a mixing of different physical components. By measuring line fluxes for the different CO transitions using the CO[4-3] emission sizes, we are primarily characterising the properties of the CO gas in the region where the galaxies are forming stars, and are able to draw conclusions on the gas excitations in the same region. This approach was also taken by \citet{ref:E.Daddi2015}. A risk of bias would arise if the CO[1-0] was much more compact than the CO[4-3] emission, in which case using extended extractions would overestimate the CO[1-0] line flux. This is however physically unlikely to be the case, as CO[1-0] is emitted by lower density gas.

As a final test of any differences in emission sizes, we perform additional CO[1-0] flux extraction using a fixed, Gaussian size, with FWHM equal to the 2$\sigma$ upper limit on the CO[4-3] sizes presented in Table~\ref{tab:sizeComp}. This was done for those galaxies that were not resolved in CO[4-3]. We verify that this does not significantly increase our measured CO[1-0] fluxes, nor does it affect our conclusions on gas excitation.
 
\begin{table*}
\begin{tabular}{ccccccc}
\hline
ID & FWHM$_{870\micron}$ & FWHM$_{150GHz}$ & FWHM$_{108GHz}$ & FWHM$_{\rm CO[4-3]}$ & FWHM$_{\rm CO[3-2]}$ & FWHM$_{\rm CO[1-0]}$ \\
\hline
A2 & - & - & - &0.58$\pm$0.11 & $<$0.66 & 0.43$\pm$0.14 \\
A1 & 0.52$\pm$0.14 & 0.72$\pm$0.16 & - & 0.51$\pm$0.08 & $<$1.17 & - \\
13 & - & - & - & $<$0.31 & - & - \\
6 & $<$0.28 & - & - & $<$0.33 & - & - \\
N7 & - & - & - & $<$0.32 & - & - \\
B1 & $<$0.54 & - & - & - & - & - \\
3 & - & - & - & - & - & - \\
S7 & - & - &- & - & - & - \\
A5 & 0.40$\pm$0.03 & 0.33$\pm$0.06 & 0.65$\pm$0.17 & N & N & N \\
A4 & 0.35$\pm$0.10& 0.62$\pm$0.09 & $<$0.65 & N & N & N\\
\end{tabular}
\caption{Continuum and line-emission sizes for the 8 galaxies with gas detections discussed in this paper, plus two additional bright continuum sources in the cluster field of view. We present continuum emission sizes at 870$\micron$, 150~GHz and 108~GHz, and sizes for the CO[4-3], CO[3-2] and CO[1-0] emission lines. The CO[4-3] sizes have been used for the uv-plane analysis in this paper; the rest are shown for comparison to demonstrate consistency. The 870$\micron$ emission sizes have been circularised for comparison, where more extended sources were used for the flux extraction itself. It can be seen that in all cases, the sizes at the different continuum wavelengths and different excitation transitions of CO are consistent within 2$\sigma$. We do not see evidence for extended cold gas reservoirs that would be missed using the CO[4-3] emission sizes. Where the CO[4-3] emission is resolved, flux was measured over the resolved CO[4-3] shape for all CO transitions and associated continuum. Flux was extracted over the same shape at both 870$\micron$ and 3~GHz within GALFIT. Errors on the 870$\micron$ size have here been calculated using Equation~\ref{eqn:sizeerror}, assuming the normalisation factor to be the average between that for the 150~GHz and 108~GHz continua. 2$\sigma$ upper limits are shown on the emission sizes for those sources that aren't resolved here, but have a flux SNR$>$5. The upper limits on the size of unresolved sources have also been derived from Equation~\ref{eqn:sizeerror}, having calculated the normalisation using the resolved galaxies in the same dataset. Those entries marked with an `N' highlight galaxies for which we detect no lines, to distinguish them from those in which the line emission is unresolved.}
\label{tab:sizeComp}
\end{table*}

\subsection{Continuum flux extraction}
\label{sec:contextract}
In order to image all datasets (ALMA band 7, band 3, band 4, JVLA Ka-band, S-band), the CASA routine \texttt{CLEAN} was used. This was performed for continuum mode=`mfs' with natural weighting to maximise sensitivity, except for the 3~GHz data for which Briggs robust=0.0 weighting was used, as it gave the best balance between the resolution and sensitivity for the cluster field at this wavelength. This allowed us to derive primary-beam corrections and dirty-beam patterns for the instruments. With the exception of the 3~GHz map, the clean images were not used for data analysis, but are shown for visualisation in Fig.~\ref{fig:A126Colour}.
The measurements for continuum fluxes at 150~GHz, 108~GHz and 38~GHz were taken from the ALMA band 4, band 3 and JVLA Ka-band observations respectively. These were extracted in the uv-plane, using the collapsed line-free spectral windows, and extrapolating to the desired frequency. The procedure for this is described in Section~\ref{sec:method:extrac}.

Although we favour the method of flux extraction in the uv-plane, as it does not have the disadvantage of correlated noise introduced by the imaging process, it is not always possible. It was necessary to measure continuum fluxes in the image plane for the 870$\micron$ data, as the constructed mosaic could not be easily transported into the correct format for \texttt{uvfit}. Similarly, the large size of the 3~GHz dataset made conversion and analysis in the GILDAS format impractical.

In order to measure the 870$\micron$ continuum fluxes of the galaxies, the software GALFIT \citep{ref:C.Peng2010} was used on the dirty, calibrated image. For those readers familiar with classical tools, GALFIT is an image plane flux-extraction software, similar in purpose to the \texttt{IMFIT} procedure in CASA, for example. We choose to use GALFIT on the images without first applying \texttt{CLEAN}, as we have full knowledge of the shape of the dirty beam with associated PSF side-lobes, derived directly from the known antenna positions and baselines. In this way, our simultaneous modelling of all sources in the field of view using GALFIT was not subject to contamination between sources or side-lobes. We do however test the consistency between these two potential image plane flux extraction techniques: using \texttt{IMFIT} in CASA on \texttt{CLEAN}ed data (shown in Fig.~\ref{fig:ContData}), and using GALFIT with the dirty beam on dirty data. To do this, we perform continuum flux extraction on the same cluster sources using each of these techniques for the 870$\micron$ data. We recover statistically indistinguishable fluxes between the two image plane methods, and present our GALFIT-measured continuum fluxes for the analysis in this paper, for both the 870$\micron$ and 3~GHz data.

Having quantified that \texttt{CLEAN} + \texttt{IMFIT} and GALFIT reproduce consistent fluxes in our image plane data, we perform a small simulation to compare the fluxes measured by GALFIT with those measured by \texttt{uvfit}. In doing this, we aimed to evaluate any systematic differences that might arise from extracting flux in the uv-plane vs. the image plane. We simulated sources in our 2mm continuum data, with sizes between a PSF and a circular Gaussian with FWHM=2". We chose these sizes as they cover the representative range of the sizes measured for the resolved cluster galaxies, as shown in Table~\ref{tab:sizeComp}. For each method (GALFIT or \texttt{uvfit}), we insert an input source of fixed size and flux into the data, and then re-extracted the flux as a measurement. This was repeated 1000 times per size. We find that both GALFIT and \texttt{uvfit} return flux distributions centered on the correct input flux, with dispersions reflecting the RMS noise of the data. Further details can be found in the Appendix. Finally, a comparison of continuum fluxes extracted for our cluster galaxies using both \texttt{uvfit} and GALFIT in our other datasets was made, verifying that the two methods again give consistent results.

Continuum fluxes were extracted at the fixed positions of HST/WFC3 rest-frame optical cluster members (e.g. \citealt{ref:R.Gobat2011}, \citealt{ref:F.Valentino2015}, \citealt{ref:V.Strazzullo2016}) and at the positions of additional ALMA continuum detections (SNR$>$3). The positions of the strong 870$\micron$ sources seen in Fig.~\ref{fig:ContData} were free to be optimised by GALFIT. If the emission could be resolved, i.e. when sizes could be measured with SNR$>$3, the size of the emission was left free to be fit by GALFIT. If the emission was not resolved, the galaxies were modelled as PSFs. Having run GALFIT, PSF-fitting was performed on the absolute values of the residual 870$\micron$ map, at the positions of all of the detected galaxies. We found all of the residuals to be of the order 1$\sigma$ or less, with the exception of the very bright galaxy A5, for which 2$\sigma$ residuals remained ($\sim$5\% of the measured flux of A5). We do not expect these levels of residuals to affect our results. For the 3~GHz flux extraction, the A-configuration and C-configuration images were both \texttt{CLEAN}ed before using GALFIT with the synthesised beam as the PSF, as the wide field of view of the JVLA means that several extremely bright radio galaxies were present in our 3~GHz image, for which the strong side-lobes needed to be first removed. The emission positions and sizes were fixed to those of the 870$\micron$ emission. Fluxes were measured separately in the A- and C- configuration data, and then combined for each galaxy, taking into account the weight of the flux based on the image RMS. The C-configuration data do therefore not significantly contribute to the 3~GHz flux measurements.

It is recognised that the flux uncertainties returned by GALFIT are often underestimated, which we also find to be the case for our measurements. Following standard practice, we therefore give the primary-beam corrected RMS value of the 870$\micron$ and 3~GHz maps for the point-source flux uncertainties in Table~\ref{tab:contdata}. For the resolved sources, we give twice the RMS value for the flux uncertainty. It is important to note here that the uncertainties on the quantities derived from these data (SFR$_{870\micron}$, M$_{d}$, SFR$_{1.4GHz}$ respectively) are much larger than those on the measured flux. The errors on these properties are dominated by factors such as the uncertainty on the dust temperature. These systematics are discussed in Section~\ref{sec:results}.

For the majority of the galaxies presented at 870$\micron$ in this paper (8/11), we do not perform detection in the 870$\micron$ map, only flux measurement. Detection for these eight galaxies is done using the CO[4-3] line, as previously discussed. However, we also consider galaxies without a CO[4-3] line detection as a detection at 870$\micron$, if the source has an 870$\micron$ continuum flux with SNR$>$5 (using the flux uncertainty values given in Table~\ref{tab:contdata}), or an SNR$>$2.5 for those sources where flux was measured at fixed position on top of an optical counterpart. In order to quantify the number of spurious sources we might detect at fixed position with SNR$>$2.5, we perform blind continuum extraction in the inverse-870$\micron$ map, for 500 randomly selected positions. This was done at fixed position, for a PSF shape. We find an expected number of $\sim$0.15 spurious sources in our data at SNR=2.5, based on the $\sim$25 sources for which we measure 870$\micron$ flux.

\begin{figure*}
    \centering
    \includegraphics[width=\textwidth]{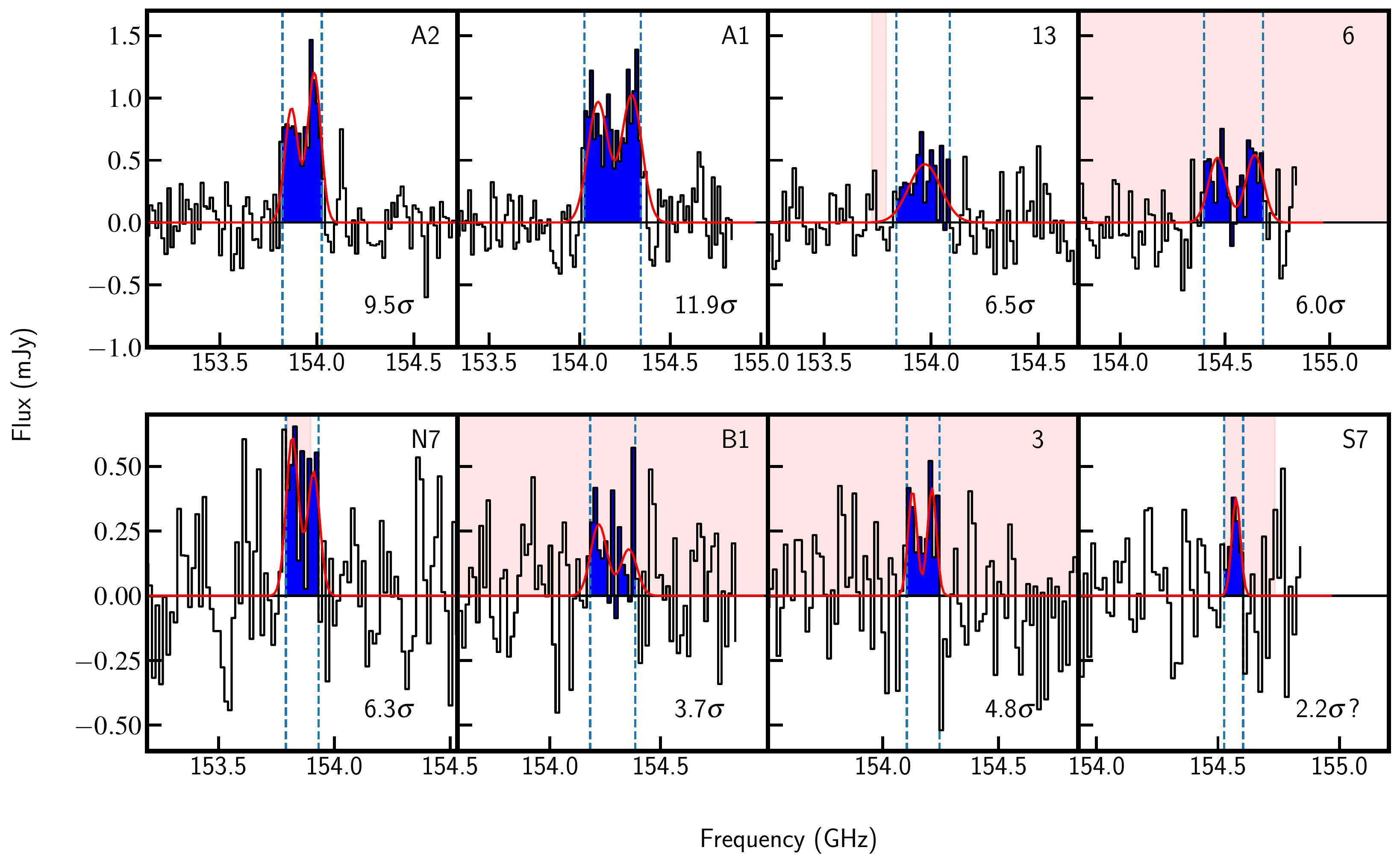}
    \caption{CO[4-3] line detections for eight galaxies in \Cl. The blue shaded regions highlight the flux over the linewidth of the CO[4-3] emission, and the blue dashed lines delimit the linewidths, as discussed in the text. The red curves show the Gaussian line profiles that were fit to the data. The red shaded regions show the 1$\sigma$ limits of the measured optical redshift, where available. The significances of the lines, shown in the bottom-right corner of each panel, were measured from the data collapsed over the linewidth indicated by the dashed lines. A1, A2, B1, N7 and S7 are all part of merging or interacting systems. IDs 13 and S7 are both AGN, ID 6 is a galaxy with a prominent compact bulge component, potentially transitioning from star-forming to quiescent, and ID 3 has previously been classified as a passive galaxy \citep{ref:V.Strazzullo2016}.}
    \label{fig:EightLines}
\end{figure*}

\begin{figure*}
    \centering
    \includegraphics[width=\textwidth]{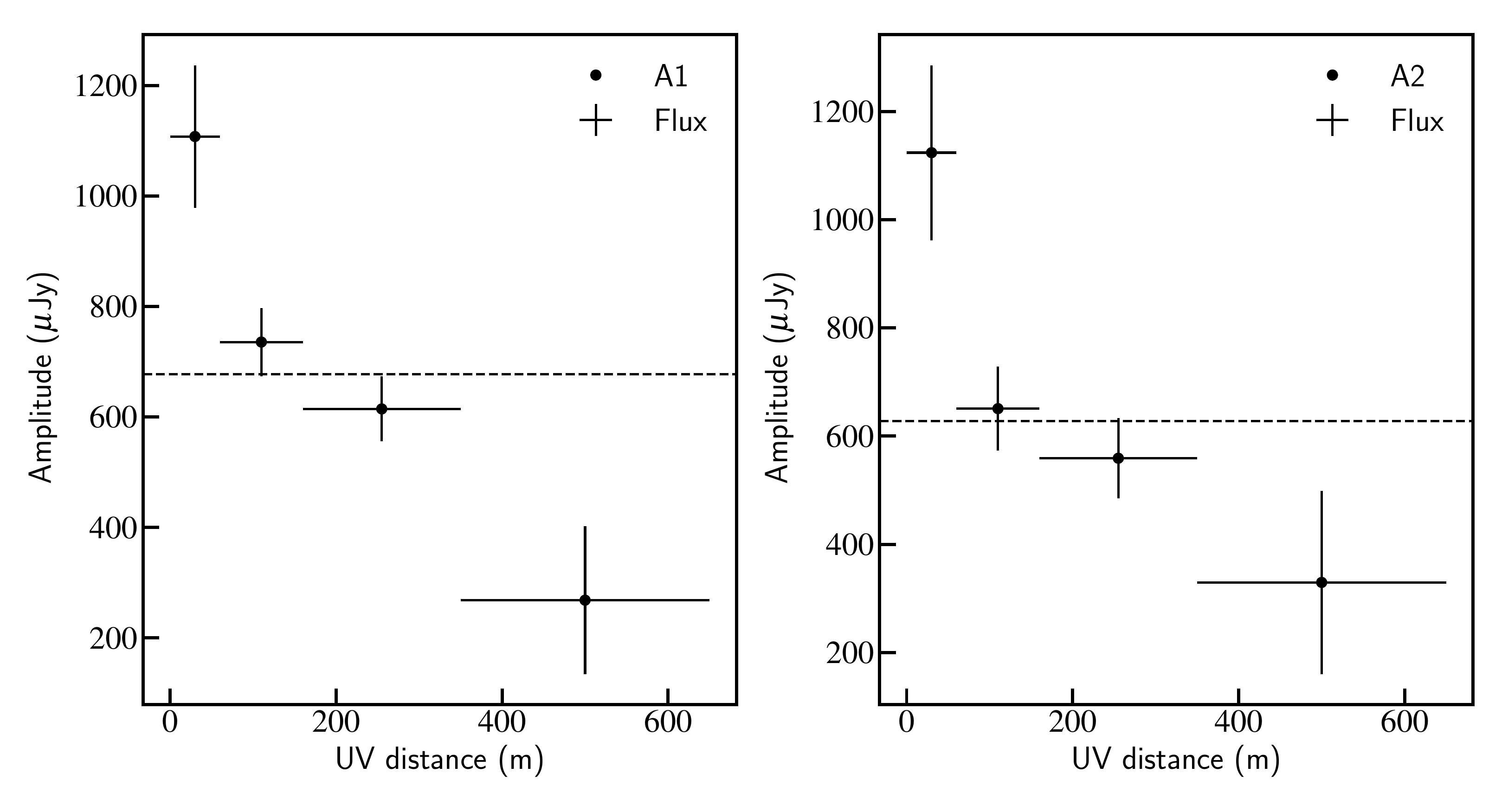}
    \caption{Binned flux amplitude vs. uv-distance for galaxies A1 and A2. The dashed horizontal line shows the best-fitting constant amplitude value for each galaxy, which would be an appropriate amplitude vs. uv-profile for a point source. However, the decrease in amplitude with uv-distance is clear evidence that the galaxies are resolved.}
    \label{fig:AmpUV}
\end{figure*}

\section{Results}
\label{sec:results}

\begin{table*}
\begin{tabular}{ccccccc}
\hline
ID & F$_{870\micron}$ ($\mu$Jy) & F$_{150GHz}$ ($\mu$Jy) & F$_{108GHz}$ ($\mu$Jy) & F$_{38GHz}$ ($\mu$Jy) & F$_{3GHz}$ ($\mu$Jy) & Description\\
\hline
A2 & 515$\pm$135 & 29$\pm$9 & $<$14 & $<$9 & $<$6 & Merger\\
A1 & 1370$\pm$140 & 87$\pm$9 & $<$15 & $<$9 & 7$\pm$3 & Merger\\
13 & 248$\pm$69 & $<$16 & 19$\pm$7 & $<$6 & 6$\pm$2 & AGN\\
6 & 709$\pm$75 & $<$18 & $<$15 & 7$\pm$3 & 5$\pm$2 & Prominent bulge\\
N7 & 217$\pm$79 & $<$17 & 28$\pm$7 & $<$6 & 5$\pm$2 & Interacting\\
B1 & 346$\pm$69 & $<$15 & $<$11 & $<$6 & 4$\pm$2 & Merger\\\
3 & $<$141 & $<$15 & $<$13 & $<$6 & $<$3 & Passive\\
S7 & $<$150 & $<$17 & $<$15 & $<$6 & $<$3 & AGN, interacting \\
A5 & 6047$\pm$150 & 359$\pm$15 & 96$\pm$8 & $<$8 & 27$\pm$3 & - \\
A4 & 1863$\pm$140 & 150$\pm$10 & 50$\pm$8 & $<$9 & $<$6 & - \\
2 & 184$\pm$73 & $<$18 &  $<$16 &  $<$6 &  $<$3 & Prominent bulge \\
\end{tabular}
\caption{Line-free continuum fluxes for the 8 galaxies with molecular gas detections, and two additional bright continuum sources in the cluster field of view. We also include a cluster member with a detection in 870$\micron$, but no CO detections (ID 2). 2$\sigma$ upper limits are shown. The final column gives a summary of the likely environment/state of each galaxy.}
\label{tab:contdata}
\end{table*}

\begin{table*}
\centering
\begin{tabular}{cccccccc}
\hline
ID & I$_{\rm CO43}$ & I$_{\rm CO43, de-boost}$ & I$_{\rm CO32}$ & I$_{\rm CO10}$ & FWZI & FWHM & CO[4-3]$_{\rm FWHM}$ size\\
 & (mJykms$^{-1}$)  & (mJykms$^{-1}$) & (mJykms$^{-1}$)  & (mJykms$^{-1}$) & (\kms) & (\kms) & (") \\
\hline
A2 & 338 $\pm$ 36 & 307 $\pm$ 54 & 340 $\pm$ 55 & 72 $\pm$ 12 & 426 & 387 $\pm$ 89 & 0.58 $\pm$ 0.11\\
A1 & 509 $\pm$ 43 & 474 $\pm$ 67 & 401 $\pm$ 115 & $<$28 & 639 & 619 $\pm$ 111 & 0.51 $\pm$ 0.08\\
13$^{*}$ & 172 $\pm$ 27 & 148 $\pm$ 38 & 146 $\pm$ 56 & $<$18 & 517 & 343 $\pm$ 93 & $<$0.31 \\
6$^{*}$ & 210 $\pm$ 35 & 178 $\pm$ 50 & - & 30 $\pm$ 9 & 578  & 541 $\pm$ 162 & $<$0.33 \\
N7$^{*}$ & 135 $\pm$ 21 &116 $\pm$ 31 & 112 $\pm$ 43 & $<$14 & 304  & 302 $\pm$ 115 & $<$0.32 \\
B1$^{*}$ & 83 $\pm$ 22 & 62 $\pm$ 28 & 113 $\pm$ 56 & 18 $\pm$ 8 & 426  & 533 $\pm$ 213 & - \\
3$^{*}$ & 86 $\pm$ 18 & 70 $\pm$ 24 & $<$89 & $<$13 & 304  & 252 $\pm$ 83 & -\\
S7$^{*}$ & 41 $\pm$ 19 & - & - & $<$11 & 183 & 80 $\pm$ 75 & - \\
\end{tabular}
\caption{Measured CO line fluxes and linewidths of the 8 cluster galaxies discussed in this paper. 2$\sigma$ upper limits on the line fluxes are shown. The CO[4-3] emission sizes are the source sizes used for flux extraction, as described in Section~\ref{sec:method}. 2$\sigma$ upper limits on the sizes are shown for those sources extracted as PSFs, based on the beam size and the S/N of the detection. For A1 and A2, we show the FWHM of the circular Gaussian used for flux extraction. No CO[3-2] flux is given for S7 or galaxy 6, as the spectral setup did not cover the appropriate frequency range at the redshift of S7, and only covered a small fraction of the CO[3-2] linewidth in ID 6. The fluxes for the sources with IDs marked by a star were extracted at fixed rest-frame optical position, for all CO transitions. The FWZI values are the linewidths used to calculate I$_{\rm CO}$, the linewidth corresponding to the flux highlighted in blue in Fig.~\ref{fig:EightLines}. The FWHM values are from the Gaussian line fitting, also shown in Fig.~\ref{fig:EightLines}. For those sources modelled as a double Gaussian, this is the sum of the FWHM of a single Gaussian, plus the separation between the two Gaussians.}
\label{tab:CO43data}
\end{table*}

\subsection{Molecular gas detections}
\label{sec:coldgas}

Images of \Cl\ in the rest-frame optical and in all five of our submillimetre-radio datasets are shown in Fig.~\ref{fig:ContData}. Prominent sources can be seen at 870$\micron$, 2mm and 3mm. In order to investigate the molecular gas properties of our galaxies we first focus on our ALMA band 4 (2mm) data containing the CO[4-3] emission of the cluster galaxies, as this dataset is expected to have the highest signal-to-noise (thermally excited CO emission has flux $\propto$J$^{2}$ \citealt{ref:E.Daddi2015}). Detecting galaxies in this way is closely equivalent to selecting galaxies in SFR, with the depth of our CO[4-3] data in the core pointing corresponding to a 5$\sigma$ SFR detection limit of $\sim$32M$_{\odot}$yr$^{-1}$. Having extracted spectra at the positions of optical cluster members and prominent sub-mm detections, we find secure CO[4-3] line detections in seven galaxies, and a further tentative detection (S7), in the core of \Cl. Galaxies 13, N7, B1, 3, 6 and S7 were extracted at fixed position corresponding to the rest-frame optical HST counterpart, increasing the reliability and robustness of these detections. By extracting spectra at fixed rest-frame optical position, we are removing the positional degree of freedom. This significantly reduces the likelihood of artificial flux boosting from the CO[4-3] data, and allows us to have more robust detections at lower signal-to-noise.

We consider different detection thresholds for those line fluxes measured at fixed optical position, and for those for which the CO[4-3] position was fit by \texttt{uvfit}. For the galaxies for which spectra were extracted at fixed optical position, we consider two factors: both the signal-to-noise of the line flux, and the consistency of the line with the previously derived optical redshift. At these fixed positions, we impose a limit of SNR$>$3.5 on the extracted CO[4-3] line fluxes. When we include the effect of flux boosting by noise (see Section \ref{sec:sims}), this limit corresponds to a probability of $\sim$0.009 of the lines being pure noise, when accounting for de-boosting. In addition, at fixed position we also include the likelihood of the redshift of the detected line being consistent with the optical redshift, if it were a serendipitous detection. For those galaxies with the largest errors on their optical redshifts, $\sim$0.02-0.03, this corresponds to a likelihood of $\sim\frac{2.5}{7.5}$=0.3, from the uncertainty on the redshift in GHz, divided by the total bandwidth of the observations. When we combine the probabilities from the SNR and the redshift, our overall detection threshold gives a probability of $<$0.0029 of our detected lines being due to noise for single-parameter, pure Gaussian-like statistics. Galaxies 13, N7, B1, 3, and 6 are all above this detection criteria, and we find no CO[4-3] emission from other galaxies at fixed optical position under this criterion. We use this probability to calculate the expected number of spurious detections above our threshold, given the number of galaxies for which we extract CO[4-3] spectra. We find a value of only $\sim$0.1 spurious detections expected from our dataset.

In the cases of galaxies for which the CO[4-3] lines were free to be spatially optimised, we instead report only detections with SNR$>$5 (after accounting for flux de-boosting, see Section 3.2.1). A1 and A2 both lie on top of very faint, red, rest-frame optical counterparts and were left free to vary. We find no evidence for other lines at generic positions in the map that exceed our detection threshold.

We report galaxy S7 for interest, despite the fact that it does not formally satisfy our detection criterion, on the basis that we find the line at fixed spatial position on the sky, and the CO redshift derived overlaps with the previously measured optical redshift (see Fig.~\ref{fig:EightLines}). We do however consider this line to be tentative, and we therefore do not include galaxy S7 when deriving physical properties of galaxies from these data.
 
In order to further quantify the number of spurious detections we might expect in the CO[4-3] data, we also performed blind spectral extraction in the ALMA band 4 data. We randomly selected 600 positions within the CO[4-3] dataset FWHM field of view, having removed all known sources from the map, and extracted a point-source spectrum at each fixed position. We subtracted the median continuum flux value, and applied our line-search algorithm over the spectral data. We find that for our lowest significance detection, B1, the number of spurious detections suggested by these simulations at this SNR is $\sim$0.3 based on the $\sim$25 galaxies for which we originally extracted spectra, if we consider the full 1.8~GHz sideband.  This is similar, but slightly higher, than the previous estimate. However, we notice that the redshift of this source is very close (to within 0.001) to the systemic redshift of this multiple-galaxy system. In fact, if we limit our simulation search even to just 1.98 $<$ z $<$ 2.0, conservatively, we would only expect $\sim$0.04 spurious sources. This further confirms the robustness of our detections down to the level of B1.

We find no significant detections in any of our datasets for the low-metallicity galaxies towards the outskirts of the cluster \citep{ref:F.Valentino2015}, and therefore return to further analysis of these galaxies in a future publication. As discussed in Section~\ref{sec:method}, these galaxies were observed with a shallower pointing at 2mm than the core of the cluster.

We detect CO[4-3] emission in all of the cluster members that we see in the 870$\micron$ emission map, with the exception of galaxy ID 2 shown in Fig.~\ref{fig:ContData}. We inspect all of our datasets for evidence of additional strong spectral lines or continuum sources, to ensure that we are not missing galaxies that are not seen in CO[4-3]. We do not find any such source, and - due to the high sensitivity of the CO[4-3] data - a galaxy displaying bright CO[1-0] but not CO[4-3] emission would arguably have highly unusual excitation properties. However, as discussed, our CO[4-3] selection technique is biased, in that it is very similar to a SFR selection method. We therefore fully account for this when interpreting our results. Our key comparisons with previous studies, made in Section~\ref{sec:quantEnviron}, are made with the same SFR selection limits in place.

Rest-frame optical images of the eight detected galaxies, overlaid with CO[4-3] line contours, can be seen in Fig.~\ref{fig:A126Colour}. We verify that the astrometry is consistent between our different datasets, using bright sources that appear in more than one image, including galaxies that are not cluster members. We find small offsets between the peak fluxes, with average $\Delta$RA$_{peak}\sim6.6\times10^{-3}$" and $\Delta$Dec$_{peak}\sim3.4\times10^{-2}$", much smaller than the positional accuracy. The CO detections in galaxies A1 and A2 are both clearly offset from the optical core of a massive galaxy nearby, separated by $\sim$0.95" and $\sim$1" respectively. It appears that although both galaxies A1 and A2 have strong CO emission, they are heavily dust-obscured. The bright optical neighbours of both galaxies are also spectroscopically confirmed cluster members, indicating that both galaxies are most likely part of gas- and dust-rich merging systems. The bright optical core near to galaxy A2 is one of the highly star-forming, low-metallicity galaxies presented in \citet{ref:F.Valentino2015}. Conversely, we see that for the other systems in which we detect CO-emission, the CO[4-3] and the optical counterparts for the galaxies peak on top of one another, such as for IDs 13 and 6. Small offsets between the CO[4-3] and optical emission are most likely driven by noise fluctuations. For all galaxies except for A1 and A2, the offset between the peak CO[4-3] and optical counterparts reaches only $\lesssim$0.16", approximately one tenth of the size of the synthesised beam (with the exception of our tentative detection S7, $\sim$0.3"). These offsets are typically at $<$2$\sigma_{pos}$ significance with respect to the uncertainty on the CO[4-3] position returned by \texttt{uvfit} when left free to vary. This positional uncertainty was verified through simulations of injected point sources\footnote{Representative, S/N-dependent positional uncertainty values $\sigma_{\rm pos}$ were derived by injecting and re-extracting simulated point sources.}.  We therefore choose to extract the spectra and line flux at fixed rest-frame optical position for all galaxies apart for A1 and A2, in order to minimize the effects of noise fluctuations. If we were to instead extract fluxes at the CO[4-3] peak position, the increase in flux would be $<$1$\sigma$ in all cases.

\subsubsection{Galaxy Characteristics}

As can be seen in Fig.~\ref{fig:A126Colour}, we are detecting star-forming gas in several different types of galaxy. Two detected galaxies are part of the interacting system currently forming the future brightest cluster galaxy (BCG, A1 and B1), and it can be seen that the emission from B1 in this system is somewhat broad and potentially diffuse. It should be noted that B1 is galaxy ID~1 in \citet{ref:V.Strazzullo2016} and \citet{ref:V.Strazzullo2017}, but we use B1 here for clarity between A1 and B1. As discussed above, we also detect gas in another probable merger with a nearby low-metallicity optical component (A2), galaxies with prominent compact bulge components (galaxies 6 and 3), active galactic nuclei (galaxies 13 and S7), and an additional star-forming galaxy (N7) that is interacting with galaxy S7 at close projected separation. These characteristics are also summarised in Table~\ref{tab:contdata}.
    
Due to the dense environment of the cluster core, we must consider not only the projected separations of our merging sources, but also the velocity distances between them, to ensure that they are not simply projection superpositions. We compare the CO redshifts of our merging sources with the optical redshifts of their close companions, and find that the velocity offsets are also consistent with real merging or interacting systems. We find velocity offsets of $\sim$15\kms\ between A1 and its companion, and $\sim$15\kms\ between B1 and the same companion, confirming that A1 and B1 are indeed part of the merging system assembling the future BCG. We find a velocity distance of $\sim$590\kms\ between A2 and its bright companion, again confirming the merging nature of this galaxy pair. Finally, we find the velocity separation between the very close pair N7 and S7 to be $\sim$1380\kms, consistent with an interacting pair of galaxies at a projected separation of $<$0.5". We can therefore assert that the large number of merging galaxies that we detect is not biased by projection effects.

\subsubsection{CO Line fluxes}

The 1D spectra of these line detections are shown in Fig.~\ref{fig:EightLines}. The spectra for the CO[3-2] lines and CO[1-0] lines are given in the Appendix. Line detection was not performed for the CO[3-2] and CO[1-0] lines, we simply highlight the velocity range over which the CO[3-2] and CO[1-0] line fluxes were extracted, corresponding to the CO[4-3] redshift and velocity width. From the line spectra in Fig.~\ref{fig:EightLines}, CO redshifts were determined from the flux-weighted line centres over the channels contained in the linewidth, and are shown in Table~\ref{tab:SFRs}. We show in Fig.~\ref{fig:EightLines} that the CO redshifts are in good agreement with the redshifts derived from optical/NIR spectroscopy for 6/8 of the galaxies, with offsets between $\sim$20-400\kms \mbox{\citep{ref:R.Gobat2013, ref:F.Valentino2015}}. However, A1 and A2 both appear to be heavily dust-obscured and have fainter, redder optical counterparts (Fig.~\ref{fig:A126Colour}). They therefore do not have an optically-derived redshift, but the redshift values calculated using the CO[4-3] lines are consistent with the redshift of \Cl.

The line-free continuum fluxes for each galaxy are given in Table~\ref{tab:contdata}, including two strong continuum sources without CO detections (A4 and A5), which are discussed in Section~\ref{sec:A4A5}. The integrated CO[4-3] line fluxes are tabulated in Table~\ref{tab:CO43data}, measured directly from the collapsed data over the linewidths, as well as the CO[4-3] emission sizes used for flux extraction.

In the case of A1 and A2, the CO[4-3] emission is resolved, and the sizes (and associated errors) given in Table~\ref{tab:CO43data} are the FWHM of the circular Gaussians measured by Gildas. To demonstrate this, we show the flux amplitude again uv-distance profiles of A1 and A1 in Fig.~\ref{fig:AmpUV}, binned due to the large number of individual visibilities. If these galaxies were point sources in the image plane, they would show a constant amplitude profile in the uv-plane. Conversely, decreasing amplitude with uv-distance demonstrates that the galaxies are indeed resolved, which is clearly the case for these galaxies. To quantify this, we calculate the best-fitting constant amplitude profile for both galaxies using a least-squares approach, which are shown by the dashed lines in Fig.~\ref{fig:AmpUV}. The $\chi^{2}$ values between these fits and the binned data are 22.5 and 13.5 for A1 and A2 respectively, for 3 degrees of freedom. These give probabilities of these galaxies being unresolved of $<$5.1$\times$10$^{-5}$ and $<$3.7$\times$10$^{-3}$.

We also use these data to measure an alternative detection significance of A1 and A2, which is independent of their resolved sizes. To do this, we add in quadrature the SNR of each binned amplitude in Fig.~\ref{fig:AmpUV}, defined as the amplitude divided by the error on the amplitude. This gives us SNRs of 18.1$\sigma$ and 13.4$\sigma$ for A1 and A2 respectively - even higher than those found using the flux measurements within their Gaussian sizes. 

The errors on the sizes measured using \texttt{uvfit} are well-defined. Through interferometry, the ability to resolve a source is dependent on both the beam size and the signal-to-noise of the source, meaning that bright sources can be resolved down to sizes several factors smaller than the beam. This is made possible by the well-understood beam pattern. In order to characterise the relationship between the error on the size of a galaxy and its brightness, we simulate $\sim$70 galaxies, by inserting sources into empty regions of our 2mm dataset at fixed circular Gaussian size with a range of input fluxes and sizes. We then measure the fluxes and sizes of the simulated sources using \texttt{uvfit}. We confirm that the size errors as given by \texttt{uvfit} are sensible. We find that the relationship between flux SNR and the error on the size follows the trend $FWHM_{err} \propto SNR^{-1}$, as might be expected based on literature studies (e.g. \citealt{ref:J.Condon1997}). The errors on the sizes of A1 and A2 can be described by a relationship of the form:

\begin{equation}
\label{eqn:sizeerror} 
FWHM_{err} = 0.93\times\frac{FWHM_{beam}}{SNR}
\end{equation}

where $FWHM_{err}$ is the 1$\sigma$ error on the FWHM Gaussian size, $FWHM_{beam}$ is the circularised FWHM of the ALMA synthesised beam, SNR is the SNR of the CO[4-3] line detection, and the factor of 0.93 was calibrated by comparing the output of Eqn.~\ref{eqn:sizeerror} with the size errors reported by Gildas. It is the size errors given by Gildas that are shown in Table~\ref{tab:CO43data} for A1 and A2. We cannot constrain the sizes of the lowest SNR sources, as their low SNR brings into doubt our ability to detect them if they are resolved and diffuse. We therefore put upper limits on the sizes of the SNR$>$6 unresolved sources using Eqn.~\ref{eqn:sizeerror}. We see that our CO[4-3] emission sizes are likely to be relatively compact.

For comparison with the integrated CO[4-3] line fluxes over the collapsed data, we fit Gaussian line profiles to each of the detections using least-squared minimisation, which are shown over-plotted in Fig.~\ref{fig:EightLines}. For all galaxies except A1 and A2, these are simply included for guidance and are not used in further analysis. For some galaxies, a double-peaked profile appears to be an appropriate fit to the data, while for others we show a single Gaussian. For the single Gaussian fits, the normalisation, centre and FWHM were free to vary. For the double Gaussians, the normalisations of the two peaks were free to vary independently. We imposed the constraint that the FWHM of the two Gaussians should be equal to one another, but this FWHM and the separation between the peaks were left as free parameters. The errors on these Gaussian FWHM are derived from the simulations described in Section~\ref{sec:sims}, unless the fitting procedure returned larger errors than the simulations (this was the case for B1 and S7). The shapes of the CO[4-3] lines in galaxies A1 and A2 are clearly suited to a double-peaked profile. We use a least-squared method to asses the goodness-of-fit, and find probabilities for the single Gaussian fits to the data of 1.5$\times$10$^{-5}$ and 2.4$\times$10$^{-3}$ for A1 and A2 respectively. For the double Gaussian profiles, we find probabilities of 0.031 and 0.52 respectively. This statistically confirms that the double Gaussian profiles are more appropriate fits to the data. However, the relatively low probability of A1 being described by a double Gaussian also indicates that a more complex line profile is required to well-fit the data. For the remainder of our galaxies, we find that both single and double Gaussian profiles give reasonable fits to the data, so we show the profile with the greatest probability to guide the eye in Fig.~\ref{fig:EightLines}. It is important to note here that we find the linewidths between the single and double Gaussian fits to be consistent. We also compare the line fluxes integrated under both single and double Gaussian profiles with the line fluxes from the collapsed data, and find that the different methods again give consistent results within 1$\sigma$. We therefore use the line fluxes measured directly from the collapsed data for all galaxies, over the velocity range shown in blue in Fig.~\ref{fig:EightLines} (Full Width Zero Intensity, FWZI). We do however also tabulate the Gaussian-fit linewidths in Table~\ref{tab:CO43data}, as these widths are more appropriate for dynamical arguments, particularly for our brightest galaxies, A1 and A2 (see Section~\ref{sec:dynmass}). The linewidths of our detections vary between $\sim$100kms$^{-1}$ and $\sim$650kms$^{-1}$, within the range of expected values for star-forming galaxies at this redshift.

\begin{figure}
    \centering
    \includegraphics[width=0.5\textwidth]{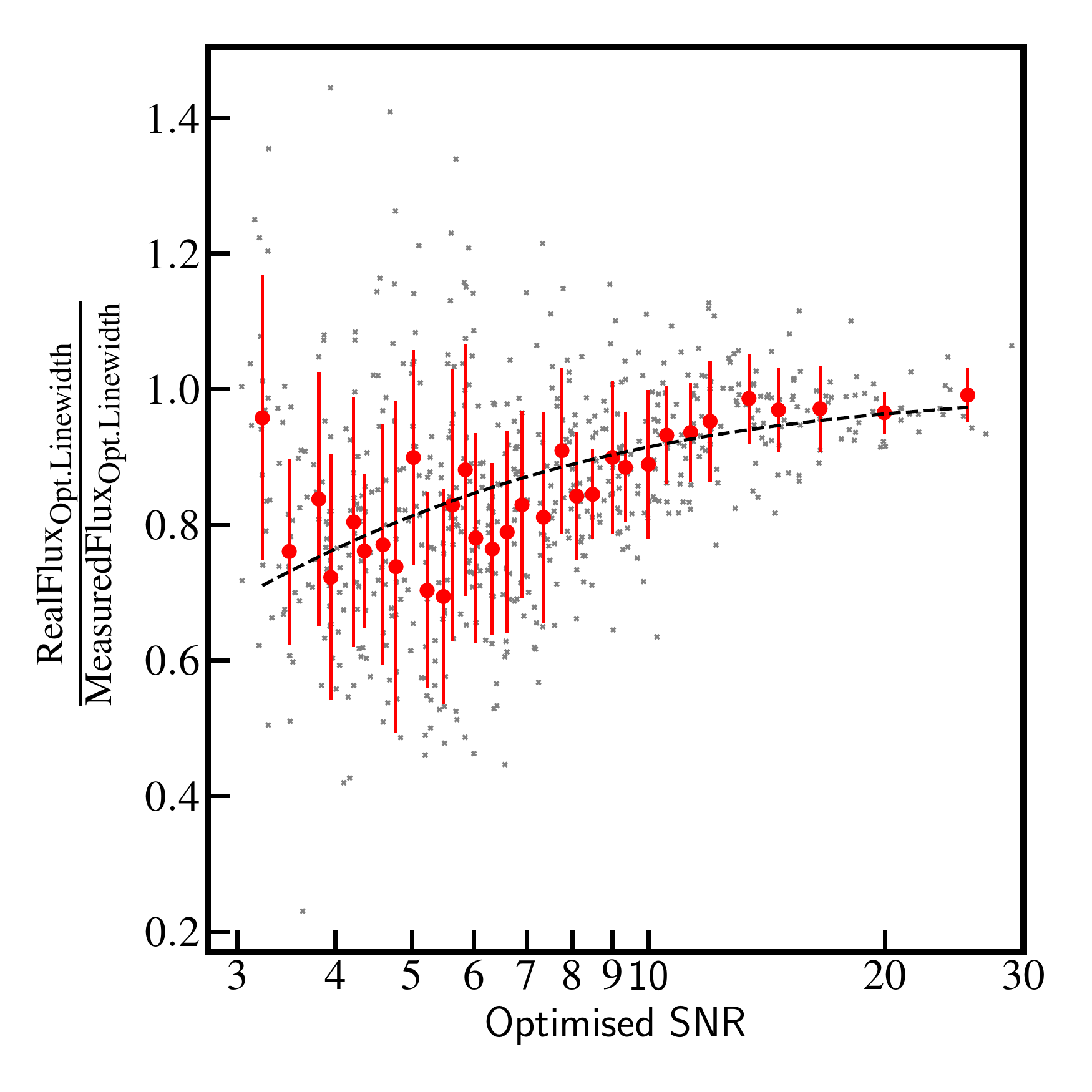}
    \caption{The ratio between the simulated input flux and the flux recovered by our line-search algorithm over the optimised linewidth, as a function of recovered (output) signal-to-noise for a simulation of double-Gaussian line profiles. This ratio is used to de-boost the CO[4-3] line fluxes of our galaxies. In this way, we correct for any artificial noise boosting in the spectra, arising from the optimisation of the CO[4-3] linewidth. The grey crosses are the individual data points, and the red circles are the medians of the data in bins of 20. The red error bars show the standard deviation within each bin. The black dashed line is the polynomial fit to the data, from which we have derived the flux corrections. A least-squared approach was used for this fitting.}
    \label{fig:fluxboost}
\end{figure}

\begin{table*}
\begin{tabular}{cccccccccc}
\hline
ID & RA & Dec & z$_{\rm CO}$ & z$_{\rm opt}$ & SFR$_{\rm CO43}$ & SFR$_{870\micron}$ & SFR$_{1.4GHz}$ & log(M$_{*}$) \\
 & (deg) & (deg) & - & - & $(M_{\odot}yr^{-1})$ & $(M_{\odot}yr^{-1})$ & $(M_{\odot}yr^{-1})$ & log$(M_{\odot}$) \\
 \hline
A2 & 222.30710 & 8.93951 & 1.9951 $\pm$ 0.0004 & - & 102 $\pm$ 24 & 85 $\pm$ 22 & $<$157 & -  \\
A1 & 222.30872 & 8.94037 & 1.9902 $\pm$ 0.0005 & - & 128 $\pm$ 31 & 227 $\pm$ 22 & 113 $\pm$ 46 &  - \\
13 & 222.30856 & 8.94199 & 1.9944 $\pm$ 0.0006 & 1.99851 $\pm$ 0.00066 & 40 $\pm$ 13 & 38 $\pm$ 11 & 91 $\pm$ 24 & 10.46 $\pm$ 0.3 \\
6 & 222.30991 & 8.93779 & 1.9832 $\pm$ 0.0007 & 1.98 $\pm$ 0.02 & 59 $\pm$ 19 & 118 $\pm$ 12 & 84 $\pm$ 24 & 10.71 $\pm$ 0.3 \\
N7 & 222.31029 & 8.93989 & 1.9965 $\pm$ 0.0004 & 1.997 $\pm$ 0.001 &  31 $\pm$ 10 & $<$26 & 70 $\pm$ 24 & 10.07 $\pm$ 0.3 \\
B1 & 222.30891 & 8.94071 & 1.9883 $\pm$ 0.0070 & 1.99 $\pm$ 0.03 & 20 $\pm$ 10 & 57 $\pm$ 12 & 64 $\pm$ 24 & 10.81 $\pm$ 0.3 \\
3 & 222.30910 & 8.93951 & 1.9903 $\pm$ 0.0004 & 2.00 $\pm$ 0.03 & 23 $\pm$ 9 & $<$23 & $<$48 & 10.31 $\pm$ 0.3 \\
S7 & 222.31021 & 8.93980 & $\sim$1.983 & 1.982 $\pm$ 0.002 & $\sim$8 & $<$26 & $<$48 & 10.48 $\pm$ 0.3 \\
A5 & 222.30963 & 8.93690 & - & - & - & 1004 $\pm$ 24 & 410 $\pm$ 46 & - \\
A4 & 222.30648 & 8.93778 & - & - & -& 309 $\pm$ 22 & $<$174 & - \\
2 & 222.30586 & 8.94297 & - & 1.98 $\pm$ 0.02 & - & $<$24 & $<$48 & 10.81 $\pm$ 0.35 \\
\end{tabular}
\caption{Properties and SFRs of the 8 galaxies with CO[4-3] detections, as well as two additional bright continuum detections in the cluster field of view (A4 and A5), and a galaxy only detected at 870$\micron$ (ID 2). 2$\sigma$ upper limits are shown. The SFR$_{\rm CO[4-3]}$ values were derived from the de-boosted line fluxes. The errors given on the SFRs from CO[4-3] and 870$\micron$ are displayed only with the errors corresponding to the flux measurements. These errors are smaller than the errors used for the analyses in this paper, which also take into account the errors arising from the systematics, which are discussed in the text. The optical redshifts were calculated in previous studies using HST/WF3 grism and MOIRCS spectroscopy \citep{ref:R.Gobat2011, ref:F.Valentino2015}.}
\label{tab:SFRs}
\end{table*}

\subsubsection{CO[4-3] flux boosting}
\label{sec:sims}
As described in Section~\ref{sec:method:extrac}, the channels over which we extract the CO[4-3] flux, shown in Fig.~\ref{fig:EightLines}, were taken as the channels that gave the highest S/N, thus constraining the line. However, although common practice, it is possible that the choice of velocity range in this way could be influenced by the channel-to-channel noise, thus artificially boosting the total measured flux over this linewidth. In order to quantify this effect, we created a simulated sample of CO[4-3] double-peaked Gaussian line profiles, with recovered SNRs between 3 and 30, given the representative noise on our measured 1D spectra. We added these simulated line profiles to simulated random noise, and then applied the same line-finding algorithm as was used for our real data. We compared the output flux (signal+noise, measured flux) recovered over the linewidth found by the algorithm, with the pure input flux (signal, real flux) over the same linewidth. This gave us a direct measure of the effect of potential noise boosting. The results of these simulations are shown in Fig.~\ref{fig:fluxboost}. We bin our data by SNR, and calculate the median flux correction factor, the ratio of Real Flux/Measured Flux over the optimised linewidth, in each SNR bin. We then fit this binned data with a polynomial function of the form $y = ax^{b}$. This was done for simulations of double Gaussian profiles with FWHM of a range of linewidths between $\sim$200\kms and $\sim$700\kms, spanning the full range of linewidths for our observed galaxies. We found the flux boosting correction to be independent of linewidth.

It can be seen from Fig.~\ref{fig:fluxboost} that the effect of noise boosting is minimal at high SNR, shown by Real Flux/Measured Flux~=~1. The  Real Flux/Measured Flux decrease smoothly with decreasing SNR, implying an increasing effect of noise boosting. The effect ranges from a minimum of $\sim$7\% ($<$1$\sigma$) for our brightest source (A1), up to $\sim$25\% for our faintest secure source (B1). We therefore use the polynomial fit to the data to calculate appropriate flux correction factors for each of our galaxies. At S/N$<$3, it is difficult to define a robust correction factor from these simulations. We therefore do not perform flux de-boosting for S7, or derive physical parameters from the line flux, due to the tentative nature of the detection. For our other detections, this correction factor was applied to the CO[4-3] line fluxes, and is essentially a de-boosting, ensuring that our results are robust. The errors on the correction factors are derived from a polynomial fit to the error bars on the binned Real Flux/Measured Flux ratios themselves, and are added in quadrature to the errors on our I$_{\rm CO[4-3]}$ values. The de-boosted line fluxes are also given in Table~\ref{tab:CO43data}. Is it these de-boosted line fluxes and errors that we use for the remainder of our analysis, to derive all physical properties of the galaxies. This correction does not need to be applied to the line fluxes at lower excitations because the velocity range of the line was optimised by the CO[4-3] flux, and the channel-to-channel noise distribution is independent in the different data sets. We therefore do not suffer from the same noise bias at lower J-transitions.

\begin{figure*}
    \centering
    \includegraphics[width=\textwidth]{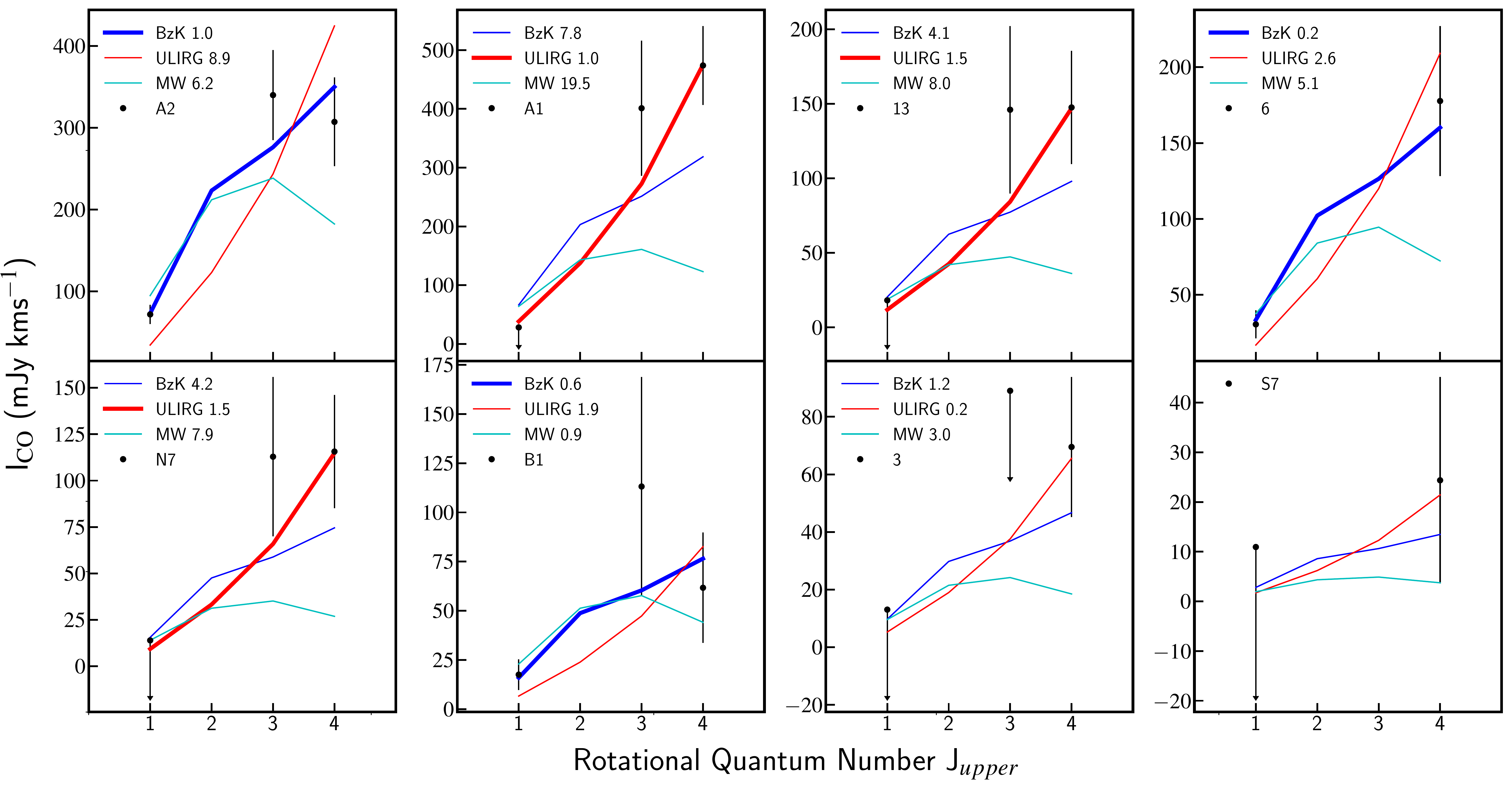}
    \caption{CO SLEDs of 8 cluster galaxies detected in CO[4-3]. The legends show the reduced $\chi^{2}$ value for the fits of the observed galaxy CO SLED with the three excitation models: BzK, ULIRG/SB and MW. The SLEDs of some galaxies are best-fit by a BzK excitation model, and some are best-fit by a ULIRG/starburst-like model. The preferred template for each galaxy is plot with a thicker line, where possible. We do not draw any conclusions about the gas excitation of galaxies ID 3 and S7, due to the limited data available. A least-squared analysis was used for the fitting, and 2$\sigma$ flux upper limits are shown.}
    \label{fig:SLEDs}
\end{figure*}

In addition to the flux boosting correction factors, we also use our simulations to derive diagnostics on the error on the line FWHM, as well as the potential flux loss due to `clipping' the line in velocity space. We fit a double Gaussian to each of our simulated line profiles, before calculating the standard deviation of the output FWHM in bins of flux SNR. We find the error on the FWHM to be of the order 15-40\%, depending on the SNR of the detection. The Gaussian fits to the individual line profiles in Fig.~\ref{fig:EightLines} returned smaller errors on the FWHM (except for galaxies B1 and S7), so in these cases we use the uncertainty from our simulations as a more conservative measure of the FWHM error, shown in Table~\ref{tab:CO43data}.

We find the flux loss due to the linewidth optimisation to be $\sim6\%$. However, this value has a certain amount of uncertainty when applied to our cluster galaxies. The flux loss due to clipping is highly dependant on the underlying profile of the CO lines of our galaxies (e.g. double or single Gaussian, or even steeper profiles), which are not well constrained. We therefore consider the loss between the measured and intrinsic line fluxes to be 6$\pm$6\%. Any potential flux loss will affect each of the CO transitions equally, assuming that the lines have the same width and profile. It therefore will not affect our results on gas excitation, and we tabulate the measured line fluxes over our defined FWZI in Table~\ref{tab:CO43data}. We incorporate the effect of this possible flux loss when deriving the luminosity of the CO[1-0] transition, discussed in Section~\ref{sec:sfrs}, and the SFRs derived from CO[4-3].

Finally, we also use our simulations to quantify the error on the redshifts derived from the CO[4-3] lines, shown in Table~\ref{tab:SFRs}. We wish to compare the error on the redshift calculated from the flux-weighted central frequency of the line (as tabulated), with the intrinsic dispersion on the derived redshifts from our simulations. Our simulated lines have a known redshift, corresponding to the central redshift of the double-peaked profile, and we use the same methods to both find the lines and calculate the redshift of the simulated lines as we have used for our data. We find that the error on the redshift increases with increasing linewidth and with decreasing SNR, but that the output redshift dispersion is smaller than the formal error coming from the frequency uncertainty. The exception to this is for galaxy ID 3, at FWHM 250\kms and SNR=4.8. In this case, the two errors are the same. We therefore show only the errors from the original redshift derivation in Table~\ref{tab:SFRs}.

\subsection{CO Spectral Line Energy Distributions}

In addition to the CO[4-3] emission, we also find several significant line flux measurements for both the CO[3-2] and CO[1-0] transitions in these eight galaxies, having extracted line fluxes over the velocity range corresponding to that of the CO[4-3] transition. These are shown in Table~\ref{tab:CO43data}. Incorporating the integrated line flux measurements of CO[3-2] and CO[1-0] with those of CO[4-3], we are able to construct CO spectral line energy distributions for the cluster galaxies, shown in Fig.~\ref{fig:SLEDs}. We compare the excitation properties of the ISM gas in our eight cluster galaxies with known SLEDs: `normal' z$\sim$2 star-forming galaxies (BzK), ULIRG/starburst galaxies, and the Milky Way (MW) \citep{ref:P.Papadoupolos2012, ref:E.Daddi2015}.

Studying the gas excitation properties of a galaxy (the ratio of the excited, high-J gas to the total gas reservoir at lower J-transitions) gives us insight into the relative physical states of the gas. Observations have shown that some populations of galaxies, such as ULIRGs and starburst galaxies, have a large amount of their molecular gas in the densest states traced by high J-transitions, compared to the more extended gas traced by lower J-transitions. The flux in the higher J-transitions is tightly correlated with the SFR of a galaxy, and these excitation properties therefore give us insight into the mode of star-formation that is taking place in the galaxy \mbox{\citep{ref:E.Daddi2015}}. ULIRG- or starburst-excited galaxies typically form stars in a violent, short-lived fashion, whilst BzK galaxies fall along the more secularly star-forming main-sequence (MS) of galaxies \citep{ref:E.Daddi2004, ref:D.Elbaz2011}. We do not try to find the best-fit SLED for each galaxy, but wish to characterise our galaxies with respect to the limiting cases most commonly found among star-forming galaxies. Moreover, the method of comparing our CO data with SLED templates allows us to use all three CO transitions simultaneously, and is therefore less limited by measurement upper limits than other types of excitation analysis, such as the ratios between different transitions.

From Fig.~\ref{fig:SLEDs}, we see that the CO excitation properties vary between galaxies in the cluster. Least-squared fitting was used to compare the data to the templates, where the normalisation of each template was allowed to vary. In the fitting process, we used the measured line flux values for each transition, not the upper limits. We are able to characterise the excitation state of the gas in six of our eight galaxies using $\chi^{2}$ analysis. The reduced $\chi^{2}$ values for each of the fits reveal that for 3/6 of these cluster galaxies, a starburst-like gas excitation better fits the data, with typical $\chi^{2}_{red}$ values between 1 and 1.5. The other three galaxies are best-fit by BzK excitation properties, with $\chi^{2}_{red}$ between 0.2 and 1.0. One might expect that not having significant CO[1-0] line flux would make a galaxy more likely to show starburst-like excitation properties, and we do appear to see this trend in our data. No conclusions can be drawn with these data on the excitation properties for S7 and ID 3. For ID 3, we find that both ULIRG and BzK templates give a reasonable fit, and there were no data available for the CO[3-2] transition at the redshift of S7 with which to characterise the excitation. From Fig.~\ref{fig:SLEDs}, it is immediately striking that 50\% of these galaxies show a starburst-like, rapid star-formation mode based on the gas excitation properties, as this is a significantly higher fraction than observed in the field (see Section~\ref{sec:quantEnviron}).

\subsection{Star-formation rates}
\label{sec:sfrs}
Although we cannot spatially distinguish between the 870$\micron$ and 3~GHz fluxes of N7 and S7, due to their close separation on the sky, we see in \citet{ref:V.Strazzullo2017} that the infrared (IR) luminosity derived from the CO[4-3] emission of N7 is consistent with that derived from the 870$\micron$ flux emission of the combined N7-S7 system. We also see that the 3~GHz continuum flux, shown in Fig.~\ref{fig:ContData}, is much brighter at the rest-frame optical position of N7 than S7. As shown in Table~\ref{tab:SFRs}, the SFR derived from CO[4-3] for S7 is very low, and we therefore believe that the continuum flux measurements at the position of this pair of sources can be associated with N7. Due to the small amount of gas measured in galaxy S7, considering N7 and S7 as a double source both contributing to the 870$\micron$ and 3~GHz continua, or considering them as separate sources and associating the continua with N7, has a negligible effect on our results. An exception to this is in the SFR-M$_{*}$ plane, where we are able to investigate the nature of the two galaxies separately. This is shown in Section~\ref{sec:ssfrs}. The continuum flux values underneath each of the CO lines were measured separately for S7 and N7, given in Table~\ref{tab:contdata}.

It has previously been shown that there is a tight, virtually linear correlation between the CO[5-4] line luminosity and infrared luminosity (and therefore SFR) of both MS and SB galaxies, across a range of redshifts \mbox{\citep{ref:E.Daddi2015}}. In order to estimate an IR luminosity from the CO[4-3] line fluxes, we therefore derive CO[5-4] line fluxes for our galaxies. We adopt the preferred excitation template shown in Fig.~\ref{fig:SLEDs}, using the appropriate flux ratio between the two transitions \mbox{\citep{ref:P.Papadoupolos2012, ref:E.Daddi2015}}. For those galaxies without a preferred template, we assume a BzK transition ratio. In all cases, the uncertainty introduced by the choice of SLED template is much smaller than the systematic errors on the SFRs, which are discussed below. From the CO[5-4] line flux we derive an infrared luminosity and thence a SFR for each galaxy. The SFR values derived in this way are shown in Table~\ref{tab:SFRs}, SFR$_{\rm CO43}$. As discussed in Section~\ref{sec:sims}, these SFRs and error have been increased by 6\%, to account for possible flux loss from clipping of the CO lines in velocity space. Alongside these, we are also able to derive other estimates of the SFR, using the 870$\micron$ and 3~GHz continuum fluxes described in Section~\ref{sec:method}. In the case of SFR$_{870}$, the SFRs were calculated taking the average 870$\micron$ to L$_{IR}$ conversion factor between the MS and SB templates in \citet{ref:M.Bethermin2015} at z=2, and the systematic uncertainties introduced here are discussed below. We find several robust detections in the cluster core at 870$\micron$ (shown in Table~\ref{tab:contdata}), which allow us to derive these SFRs. For the SFRs derived from the 3~GHz data, we assume a synchrotron contribution to the radio data with a slope of $\nu^{-0.8}$, and thus map the 3~GHz fluxes to 1.4~GHz equivalent flux. We then use the templates in \citet{ref:M.Bethermin2015} at z=2.0 to convert to L$_{IR}$, based on the well-known L$_{1.4GHz}$-L$_{FIR}$ (far-IR, FIR) correlation \citep{ref:J.Condon1991a, ref:J.Condon1992}. This conversion is invariant for MS and SB galaxies.

We find that the different estimates for SFR are generally consistent. An exception to this may be the CO[4-3] vs. 870$\micron$ SFRs for galaxy A1, and a potential explanation for a discrepancy between the gas and dust properties of this galaxy is discussed in Section~\ref{sec:timescales}. We note that galaxy 13 contains a radio-quiet AGN, with a bolometric X-ray luminosity placing it in the quasar regime \citep{ref:F.Valentino2016}. We therefore consider that the contribution from X-ray Dissociation Regions (XRDs) to the higher excitations of CO in this galaxy may not be negligible. However, the FIR regime is expected to be largely unaffected by AGN \citep{ref:J.Mullaney2012a}, and so the agreement between the CO[4-3] SFR and the 870$\micron$ SFR for galaxy 13 suggests that the effect of XDR emission is not an important factor here. Additionally, we see disagreement between the SFR$_{\rm 870\micron}$ and SFR$_{1.4GHz}$ for galaxies A4 and A5. As discussed in Section~\ref{sec:A4A5}, we do not have redshift measurements for these galaxies, and the assumption of z=2 for the calculation of SFR$_{1.4GHz}$ may have given rise to this inconsistency.

As also discussed in \citet{ref:V.Strazzullo2017}, the CO[4-3] and 870$\micron$ SFR measurements are always consistent given the uncertainties involved in estimating L$_{IR}$ from the 870$\micron$ continuum and CO[4-3] line fluxes. It is a useful consistency check to be able to compare these estimates of SFR, because neither CO[4-3] nor 870$\micron$ flux allows us to trace the star-formation directly. We include in our error budget $\sim$0.2~dex for the CO[5-4] vs. L$_{IR}$ conversion (\citealt{ref:E.Daddi2015, ref:D.Liu2015}, E. Daddi et al. 2018, in prep) and $\sim$0.15~dex for the 870$\micron$ to L$_{IR}$ conversion. Here we are taking an average of the MS and SB conversion factors from \citet{ref:M.Bethermin2015}, which differ by a factor of 2. Understanding these areas of uncertainty, and finding consistent results given by both independent methods, gives us confidence that our SFR estimates are robust and not heavily biased by the assumptions made. In the remaining analyses, we therefore use the average of the SFRs from CO[4-3] and 870$\micron$ tracers as our preferred SFR. This is simply referred to as the SFR for the rest of the paper, and all of the measurement plus systematic errors have been included in the calculation of the SFR error.

\subsection{Star-formation efficiencies}
\begin{figure}
    \centering
    \includegraphics[width=0.5\textwidth]{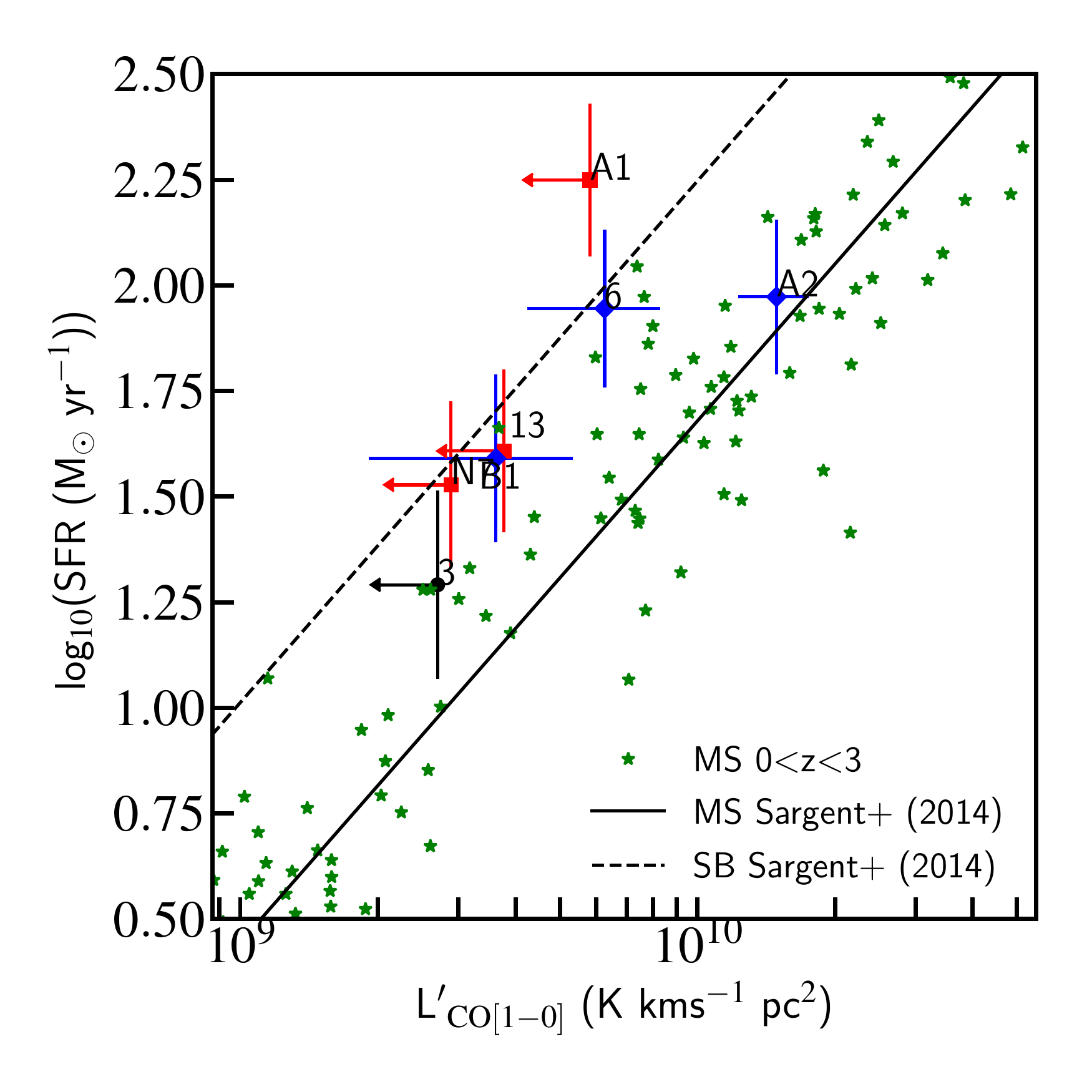}
    \caption{The SFR - L$^{\prime}_{\rm CO[1-0]}$ relation. The galaxies are coloured according to their excitation properties (red=SB, black=either, blue=BzK). SFRs are the average SFR derived from the CO[4-3] line flux and the 870$\micron$ continuum flux. Seven of the eight cluster galaxies lie at enhanced SFE, compared to the MS \citep{ref:M.Sargent2014}. A sample of MS galaxies between 0$<$z$<$3 are shown by the green stars for comparison \citep{ref:E.Daddi2010b, ref:J.Geach2011, ref:G.Magdis2012, ref:A.Bauermeister2013, ref:L.J.Tacconi2013}. The errors on the average SFRs include a 0.17~dex systematic uncertainty, as well as the measurement errors. The MS and SB relations are shown by the solid and dashed black lines respectively. 2$\sigma$ upper limits are shown.}
    \label{fig:LIR_LCO}
\end{figure}

An indirect measure of the star-formation efficiency of the cluster galaxies can be seen in Fig.~\ref{fig:LIR_LCO}, the SFR - L$^{\prime}_{\rm CO[1-0]}$ relation. We define SFE as the SFR per unit molecular gas mass. We compare the SFR with the luminosity of the CO[1-0] line - a proxy of the total cold molecular \Htwo\ reservoir. The CO[1-0] line flux (or upper limit) was measured directly using the JVLA Ka-band data described in Section~\ref{sec:method}, not extrapolated from the best-fit SLED. Galaxy A2 is consistent with the SFR - L$^{\prime}_{\rm CO[1-0]}$ relation for MS galaxies, and all of the other cluster galaxies are shifted to the left of the MS locus, indicating that they have an increased SFE with respect to MS galaxies that are selected independent of environment. The median offset of the cluster galaxies from the MS locus is 0.37~dex. This shift does not appear to depend on each galaxy's preferred excitation template. Interestingly, we don't see clear evidence for the starburst-excited galaxies (shown by red squares) having much increased SFEs compared to the BzK-excited galaxies (blue diamonds), similar to what was recently concluded by \citet{ref:H.Dannerbauer2017}.

It can be seen in Fig.~\ref{fig:SLEDs} that less than half of the cluster galaxies have significant CO[1-0] flux measurements in our deep Ka-band data. As can be seen in Fig.~\ref{fig:LIR_LCO}, based on the positions of these upper limits, we would have been able to measure significant CO[1-0] flux in these galaxies if they had been forming stars on the MS regime of the SFR - L$^{\prime}_{\rm CO[1-0]}$ plane. This further hints that the significant star-formation of our galaxies is being fuelled by small amounts of \Htwo\ gas. We have not included the effect of the CO-to-\Htwo\ conversion factor in Fig.~\ref{fig:LIR_LCO} as we are only plotting the observed L$^{\prime}_{\rm CO[1-0]}$. Using a lower CO-to-\Htwo\ conversion factor for galaxies A1 and A2 for example, due to their highly merging nature, would increase their SFE with respect to the MS. These conversion factors, as well as the integrated Kennicutt-Schmidt (KS) plane, are discussed in Section.~\ref{sec:mergers}.

\begin{figure}
    \centering
    \includegraphics[width=0.5\textwidth]{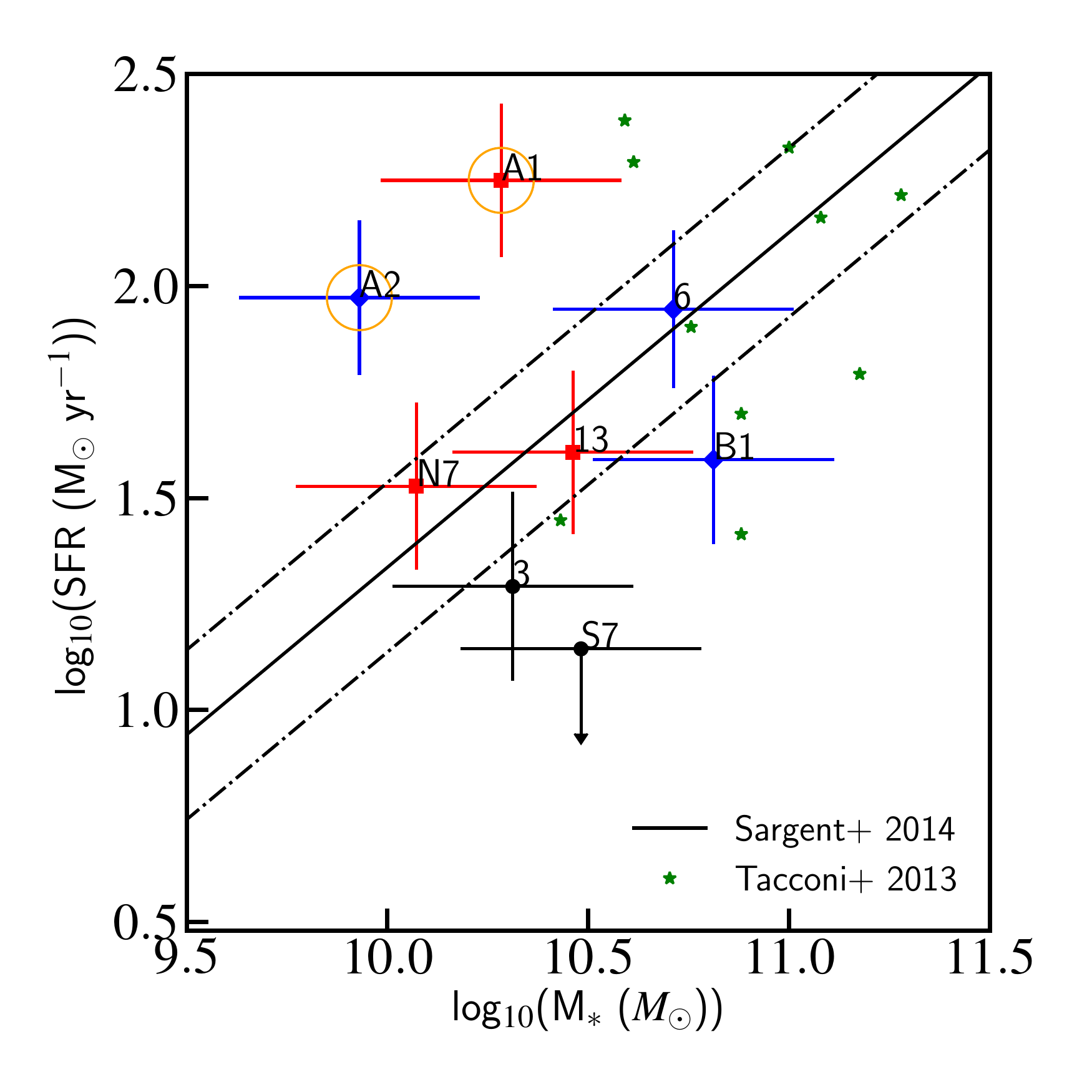}
    \caption{The SFR-M$_{*}$ plane. The MS of star-forming galaxies at z=2 is shown by the solid black line \citep{ref:M.Sargent2014}, and a sample of MS galaxies at z$\sim$2 are shown by the green stars \citep{ref:L.J.Tacconi2013}. The dash-dotted lines represent the 0.2~dex scatter on the MS relation. The cluster galaxies do not show systematic evidence for an increased sSFR with respect to the MS. The errors on the average SFRs include a 0.17~dex systematic uncertainty, in addition to the measurement errors. The colour-coding is as in Fig.~\ref{fig:LIR_LCO}. The orange empty circles highlight galaxies that have had their stellar masses estimated from their dynamical mass, A1 and A2. 2$\sigma$ upper limits are shown. For S7, the SFR has been calculated using the CO[4-3] line flux only, as we do not believe the measured 870$\micron$ flux is associated with S7.}
    \label{fig:SFR_Mstar}.
\end{figure}

\subsection{Dynamical masses of A1 and A2}
\label{sec:dynmass}
In order to discuss several properties of the cluster galaxies, we need a measure of their stellar mass. However, as can be seen from Fig.~\ref{fig:A126Colour}, galaxies A1 and A2 are highly dust obscured at restframe optical wavelengths. This means that stellar masses derived from SED modelling are potentially unreliable for these galaxies. We find that the sizes of the CO[4-3] line emission can be resolved in both A1 and A2, with 2D circular Gaussian FWHM sizes of 0.51" and 0.58" respectively. We do not significantly resolve different sizes for the major and minor axes. We can therefore derive dynamical masses for galaxies A1 and A2, using the resolved CO[4-3] emission size and the linewidths derived from the double Gaussian fitting of the CO[4-3] lines shown in Fig.~\ref{fig:EightLines}. We use the following relation from \citet{ref:E.Daddi2010a}:

\begin{equation}
\label{eqn:dynmass}
M(r<r_{e}) = 1.3 \times \frac{r_{e} \times (v_{fwhm}/2)^{2}}{Gsin^{2}i}
\end{equation}

where $v_{fwhm}$ is the FWHM velocity of the CO double Gaussian, $r_{e}$ is the effective radius of the CO, and $M(r<r_{e})$ is the dynamical mass within that radius. This relation is not expected to vary significantly for mergers near coalescence. We do not have ellipticity information for A1 and A2, so we take the average inclination angle of 57$\pm$21$\degree$, from statistical averages of randomly orientated galaxies. In order to calculate the dynamical mass, we consider the minimum and maximum inclinations from the 1$\sigma$ error on this average, and incorporate a correction factor on the measured $r_{e}$ into Eqn.~\ref{eqn:dynmass}. This correction factor accounts for the fact that the circular, resolved $r_{e}$ that we have measured will underestimate the intrinsic $r_{e}$, depending on the inclination angle of the galaxy. If the galaxies were at high inclination angles they would be almost edge-on, with a high b/a axis ratio. This means that the measured circular Gaussian would return an intermediate size. We therefore take a correction factor of 2 on the radius at high inclination angles. At low inclinations, the galaxies would be almost face-on, so we take a correction factor close to 1. The dynamical mass is therefore calculated taking the mid-point of the mass estimates at high and low inclination, with the error given by half of the spread between the two. The values for $M_{dyn, 2r_{e}}$=~2$\times M(r<r_{e})$, the dynamical mass contained within the whole diameter of the galaxy, are shown in Table~\ref{tab:dyn}.

For the analysis in this paper, we assume that half of the total dynamical mass of galaxies A1 and A2 is coming from the stellar mass, $M_{*, dyn}=\frac{M_{dyn, 2r_{e}}}{2}$, and we assume that the other half of the dynamical mass can be attributed to the molecular gas mass \citep{ref:E.Daddi2010a, ref:M.Aravena2016}. Reasonable upper and lower limits on the relative contributions from molecular gas and stellar mass, M$_{mol,dyn}$:M$_{*,dyn}$ between 1:3 and 3:1, are consistent with the errors on the derived masses. Comparing this assumption with the analysis in \citet{ref:E.Daddi2010a}, we see that the three galaxies for which \citet{ref:E.Daddi2010a} derived dynamical masses had molecular gas contributions between 41\% and 47\% to the total dynamical mass. The galaxies in that study were star-forming isolated BzK galaxies, unlike A1 and A2, and \citet{ref:E.Daddi2010a} assume a 25\% contribution from dark matter. A 50\% contribution from molecular gas therefore seems reasonable for our galaxies, and implies that we are observing our merging systems in an intermediate stage of the merging sequence.

We assume that the contribution from dark matter is negligible. There is evidence that the contribution from dark matter to the dynamical mass at these redshifts is small \citep{ref:E.Daddi2010a, ref:S.Wuyts2016, ref:R.Genzel2017}, and we cannot further refine the contributions from molecular gas, stellar mass and dark matter to the dynamical mass using our data. Additionally, the empirically calibrated stellar mass estimates of A1 and A2 are consistent with the stellar masses that we derive from the dynamical mass within 1$\sigma$.

Values for molecular gas mass derived from the dynamical mass are also shown in Table~\ref{tab:dyn}. Deriving the gas mass using dynamical arguments allows us to estimate values for both $\alpha_{\rm CO}$ and gas-to-dust ratio (G/D), given the measured L$^{\prime}_{\rm CO10}$ and M$_{d}$ derived from observations. This is discussed further in Section~\ref{sec:dust}.

\subsection{Specific star-formation rates}
\label{sec:ssfrs}
We examine the SFR-M$_{*}$ plane in Fig.~\ref{fig:SFR_Mstar}. Multiwavelength data available for the cluster were used for SED fitting with Fitting and Assessment of Synthetic Templates (FAST) \citep{ref:M.Kriek2009}, in order to derive the galaxy stellar masses shown in Table~\ref{tab:SFRs}, as described in \citet{ref:V.Strazzullo2016}. Stellar masses for the sources S7 and N7, which are very close to each other and thus blended in all imaging besides HST, were derived from the F140W flux density and F105W-F140W colour by empirical calibration obtained from field (GOODS-S) galaxies in a similar redshift and magnitude range. We use photometry from \citet{ref:Y.Guo2013} and photometric redshifts, stellar mass estimates, and modelled SEDs from \citet{ref:C.Schreiber2015} and \citet{ref:M.Pannella2015}, following the same procedure detailed in \citet{ref:V.Strazzullo2016}. The stellar mass estimates for A1 and A2 were derived from dynamical arguments, described in Section~\ref{sec:dynmass}.
              
Interestingly, although a high fraction of these galaxies show evidence for a starburst-like star-formation mode in terms of their gas excitation, they do not appear to preferentially reside in the starburst regime on the SFR-M$_{*}$ plane, shown in Fig.~\ref{fig:SFR_Mstar}. Typically, dusty starburst galaxies that have a high SFE are also found to be elevated in specific star-formation rate (SFR per unit stellar mass) - therefore residing above the MS of star-formation (e.g. \citealt{ref:E.Daddi2010b, ref:J.Silverman2015, ref:A.Puglisi2017}). However, for our cluster galaxies, both the SB-excitation and the BzK excited galaxies scatter around the MS of star-formation at z=2, without any systematic evidence of a heightened sSFR. It can be seen in Fig.~\ref{fig:SFR_Mstar} that only galaxies A1 and A2 lie above the scatter of the MS, with some of the less highly star-forming galaxies such as IDs 3, B1 and S7 lying below the MS. It should be noted that the SFR of S7 is from CO[4-3] only, as we do not associate any 870$\micron$ continuum flux with S7, as discussed above.

It is somewhat surprising that a galaxy best fit by a BzK excitation, A2, has one of the highest sSFRs. It should be noted that A1 and A2 appear to be highly dust obscured in the optical bands, and thus the stellar masses shown here have been derived from the galaxies' dynamical masses (see Section~\ref{sec:dynmass}). In addition, our cluster galaxies do not appear to be typical isolated star-forming disks, and therefore may not follow the expected trends of other star-forming MS and SB galaxies at z=2. This is discussed further in Section~\ref{sec:discussion}.

\begin{table}
\centering
\begin{tabular}{ccccc}
 \hline
ID & $M_{dyn,2r_{e}}$ $(M_{\odot})$ & M$_{mol, dyn}$  $(M_{\odot})$ & M$_{*, dyn}$ $(M_{\odot})$ \\
 \hline
A1 & 10.58$\pm$0.3 & 10.28$\pm$0.3  &  10.28$\pm$0.3 \\
A2 & 10.23$\pm$0.3 & 9.93$\pm$0.3 &  9.93$\pm$0.3 \\
\end{tabular}
\caption{The dynamical masses and the inferred molecular gas and stellar masses for galaxies A1 and A2. We have assumed that the gas and stellar masses are both equal to half of the dynamical mass of the galaxies, but the 0.3~dex errors allow for the gas mass to be between 25-100\% of the dynamical mass. We have neglected the small contribution from dark matter.}
\label{tab:dyn}
\end{table}

\begin{figure*}
    \centering
    \includegraphics[width=\textwidth]{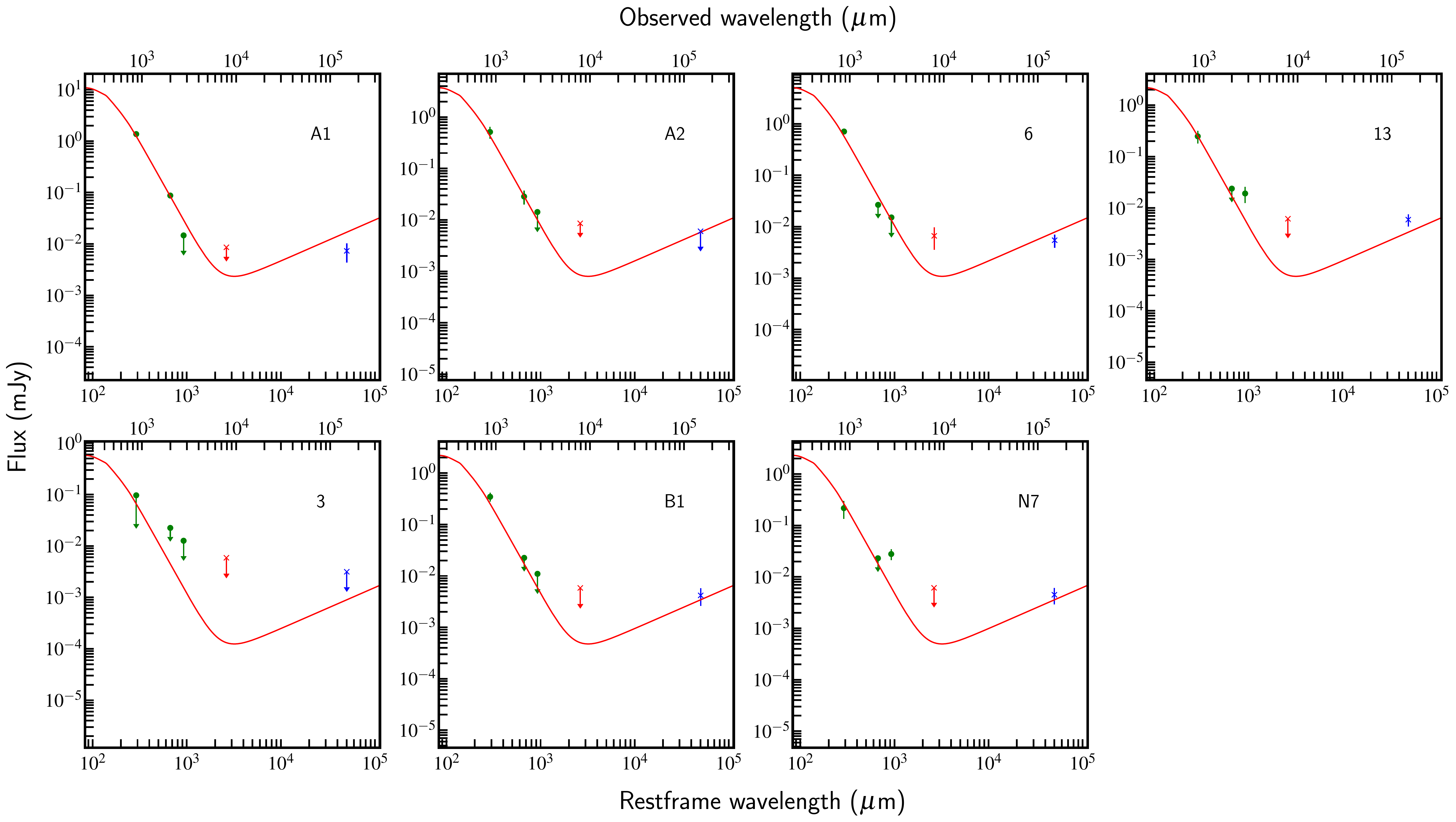}
    \caption{The spectral energy distributions of the line-free continuum data from 870$\micron$ to 3~GHz observed frequencies. The data primarily constraining the dust masses are shown by green dots (870$\micron$, 2mm and 3mm observed). The observed 7mm continuum underneath the CO[1-0] line is shown by the red crosses, and the 3~GHz continuum is shown by the blue crosses. 2$\sigma$ upper limits are shown. The starburst template SED from \citet{ref:M.Bethermin2015} is shown by the red solid line, normalised by the green data on the RJ tail.
    }
    \label{fig:SEDs}
\end{figure*}

\begin{figure*}
    \centering
    \includegraphics[width=\textwidth]{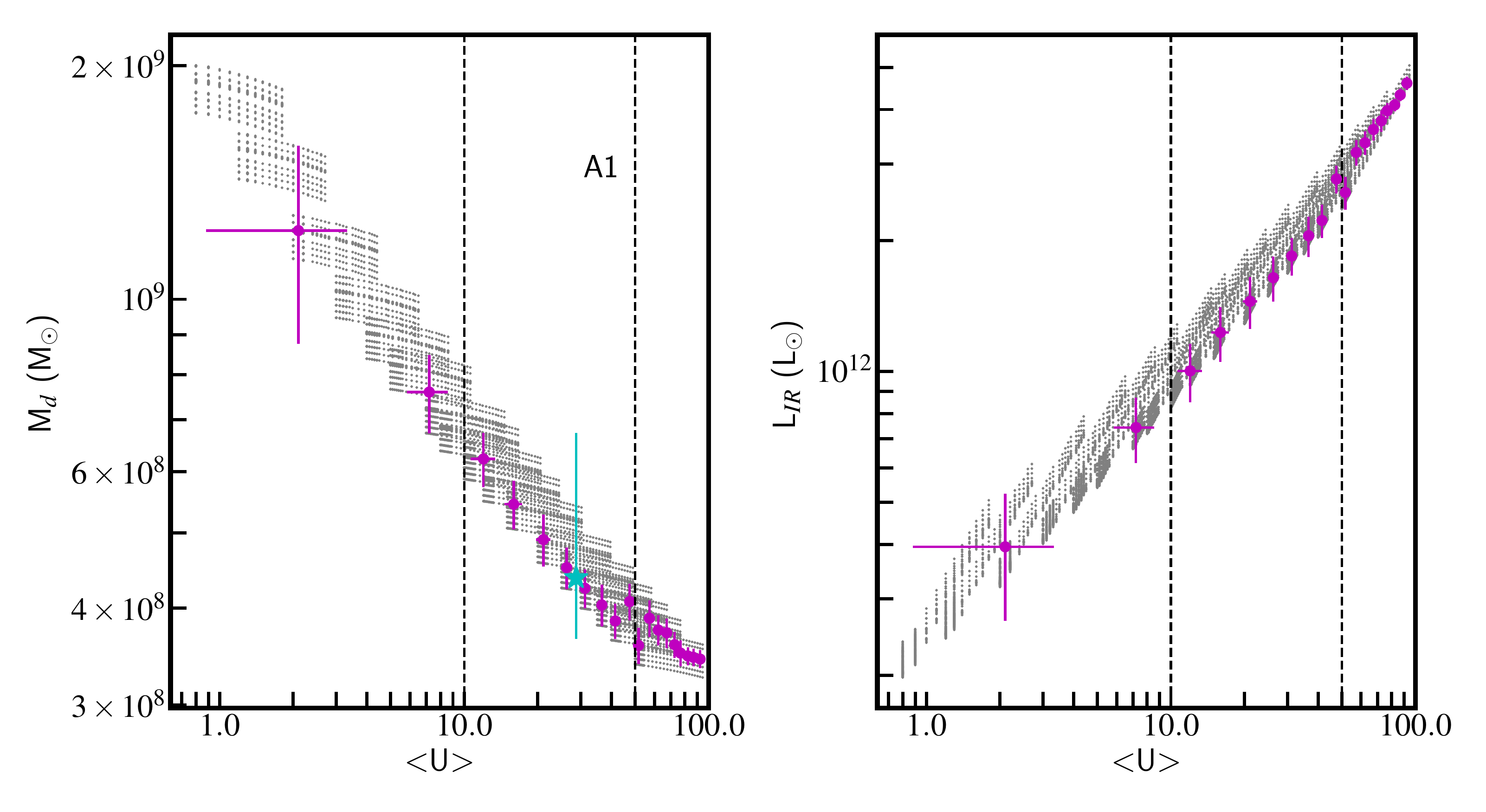}
    \caption{The variation of dust mass (left) and L$_{IR}$ (right) with the average radiation field energy density, $<$U$>$, for galaxy A1. The grey crosses show all of the permutations for more minor fit parameters, and the magenta circles show the median values in bins of $\Delta$($<$U$>$) = 5. The error bars on the median values give the standard deviation of each quantity in the bin. For our analysis, we consider the range 10$<$ $<$U$>$ $<$50, which is contained within the two black dashed vertical lines. Left: the cyan star within this range is the median dust mass used for our analysis, at the median $<$U$>$=28.75. The cyan error bars show the full spread of the binned magenta data, including the error bars, within this acceptable $<$U$>$ range. This is approximately a factor of two, and is taken to be the error on the derived dust mass. Right: we illustrate the dependence of L$_{IR}$ on $<$U$>$ over the same $<$U$>$ range. We see a factor of 3.5 difference in the possible values of L$_{IR}$ between 10$<$ $<$U$>$ $<$50.} 
    \label{fig:MdustDerivation}
\end{figure*}

\begin{figure*}
    \centering
    \includegraphics[width=\textwidth]{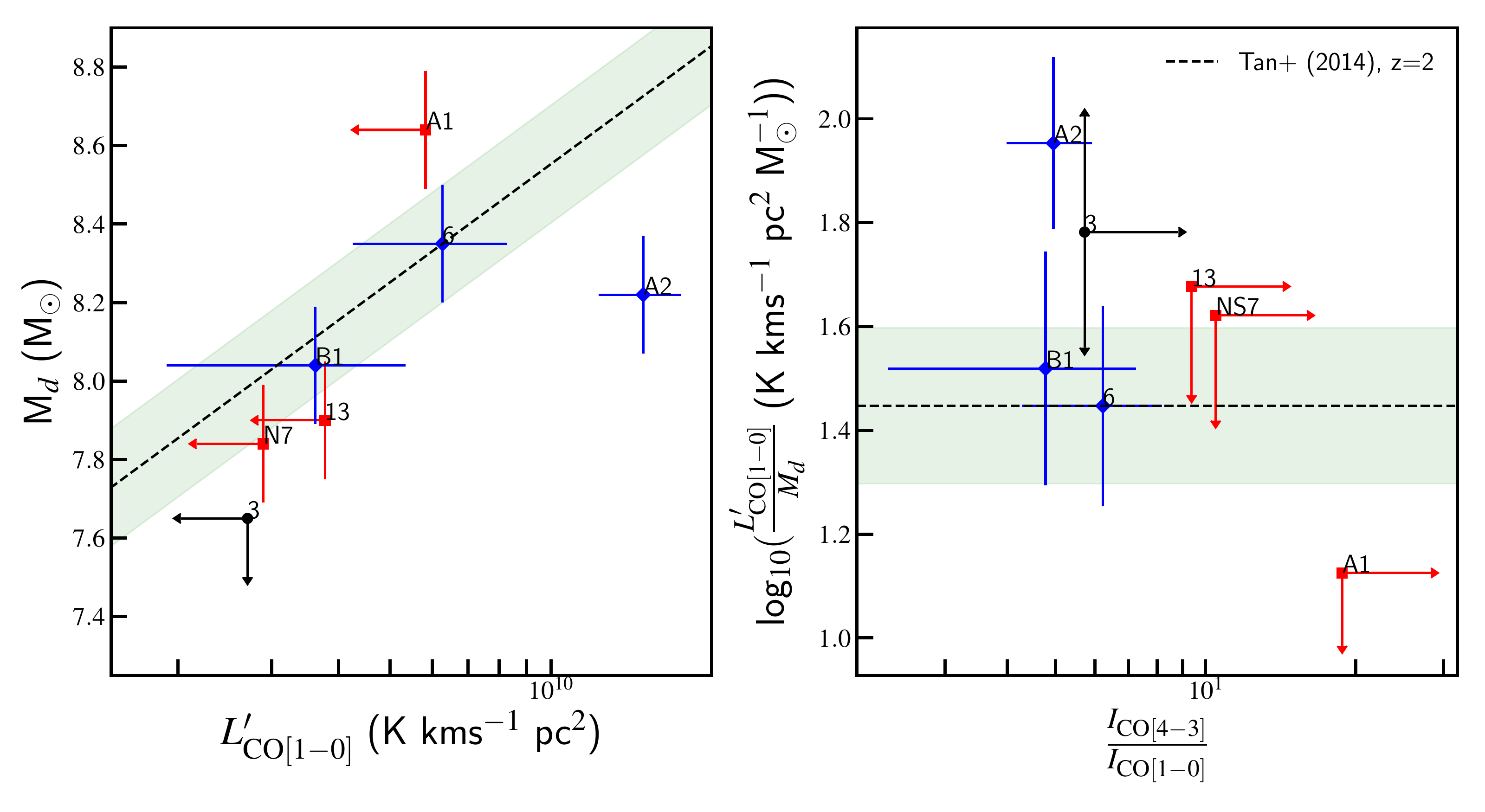}
    \caption{Left: Dust mass against L$^{\prime}_{\rm CO[1-0]}$. The expected MS trend from \citet{ref:Q.Tan2014} is shown by the black dashed line in both panels. The green bands represent the $\pm$0.15~dex scatter on the relations, from \citet{ref:E.Daddi2010a}. Right: $\frac{L^{\prime}_{CO[1-0]}}{M_{d}}$ as a function of excitation ratio. The galaxies scatter around the expected relation, with the possible exception of A1, discussed in the text. The symbol colour-coding is as in Fig.~\ref{fig:LIR_LCO}. 2$\sigma$ upper limits are shown.}
    \label{fig:Mdust_excitation}
\end{figure*}

\subsection{Dust and \Htwo\ masses}
\label{sec:dust}

\begin{table*}
\begin{tabular}{cccccccccc}
\hline
ID & $\alpha_{\rm CO}$ & G/D & log(M$_{d}$) & log(M$_{mol, dyn}$)  & log(M$_{mol, CO}$) & log(M$_{mol, d}$) & $\mu_{g}$ & log($\tau_{dep}$) \\
 & $(K kms^{-1} pc^{2} M_{\odot}^{-1})^{-1}$ & & $(M_{\odot})$ & $(M_{\odot})$ & $(M_{\odot})$ & $(M_{\odot})$ &  & (yrs)\\
\hline
A2 & 0.6$^{*}$ & 51$^{*}$ & 8.22$\pm$0.15 & 9.93$\pm$0.30 & - & - & 1.00$^{*}$$^{+1.66}_{-0.62}$ & 7.98$\pm$ 0.35 \\
A1 & 3.6$^{*}$ & 44$^{*}$ & 8.64$\pm$0.15 & 10.28$\pm$0.30 & - & - & 1.00$^{*}$$^{+1.66}_{-0.62}$ & 8.03$\pm$ 0.35 \\
13 & 4.5 & 168 & 7.90$\pm$0.15 & - &$<$10.25 & 10.12$\pm$0.15 & 0.46$^{+0.54}_{-0.25}$ & 8.53$\pm$ 0.24 \\
6 & 4.2 & 149 & 8.35$\pm$0.15 & - & 10.44$\pm$0.14 & 10.52$\pm$0.15 & 0.64$^{+0.75}_{-0.35}$ & 8.58$\pm$ 0.24 \\
B1 & 3.9 & 128 & 8.04$\pm$0.15 & - & 10.18$\pm$0.21 & 10.15$\pm$0.15 & 0.22$^{+0.25}_{-0.12}$ & 8.56$\pm$ 0.25 \\
3 & 4.7 & 180 & $<$7.65 & - & $<$10.13 & 9.90$\pm$0.15 & 0.39$^{+0.46}_{-0.21}$ & 8.61$\pm$ 0.27\\
N7 & 4.2 & 146 & 7.84$\pm$0.15 & - & $<$10.10 & 10.00$\pm$0.15 & 0.86$^{+1.00}_{-0.46}$ & 8.49$\pm$ 0.25 \\
\end{tabular}
\caption{$\alpha_{\rm CO}$, G/D ratios, dust masses, molecular gas masses from the dynamical mass, molecular gas masses from CO, molecular gas masses from dust, gas to stellar mass ratios, $\mu_{g}$, and gas depletion timescales $\tau_{dep}$. The $\alpha_{\rm CO}$ and G/D ratios of A1 and A2, denoted by a star, have been directly derived assuming a molecular gas mass from the dynamical mass, and therefore should not be used to derive molecular gas masses. For the rest of the galaxies, $\alpha_{\rm CO}$ and G/D have been calculated using the metallicity derived from the FMR. The gas to stellar mass ratios and gas depletion timescales have been calculated using the gas mass from dust, using the G/D given. In the cases of A1 and A2, again denoted by a star, this ratio is 1 by construction from the dynamical mass calculation, as discussed in Section~\ref{sec:dynmass}. 2$\sigma$ upper limits are given.} 
\label{tab:Dustmasses}
\end{table*}

\subsubsection{Dust masses}
As discussed in Section~\ref{sec:sfrs}, in addition to the detection of molecular gas in the cluster galaxies, we also detect significant 870$\micron$ continuum flux. This is a direct detection of the galaxies' dust emission. By combining the flux measurements at 870$\micron$, 3~GHz and continuum flux underneath the three CO lines, we are able to construct the submillimetre-to-radio portion of each of the galaxies' SEDs. These are displayed in Fig.~\ref{fig:SEDs}. It can be seen that this cluster is not particularly prominent at 3~GHz, unlike the high-z cluster presented in \citet{ref:T.Wang2016}. It is therefore not a strong over-density of radio sources, which are expected to become more common at z$>$2 \citep{ref:E.Daddi2017}. The slope of the Rayleigh-Jeans (RJ) tail of the SED is not expected to vary significantly with dust temperature, and we choose to compare our galaxy SEDs with a warm starburst SED template in Fig.~\ref{fig:SEDs}, based on the galaxy properties displayed in Figs~\ref{fig:SLEDs} and \ref{fig:LIR_LCO}. The RJ slope of the starburst template was fixed at $\alpha$=-3.24, measured from the template itself, and the best-fitting normalisation of the template to the data was calculated using a least-squares fitting procedure. For this fitting, only the observational data in the RJ tail, shown by green circles in Fig.~\ref{fig:SEDs}, were used. In all cases, the data themselves are used for the fitting, not the measurement upper limits. As can be seen from Fig.~\ref{fig:SEDs}, the slope of this template is a good fit to the data in most cases.

Dust masses were derived using a range of templates to calculate the bolometric luminosity under the RJ tail of the galaxies' SEDs. We modelled the data using 2-component dust models from \mbox{\citet{ref:DraineLi2007}}, and calculated the goodness-of-fit of each combination of parameters. The primary parameters that drive the dust mass are the average radiation field energy density, $<$U$>$, and the fractional contribution of the coldest dust to the model, $\gamma$. We then considered only values for dust mass, M$_{d}$, with parameter solutions in the range $<$U$>$=10-50, and $\gamma$<0.2. Typical MS-galaxies tend to have $<$U$>$$\sim$20, and $\gamma$ close to zero \mbox{\citep{ref:G.Magdis2012, ref:M.Bethermin2015}}. Highly star-forming galaxies, such as GN20, have $<$U$>$$\sim$35 and $\gamma$$\sim$0.18. We therefore choose these parameter constraints because they encompass a realistic but open range of galaxy properties. Having excluded $\gamma\geq$0.2, we bin the solutions of M$_{d}$ as a function of $<$U$>$ into bins of $\Delta$($<$U$>$)~=~5 for each galaxy, and examine the resulting trend. This is shown in Fig.~\ref{fig:MdustDerivation} for galaxy A1, alongside the dependence of L$_{IR}$ over the same range of $<$U$>$. Looking within our accepted range of $<$U$>$, we find that the derived M$_{d}$ varies by a factor of two, depending on the assumed $<$U$>$. We therefore take the median M$_{d}$ of the data binned in this range, at $<$U$>$=28.75. We take the entire spread of the binned dust masses and their errors, between $<$U$>$=10-50, as the error on the dust mass. The dust masses derived in this way are shown in Table~\ref{tab:Dustmasses}.

\begin{figure*}
    \centering
    \includegraphics[width=\textwidth]{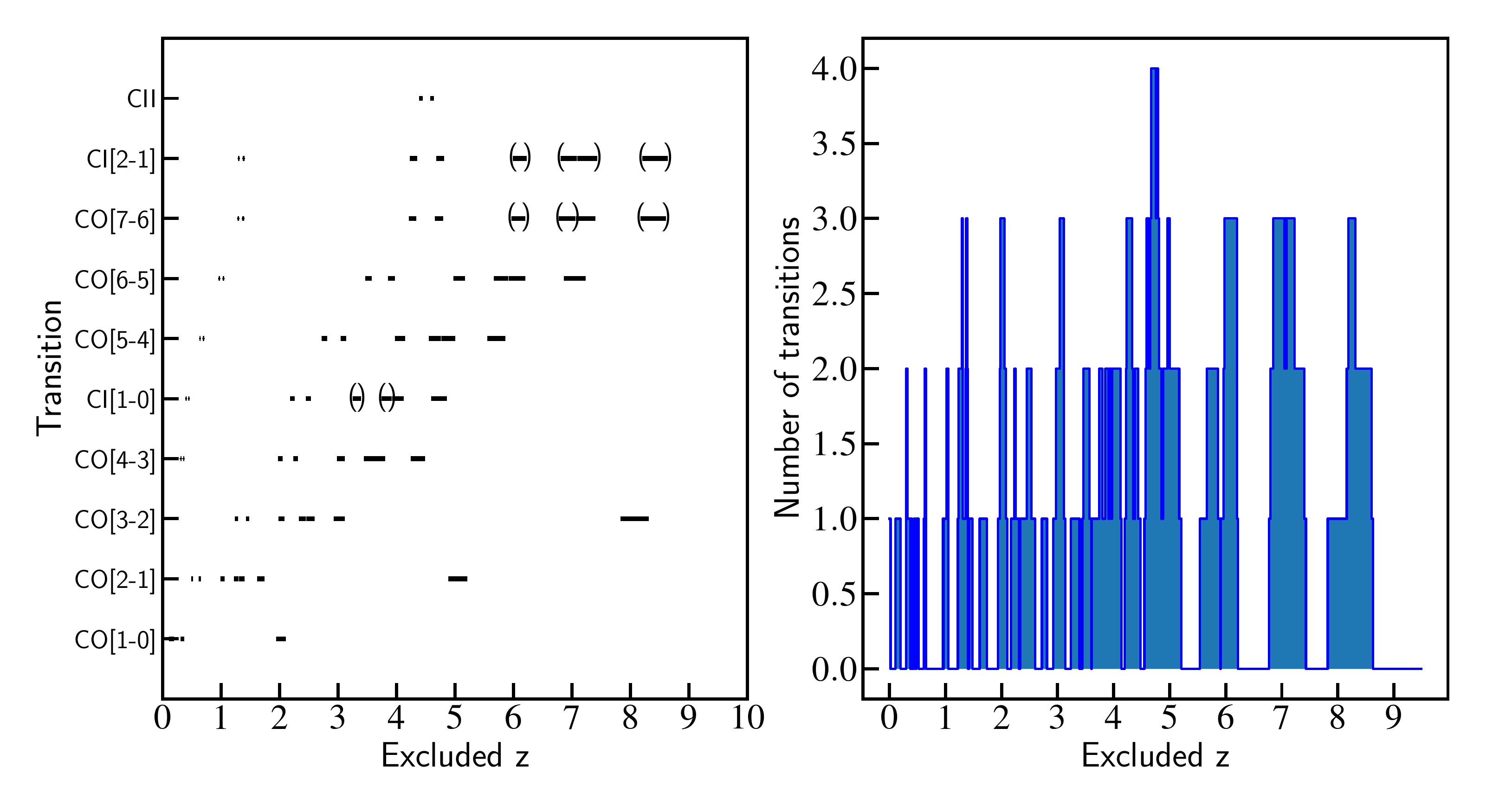}
    \caption{Left: Bright gas transitions at radio/submillimetre frequencies vs. redshift. For each transition that is not detected in our datasets for galaxies A4 and A5, we can exclude a range of redshifts based on the rest frequency of the line and the frequencies covered by our datasets. Different transitions therefore allow us to exclude different redshifts for each of our datasets. The parenthesised transitions would be detected at S/N$<$5 for the fainter galaxy, A4. Right: Histogram of the number of different transitions in the left panel that exclude each $\Delta z = 0.01$ redshift bin. The white gaps on the x-axis therefore highlight redshift ranges that have not been probed in line emission by any of our datasets.}
    \label{fig:Transition_z}
\end{figure*}

The galaxies N7, 6 and 13 show evidence for excess flux compared to the SED templates at longer wavelengths, shown in Fig.~\ref{fig:SEDs}. However, we do not comment further on this given the relatively moderate significance of the excesses. We have recently received a deeper continuum dataset at these wavelengths (currently under analysis), and so we leave interpretation of these possible excesses for a future publication. In order to ensure that the derived dust masses were not affected by these unusual SED shapes, we calculate the average dust-to-870$\micron$ flux ratio using galaxies A1, A2 and 6, our best-constrained galaxies at 870$\micron$. This gives an average value of r$_{d/870}$~=~3.19$\times$10$^{8}$M$_{\odot}$mJy$^{-1}$. When we apply these dust-to-870$\micron$ ratios to the 870$\micron$ detections of galaxies 6, N7 and 13, we recover the same values for dust mass as we found using our original method. This confirms that our dust mass derivations are robust.

Having derived dust masses for each of the galaxies, we are able to compare the dust masses with the CO[1-0] line luminosities. In Fig.~\ref{fig:Mdust_excitation}, we show the relationship between these two quantities, as well as the $\frac{L^{\prime}_{\rm CO[10]}}{M_{d}}$ ratio as a function of gas excitation. As discussed in Section ~\ref{sec:sims}, we correct the CO[1-0] luminosities of the cluster galaxies, increasing their values (and errors) by 6\%, as indicated by our simulations. Here we define the excitation as the ratio between the CO[4-3] and CO[1-0] line fluxes. As both $\alpha_{\rm CO}$ and G/D vary with metallicity in the same way, we expect a constant ratio between L$^{\prime}_{\rm CO[1-0]}$ and M$_{d}$ in order to recover consistent molecular gas masses \citep{ref:Q.Tan2014}. This is shown by the dashed line at $\frac{L^{\prime}_{\rm CO[1-0]}}{M_{d}}\sim$28 in both panels. Fig.~\ref{fig:Mdust_excitation} shows that the cluster galaxies scatter around the expected L$^{\prime}_{\rm CO[1-0]}$ - M$_{d}$ relation at z=2. Although we cannot constrain all of the $\frac{L^{\prime}_{\rm CO[10]}}{M_{d}}$ ratios due to our upper limits in CO[1-0], this confirms that our gas and dust properties are consistent with each other, and we have no reason to believe that an unusual ratio between the two is causing the heightened excitation fraction in the cluster or increased SFEs. An exception to this is galaxy A1, which appears to have a relatively large dust mass, or a relatively small L$^{\prime}_{\rm CO[1-0]}$, compared to the expected abundances. Possible interpretations of this will be discussed in Section~\ref{sec:mergers}.

\subsubsection{Molecular gas masses}
\label{sec:gas}
We now derive molecular gas masses of the galaxies using two different tracers - the dust mass and the CO[1-0] line luminosity. As previously stated, the gas-to-dust ratio is dependent on metallicity, as is the CO-to-\Htwo\ conversion factor, $\alpha_{\rm CO}$. We find no sub-mm or radio detections for the cluster galaxies that we have previously derived metallicity measurements for \citep{ref:F.Valentino2015}, and we do not have gas-phase metallicity estimations for the galaxies presented here with CO[4-3] line detections. Therefore, for all of our galaxies except A1 and A2, we use the Fundamental Metallicity Relation (FMR) \mbox{\citep{ref:C.Mannucci2010}} to derive statistical estimates of the metallicity, using the galaxy stellar masses and average SFRs. It has been suggested that the FMR is applicable for star-forming galaxies up to z$\sim$2.5, although this has proved difficult to observationally confirm and is an active area of current research (e.g. \citealt{ref:R.Sanders2015}, \citealt{ref:M.Onodera2016}). We then use the frameworks given by \citet{ref:M.Sargent2014} and \citet{ref:A.RemyRuyer2014} to derive $\alpha_{\rm CO}$ values and G/D ratios respectively. The results of these calculations are shown in Table~\ref{tab:Dustmasses}. It is important to note that these frameworks were calibrated on disk-like MS galaxies, and the galaxies in our sample are therefore likely to deviate from these relations.

For galaxies A1 and A2, we have derived the molecular gas mass from the galaxy dynamical mass, and we are therefore directly estimating values for $\alpha_{\rm CO}$ and G/D from L$^{\prime}_{\rm CO[1-0]}$ and M$_{d}$ respectively. For the rest of the galaxies, comparisons between the molecular gas masses derived using $\alpha_{\rm CO}$ and G/D are shown in Table~\ref{tab:Dustmasses}. The gas masses derived from the G/D ratio are in fact the combination of both the molecular and the atomic Hydrogen gas in the galaxies, but we are assuming that the contribution from atomic Hydrogen in the ISM at z$\sim$2 is negligible. We do not show errors on the conversion factors in Table~\ref{tab:Dustmasses}, as the errors are likely to be dominated by the systematic uncertainties in their derivation, due to the possibly non-standard nature of these sources. We discuss the implications of these uncertainties in Section~\ref{sec:discussion}.

\subsection{Unidentified bright sources}
\label{sec:A4A5}
Here we discuss the two brightest sources in our $870\micron$ continuum images, A4 and A5, and their relationship with the cluster. A4 and A5 are shown in Fig.~\ref{fig:ContData}. Their 870$\micron$ fluxes are given in Table~\ref{tab:contdata}, at 1.8mJy and 6.0mJy respectively, implying high star-formation rates (shown in Table~\ref{tab:SFRs}). In calculating these SFR estimates, we have assumed a redshift of z=2.0, and an average conversion factor between the SB and MS templates from \mbox{\citet{ref:M.Bethermin2015}}, as with the rest of the galaxies in Table~\ref{tab:SFRs}. The negative K-correction of the flux at this wavelength means that our SFR estimates are not highly sensitive to the assumed redshift \citep{ref:C.Casey2014}. Despite these bright sub-mm fluxes, we detect no spectral lines for either of these galaxies in any of the sub-mm/radio datasets discussed in this paper, which were chosen to surround the expected frequencies of lines from galaxies at z$\sim$2. Additionally, we have thus far been unable to constrain the redshifts of these galaxies using optical spectroscopy, using data from the VLT and HST/WFC3. In particular, A4 has a relatively faint optical counterpart.

However, we can begin to exclude candidate redshifts for the galaxies, based on the lack of spectral lines. Using our current data, we can already exclude a large number of redshifts for these galaxies. We have created a redshift exclusion plot in Fig.~\ref{fig:Transition_z}, based on the non-detection of the strongest lines at submillimetre-radio frequencies that we would expect to observe in such star-forming galaxies, based existing correlations with LIR  \mbox{\citep{ref:F.Walter2011, ref:J.Spilker2014}}. Based on the non-detection of each of these lines, the line restframe frequencies and the frequencies covered by our datasets, we show in Fig.~\ref{fig:Transition_z} which redshift ranges can be excluded by each transition. We show that $\sim$80\% of the redshift range 2$<$z$<$5 can already be excluded by these transitions. We would expect to detect all of the transitions given in Fig.~\ref{fig:Transition_z} at S/N$>$8 for galaxy A5 at the redshifts shown, based on their correlations with FIR. For A4, we would expect to detect all transitions at S/N$>$5 at the given redshifts, with the exception of those transitions that are parenthesised in the left panel of Fig.~\ref{fig:Transition_z}. It can be seen that this does not significantly affect the excluded redshift range. These calculations were conservatively based on a ULIRG excitation template between CO transitions, and a 400\kms\ linewidth, based on the properties of the cluster galaxies listed in Table~\ref{tab:CO43data}. It is also particularly unlikely that A4 and A5 are at z$<$1. The measured ratio of the 870$\micron$ to 3~GHz flux, the peak flux measured from previous SPIRE/PACS observations, as well as photo-z SED fitting all strongly exclude z$<$1 for A5 \mbox{\citep{ref:V.Strazzullo2013}}. Fig.~\ref{fig:Transition_z} also shows that more than 30\% of redshifts at z$<$1 are excluded, based on the non-detection of low-J CO lines. In Fig.~\ref{fig:Transition_z}, we bin candidate redshifts between z=0 and z=10 into widths of $\Delta$z=0.01, and for each redshift bin we count the number of transitions in the left panel that exclude this redshift. We see that a large range of redshifts are excluded using one or more non-detected transitions, including the redshift of \Cl, z=1.99.

It is difficult to speculate whether these galaxies are likely to be cluster members. As discussed in \citet{ref:V.Strazzullo2017}, based on the non-detection of spectral lines at z$\sim$1.99, we must presently conclude that these galaxies are not cluster members. However, given observed number counts of field galaxies at 870$\micron$ \citep{ref:A.Karim2013}, the probability of finding two galaxies this bright at 870$\micron$ by serendipitous detection alone, within 8" from the centre of the cluster, is extremely low, $<$4$\times$10$^{-5}$. This hints at cluster membership for both galaxies. This could potentially be reconciled if the galaxies are strongly lensed, but this is unlikely to be the case based on the halo mass of the cluster \citep{ref:R.Gobat2011}. We also see differences in the ratios of continuum fluxes at different wavelengths for A4 and A5, meaning that they do not have the same SED shape. This further suggests that A4 and A5 are not lensed projections of the same distant galaxy. The optical morphologies do not appear distorted in Fig.~\ref{fig:ContData}, and we don't identify suitable lensing candidates. The gravitational potential of the intra-cluster-medium of \Cl\ is not sufficient, and there are no obvious foreground galaxy candidates.

On the other hand, we would expect to see bright accompanying CO lines in A4 and A5 based on the 870$\micron$ dust emission. For example, we would expect to see the CO[4-3] transition at S/N$>$100 in A5 at z=2, in our deep 2mm dataset. The measured 3$\sigma$ upper limit on this line flux in A5 would therefore reveal a suppression of the CO emission by a factor $>$40 if A5 were in the cluster. If these bright galaxies are indeed cluster members, we would be seeing evidence of the very onset of star-formation quenching. Recent, very rapid quenching would be demonstrated by suppressed CO emission compared to the dust emission, because the gas and dust trace the star-formation on different timescales. CO is an instantaneous density tracer, whilst the dust emission traces the OB star population, and therefore takes $\sim$100~Myrs to decrease after the cessation of star-formation. This would indicate that the star-formation in these galaxies had fallen from hundreds of solar masses per year to zero in $<$100~Myrs, which would be a very extreme case. Finding galaxies in this narrow window of quenching would be extremely exciting. There are however several redshifts that are not probed by our datasets, shown in Fig.~\ref{fig:Transition_z}, including a small range near the photometric redshift of A5, z$\sim$2.8. We hope that future observations targeting these yet-unprobed redshift ranges will shed more information on the nature of these extremely bright galaxies.
   
\begin{figure}
    \centering
    \includegraphics[width=0.5\textwidth]{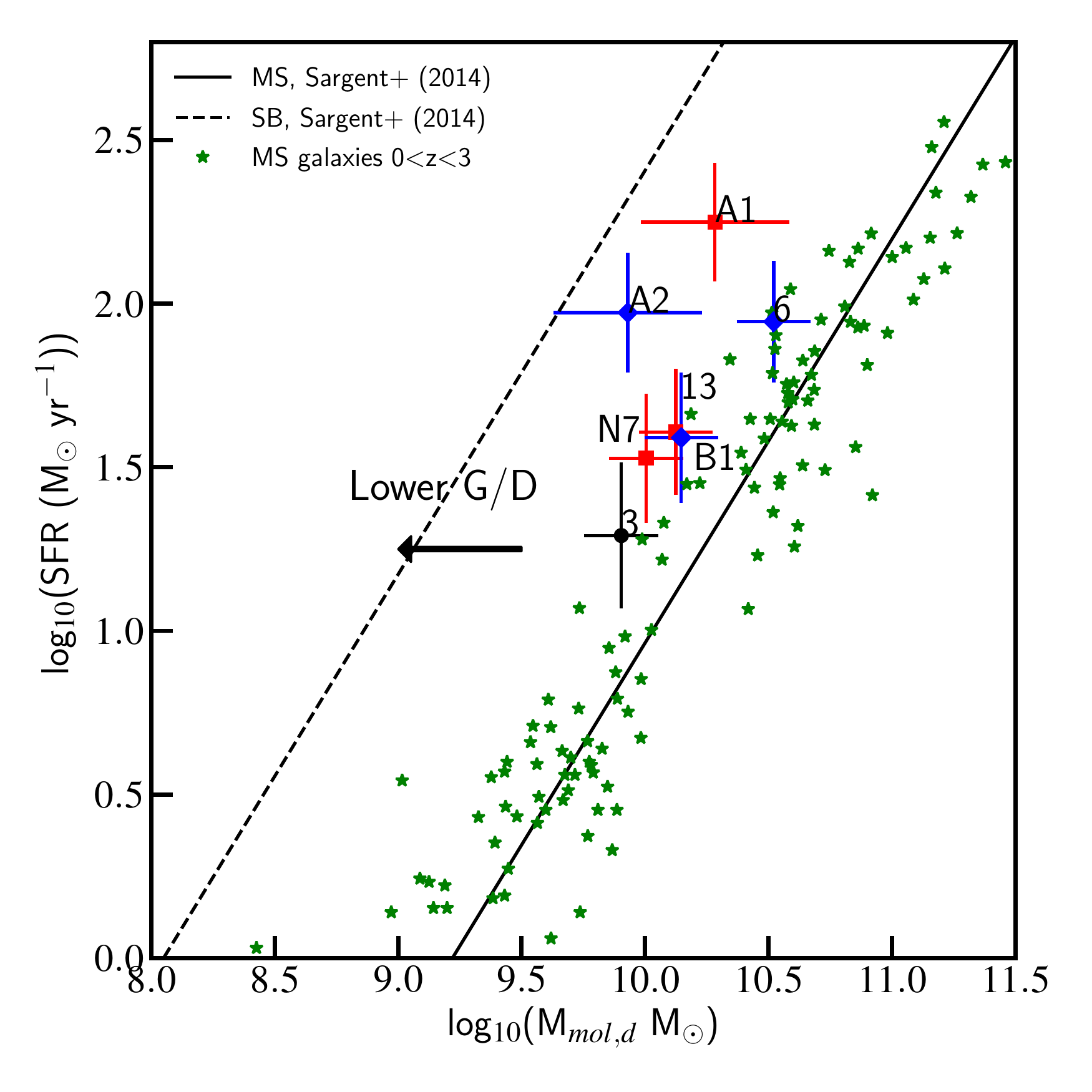}
    \caption{The integrated KS law. The colour-coding is as in Fig.~\ref{fig:LIR_LCO}. MS galaxies from previous studies are shown by the green stars \citep{ref:E.Daddi2010b, ref:J.Geach2011, ref:G.Magdis2012, ref:A.Bauermeister2013, ref:L.J.Tacconi2013}, and it can be seen that the field correlation around the locus of MS galaxies is tighter here than in the SFR-L$^{\prime}_{\rm CO[1-0]}$ plane, Fig.~\ref{fig:LIR_LCO}. The cluster galaxies all lie to the left of the MS relation, at enhanced star-formation efficiencies. The black arrow indicates the direction that the cluster galaxies would move if lower gas-to-dust ratios were used to derive their gas masses. This is highly relevant, as the G/D ratios derived from the FMR are currently all $>$100, and are not necessarily representative of a galaxy population containing a high fraction of mergers and AGN.}
    \label{fig:RealKS}
\end{figure}

\begin{figure}
    \centering
    \includegraphics[width=0.5\textwidth]{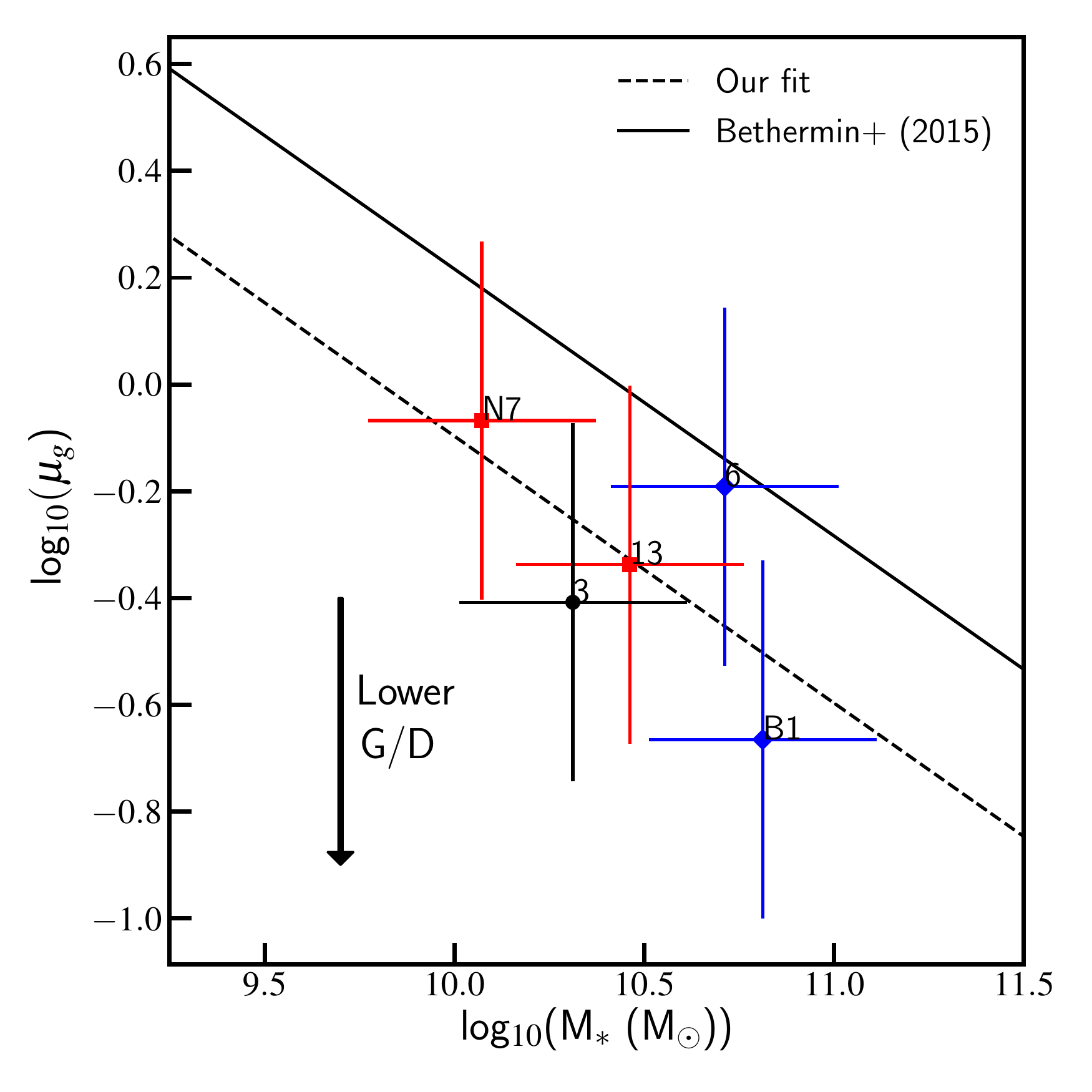}
    \caption{The gas-to-stellar mass ratio as a function of stellar mass. The black dashed line is the relation from \citet{ref:G.Magdis2012}, normalised to our data using a least-squares approach. The solid black line is the z=2 relation from \citet{ref:M.Bethermin2015}. The colour-coding is as in Fig.~\ref{fig:LIR_LCO}. A1 and A2 are not shown here, as their gas-to-stellar mass fractions have been set equal to 1 by construction. The normalisation of the trend for our cluster galaxies lies at lower gas-to-stellar mass ratio than for the \citet{ref:M.Bethermin2015} MS relation. The black arrow indicates the direction and extent that the cluster galaxies would move if lower gas-to-dust ratios were used to derive their gas masses.}
    \label{fig:gasfrac_stellar}
\end{figure}

\begin{figure}
    \centering
    \includegraphics[width=0.5\textwidth]{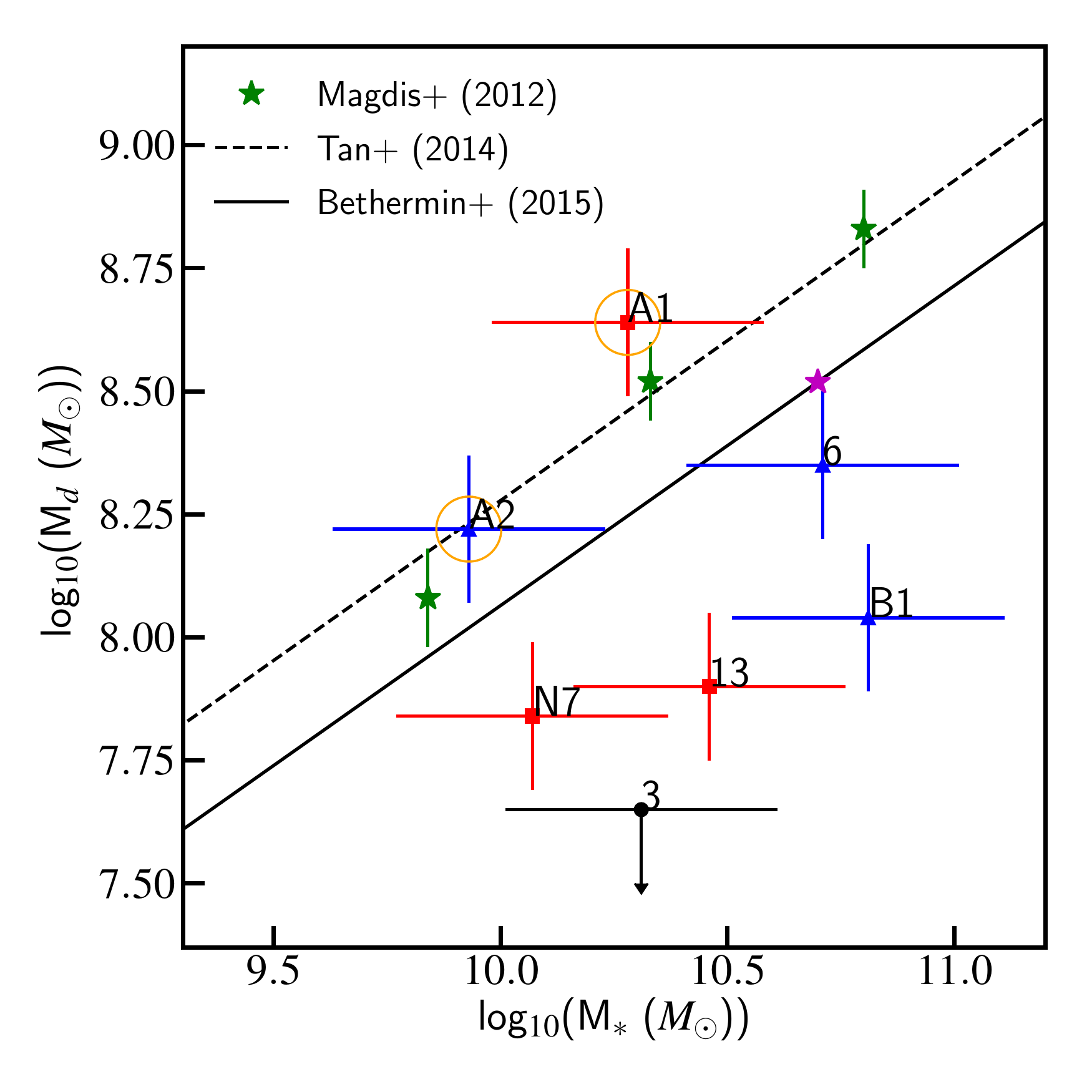}
    \caption{Dust mass vs. stellar mass. The colour-coding and symbols are as in Fig.~\ref{fig:LIR_LCO}. The green stars show the z$\sim$2 MS galaxies, stacked in bins of stellar mass, from \citet{ref:G.Magdis2012}. The black dashed line is the relation given by \citet{ref:Q.Tan2014}, normalised to the green data points. The black solid line is the normalisation given by the MS galaxies in \citet{ref:M.Bethermin2015}, with the same slope as in \citet{ref:Q.Tan2014}. The purple star is the mean mass of the \citet{ref:M.Bethermin2015} sample, used for normalisation. Several of the cluster galaxies show evidence for low dust-to-stellar mass fractions compared to these relations. 2$\sigma$ upper limits are shown.}
    \label{fig:MdustVsMstar}
\end{figure}

\section{Discussion}
\label{sec:discussion}

\subsection{Merger-driven enhancement in gas excitation and star-formation}
\label{sec:mergers}
We have discovered a population of cluster galaxies containing significant amounts of excited, dense molecular gas, shown by bright CO[4-3] line detections (Fig.~\ref{fig:EightLines}). The SLEDs of the galaxies for which we characterise the CO excitation state reveal that 50\% have starburst-like gas excitation properties. The offsets of the cluster galaxies towards the starburst SFR-L$^{\prime}_{\rm CO[1-0]}$ relation shown in Fig.~\ref{fig:LIR_LCO} also support this key conclusion. Having derived molecular gas masses for our galaxies, we can now look at their placement on the integrated Kennicutt-Scmidt plane in Fig.~\ref{fig:RealKS}. The molecular gas masses derived from the CO[1-0] gas and the dust respectively, shown in Table~\ref{tab:Dustmasses}, are found to be in good agreement, so we continue the analysis using the molecular gas mass calculated from the dust, as it is less affected by measurement upper limits than the CO[1-0] transition. The molecular gas masses for A1 and A2 are derived from the dynamical mass. We see in Fig.~\ref{fig:RealKS} that there is a tight scatter in previous field studies around the locus of MS galaxies, and all of our cluster galaxies lie to the left of this relation, at enhanced SFEs. The median offset from the MS locus is 0.4~dex, meaning that our cluster galaxies contain on average 0.4~dex less molecular gas for their SFR compared to the MS galaxies. We see that some of our galaxies lie on the edge of the MS scatter, but it is highly plausible that the molecular gas mass has been over-estimated for these galaxies. As can be seen in Table~\ref{tab:Dustmasses}, the G/D ratios derived from the FMR are all $>$100. It is unlikely that this is appropriate for our galaxies, especially considering their excited nature, with a high fraction of AGN and mergers/interactions in the sample. We indicate on Fig.~\ref{fig:RealKS} the direction and extent that these cluster galaxies would move if they were to have higher metallicities than derived from the FMR, and therefore G/D ratios more similar to those derived through dynamical arguments for A1 and A2. This would lead to a factor $\sim$3 lower gas mass, and even higher SFE offsets with respect to the MS relation.

As can be seen in Fig.~\ref{fig:A126Colour}, we find a high fraction of mergers, interactions and AGN in our star-forming cluster galaxies. Only two of our eight galaxies are isolated, non-AGN, of which one galaxy has a prominent bulge component and one was previously characterised as passive through SED fitting and colour-colour diagnostics \citep{ref:V.Strazzullo2016}. We will present further discussion on passive systems in the cluster in a future paper, as this is a very exciting and expanding area of research (e.g. \citealt{ref:M.T.Sargent2015, ref:R.Gobat2017}).

Merging systems and AGN can naturally increase the density and excitation of the molecular gas in galaxies, and there is increasing evidence to suggest that mergers play an important part in galaxy evolution in high redshift (proto-)clusters. The relative velocities of the galaxies in clusters at high redshift are relatively low (e.g. \mbox{\citealt{ref:M.Brodwin2011}}), making it more likely that galaxies will come into close contact with each other and experience either major or minor mergers. Studies of the morphologies of cluster galaxies have shown that significant numbers of cluster galaxies at z$>$1.5 have disturbed or irregular morphologies \citep{ref:S.Mei2015, ref:J.S.Santos2015}, and mergers have also been taken to explain the larger sizes of high-redshift cluster galaxies compared to otherwise similar galaxies in the field (e.g. \citealt{ref:L.Delaye2014, ref:V.Strazzullo2015}). A recent study by \citet{ref:S.Alberts2016} focussed on galaxy clusters in the range 1$<$z$<$2, and concluded that environmental interactions and an increased galaxy merger rate were triggering the heightened AGN fraction in the cores of some 11 massive clusters. Increased merger rates have also been measured in individual objects, such as the spectroscopically-confirmed proto-cluster Cl~0218.3-0510 at z=1.62 \citep{ref:J.Lotz2013}. Here, 57\% of cluster galaxies above M$_{*}>$3$\times$10$^{10}$M$_{\odot}$ were found to be in close pair/double nuclei systems compared to only 11\% in the field, giving an increase in the merger rate of between 3 and 10 times above the field value. \mbox{\citet{ref:C.Krishnan2017}} suggest that this increase in merger rate can be linked to the increase in AGN activity in this proto-cluster.

The high fraction of galaxy mergers and AGN in our star-forming cluster galaxies builds upon the above observations, and we suggest that increased merger and AGN activity in the cluster core is the key driver behind the high fraction of galaxies with starburst-like gas excitations, contributing a high fraction of the star-formation to the cluster core.

\subsection{Merger driven star-formation in the core of the cluster}
\label{sec:quantEnviron}
In order to quantify the effect of the environment on the ISM gas
excitations and star-formation in the core of the cluster, we compare our primary
findings with the expectations from field galaxy populations. We have
discovered that 50\% of our cluster galaxies have starburst-like gas
excitations. This is quite a high fraction, and possibly surprising.
For example, we can compare this with expectations from the 2-SFM model
\citep{ref:M.Sargent2014}, considering the population of galaxies in this
framework at stellar mass M$_{*}$>10$^{10}M_{\odot}$ and
SFR$>$30$M_{\odot}$yr$^{-1}$, corresponding to the minimum stellar masses
and SFRs for our galaxy sample. One might expect that starburst galaxies in the 2-SFM model also have
starburst-like excited molecular gas. In terms of the number density in
the field, there are only $\sim$6\% of starburst galaxies in the 2-SFM
framework under these selection criteria at z=2
\mbox{\citep{ref:G.Rodighiero2011, ref:M.Sargent2012}}. There is a marked
difference between this and the 50\% of excited galaxies that we find in
the cluster core.

If we now look at the star-formation rate density in the cluster, we find
even more striking results. Firstly, we find that the cluster galaxies
with highly excited gas contribute $\sim$50\% of the total SFR to the
cluster core. We again see a clear increase compared with the contribution
of just $\sim$18\% to the total SFR density that we would expect from
starburst galaxies in the 2-SFM model, in the field at z=2. However,
considering the excited galaxies in our cluster core neglects a
significant fraction of the SFR driven by the large number of mergers and
interactions. For example, we see that galaxy A2 contributes significantly
the the total SFR, despite not having starburst-like gas excitations. If
we extend our analysis and investigate the SFR contribution from mergers
and interactions in the cluster core, we find that $\sim$75\% of the total SFR
in the cluster core comes from mergers and interactions alone. It is clear
that the cluster environment of \Cl\ is having a significant impact on
both the gas excitations and star-formation within the cluster. This is a
very important result of our study, as it demonstrates and quantifies the
significant impact that galaxy mergers are having on the SFR and evolution
of this cluster in the early Universe.

\subsection{Depletion timescales and gas fractions}
\label{sec:timescales}
As discussed in Section~\ref{sec:mergers}, the galaxies in \Cl\ with dense gas detections are not disk-like MS galaxies, nor typical isolated dusty starbursts. This makes direct comparison with the literature challenging, as we are exploring properties of galaxies that have been less widely characterised to date. In this section, we compute molecular gas depletion timescales and gas-to-stellar mass ratios for our sample of merger- and AGN-dominated cluster galaxies.

The high SFEs that we see in our galaxies imply that they have a higher SFR for their total molecular gas reservoir than MS, disk-like galaxies, and will deplete their gas supplies over relatively short timescales. Calculation of the depletion timescale is of course dependent on the conversion factors adopted between the observables (L$^{\prime}_{\rm CO[1-0]}$ and M$_{d}$) and the molecular gas mass. In order to calculate gas depletion timescales, we have used the conversion factors shown in Table~\ref{tab:Dustmasses} for all galaxies except A1 and A2, and have assumed a constant SFR with no replenishment of the cold gas reservoir.

The values for $\alpha_{\rm CO}$ and G/D for A1 and A2 were calculated directly, as stated in Section~\ref{sec:dust}. It can be seen that both $\alpha_{\rm CO}$ and the G/D ratio for A2 are lower than the values usually taken for solar-metallicity, MS galaxies ($\alpha_{\rm CO}$=4.4 and G/D=100). This is somewhat unsurprising, due to the merging nature of A2. Interestingly, the G/D ratio for A1 reflects its merging-nature, but the $\alpha_{\rm CO}$=3.6 is four times higher than that usually taken for typical mergers and starburst galaxies, $\alpha_{\rm CO}$=0.8. As was also seen in Fig.~\ref{fig:Mdust_excitation}, there appears to be a discrepancy between the CO properties of A1 and the dust properties. A potential explanation for this is that the SFR of A1 has recently started to decrease. Dust is a longer-timescale tracer of star-formation than CO luminosity, so if the SFR of A1 recently peaked and has since started to decline, then this could be reflected in the instantaneous CO luminosity properties before it is reflected in the dust emission. The dynamical mass that we derive for A1 probes the total molecular gas content, and a high $\alpha_{\rm CO}$ value suggests that there is a large gas reservoir in A1. If the SFR of A1 has recently started to decrease but there is still a large molecular gas reservoir, this could imply that it is primarily the dense, star-forming gas that been removed, and the diffuse gas still remains. Alternatively, if the dust temperature of A1 is higher than the value that we have used, $<$U$>$~=~28.75, then this would reduce the derived dust content for A1, as can be seen in Fig.~\ref{fig:MdustDerivation}. However, increasing the value of $<$U$>$ would also increase the L$_{IR}$ value derived from the dust, and as can be seen in Table~\ref{tab:SFRs}, raising the SFR$_{870\micron}$ in A1 would bring it further from the SFR derived from CO[4-3]. These ideas are speculation, but the fact that these conversion factors seem to suggest different physical properties raises an interesting question.

The values for molecular gas depletion timescales for all galaxies are shown in Table~\ref{tab:Dustmasses}, again calculated using the gas masses from the dust. A typical MS galaxy, forming stars in a secular fashion, will deplete its cold gas reservoir in $\sim$0.5-1~Gyrs, whereas a violently starbursting galaxy can deplete its gas in $<$0.1~Gyrs (e.g. \citealt{ref:A.Saintonge2012, ref:A.Saintonge2016, ref:A.Leroy2013, ref:E.Daddi2010b, ref:M.Sargent2014, ref:R.Genzel2015, ref:L.J.Tacconi2013, ref:L.Tacconi2017}). We find that all of our cluster galaxies have relatively short depletion timescales, $<$400~Myrs, with 50\% of them having short timescales $<$300~Myrs, in agreement with other recent cluster studies \mbox{\citep{ref:T.Wang2016, ref:MHayashi2017, ref:S.Stach2017}}. Galaxies A1 and A2 have depletion timescales consistent with the short timescales observed in highly starbursting galaxies, $\sim$100~Myrs. We caution however that a direct comparison should not be made between our cluster galaxies and those values measured for isolated disk-like star-forming or starburst galaxies. Our galaxy sample appears heavily dominated by mergers, interactions and AGN, and is therefore not equivalent to the field samples from which these comparison values originate. Nonetheless, these short gas depletion timescales do suggest evidence for rapid star-formation quenching in this cluster environment. It is possible that this gas depletion could give rise to the massive elliptical galaxies, such as those seen in local clusters, on short timescales.

We calculate gas-to-stellar mass ratios for our galaxies, $\mu_{g}$, in Table~\ref{tab:Dustmasses}, which are shown in Fig.~\ref{fig:gasfrac_stellar}. Here, we show the relationship between $\mu_{g}$ and M$_{*}$ presented in \citet{ref:M.Bethermin2015}, $\mu_{g} \propto M_{*}^{-0.5}$, with the normalisation calibrated for a field sample of MS star-forming disks. We fit this relationship with a fixed slope to our data, using a least squared approach, and we find that the best-fit normalisation for our galaxies is slightly lower than that of MS galaxies at z=2. This indicates that the gas-to-stellar mass ratios of our cluster member galaxies are lower than those of field MS galaxies at z=2. If our cluster galaxies were a sample of isolated, disk-like galaxies, this lower gas-to-stellar mass ratio might be expected, based on the high SFEs of the cluster galaxies that we see in Fig.~\ref{fig:RealKS}. However, we cannot robustly make these direct links for our sample of cluster galaxies, which may not follow the same scaling relations as field MS disk galaxies. We believe that the lower gas-to-stellar mass ratios found for our cluster galaxies in Fig.~\ref{fig:gasfrac_stellar} are driven by the different physical natures of the galaxies in the cluster compared with MS disk galaxies in the field at z=2, an important outcome of this study.

As discussed in Section~\ref{sec:dust}, the conversion factors and therefore gas masses derived for our cluster galaxies are based on the framework of normal MS galaxies, with the exception of A1 and A2. As can be seen in Table~\ref{tab:Dustmasses}, we derive high $\alpha_{\rm CO}$ and G/D ratios from the FMR that do not seem to reflect the true nature of the galaxies. We might expect our active, interacting galaxies to have lower conversion factors, such as those derived directly for galaxy A2. In the case of lower conversion factors, the molecular gas masses would decrease, leading to lower gas-to-stellar mass ratios (shown on Fig.~\ref{fig:gasfrac_stellar} by the black arrow), and even shorter gas depletion timescales for our galaxies. This would increase the tension between the gas-to-stellar mass ratios of our galaxies and the field galaxies shown in Fig.~\ref{fig:gasfrac_stellar}, and potentially imply star-formation quenching on even shorter timescales in the cluster environment. There is currently not a great deal of literature on the gas contents of such complex merging and active systems in clusters to compare our results with, but we hope future studies on similar systems at high redshift will provide samples for comparison.

We compare the dust-to-stellar mass ratios of the cluster galaxies with previous studies in Fig.~\ref{fig:MdustVsMstar}. Here, we see that approximately half of the cluster galaxies show evidence for low dust-to-stellar mass fractions compared with the MS field relations. This is not entirely unexpected, as we have seen in Figs~\ref{fig:SFR_Mstar}~and~\ref{fig:RealKS} that the SFR in the cluster is being driven by relatively small amounts of molecular gas, for SFRs broadly consistent with the MS of star-formation. Fig.~\ref{fig:gasfrac_stellar} illustrates this low molecular gas content per stellar mass, and we have shown in Fig.~\ref{fig:Mdust_excitation} that the dust and gas properties are consistent with one another (with the possible exception of A1). We do however caution that some cluster galaxies, such as IDs 6 and 3, are bulge-dominated, and may contain a significant passive stellar population, and we do not currently have the short-wavelength data to directly probe the dust temperatures of these galaxies. As is shown in Fig.~\ref{fig:MdustDerivation}, if we were to assume a dust temperature $<$U$>$ less than 10, such as for the field elliptical galaxies studied in \citet{ref:R.Gobat2017}, this could in turn affect our derivations of properties for these galaxies. We will return to this interesting possible effect in a future work.

\section{Conclusions}
\label{sec:conc}
We have presented deep observations from JVLA and ALMA, in order to characterise the gas content of galaxies in a maturing cluster at z=1.99, \Cl. In doing this, we are able to study how dust-obscured star-formation and ISM gas content are linked to the environment and galaxy interactions, during this crucial phase of high-z cluster assembly. We detect molecular gas in eight cluster galaxies, and find that these galaxies are dominated by mergers, interactions and AGN, rather than isolated disk-like star-forming galaxies. We present 870$\micron$ continuum measurements of the galaxies as well as multiple transitions of $^{12}$CO: the CO[4-3] transition tracing the dense, star-forming gas, the CO[3-2] transition, and the CO[1-0] transition as a tracer of the total molecular gas reservoir. From these line fluxes, we are able to construct CO SLEDs for the galaxies, and find a highly increased fraction of galaxies with starburst-like gas excitations, $\sim$8 times greater than in the equivalent field population. We also find an enhanced contribution from the starburst-excited galaxies to the total SFR in the cluster core, $\sim$3-4 times greater than the starburst galaxy contributions to SFR in the field.

We calculate SFRs for the cluster galaxies from the CO data, and compare these with estimates from other SFR tracers such as our ALMA 870$\micron$ continuum observations. We find the tracers to be consistent, and use these SFRs to investigate the star-formation efficiencies of the galaxies, which are shown to be globally increased with respect to non-environment-selected samples. We do not however see a trend of increased sSFR. We derive dynamical masses for our two brightest, resolved galaxies in CO[4-3], and directly calculate values for $\alpha_{\rm CO}$ and G/D based on their inferred molecular gas masses. For the rest of our galaxies, we calculate estimates of gas-phase metallicity using the FMR, and subsequently values for $\alpha_{\rm CO}$ and G/D.

We use ALMA continuum flux measurements to construct the submillimetre-radio portion of the galaxies' SEDs. We model the RJ tail of the SEDs, and in this way derive dust masses for each of the cluster galaxies. We compare molecular gas masses calculated from the galaxies' dust with gas masses and limits from the CO[1-0] line luminosities, and find the results to be in good agreement. We use the molecular gas masses derived from the dust mass for the analysis in this paper.

We do not measure significant CO[1-0] flux for 5/8 galaxies despite our deep JVLA data, further indicating that the star-formation in these galaxies is being fuelled by small amount of molecular gas. We calculate gas depletion timescales using the galaxies' SFRs and molecular gas contents, and find that the gas will be depleted on short timescales, between $\sim$100-400~Myrs. The gas-to-stellar mass ratios of the cluster galaxies appear to be lower than disk-like MS galaxies at z=2, but we caution against direct comparison with field samples due to the merging and active nature of the cluster galaxies.

Our results suggest that environmental effects in this dense cluster core have given rise to a galaxy population with a high fraction of excited, starburst-like galaxies, that will deplete their star-forming gas on short timescales. We conclude that the high gas excitations and star-formation activity are primarily driven by mergers and interactions in the core of \Cl. We hope that increasing amounts of data will become available to facilitate further studies of the relationships between galaxy properties in clusters, continuing to build a more complete picture of the gas content and star-formation drivers of galaxies in dense environments. Future near-IR wide field surveys from space (e.g. Euclid; \citet{ref:R.Laureijs2011}, and WFIRST) will detect a large number of galaxy clusters at z$>$1.5-2. This will allow their synergistic study with ALMA, in order to investigate with unprecedented statistics the evolution of gas and galaxies in the most dense environments during the `cosmic noon' epoch.

\section*{Acknowledgements}
We thank the referee for their suggestions. We also thank Frederic Bournaud for his help and discussion on deriving dynamical masses. RTC acknowledges support from STFC G1687 grant ST/N504452/1. MTS was supported by a Royal Society Leverhulme Trust Senior Research Fellowship (LT150041). GEM acknowledges support from the Carlsberg Foundation, the ERC Consolidator Grant funding scheme (project ConTExt, grant number No. 648179), and a research grant (13160) from VillumFonden. AC acknowledges the grants ASI n.I/023/12/0 `Attività relative alla fase B2/C per la missione Euclid' and PRIN MIUR 2015 `Cosmology and Fundamental Physics: illuminating the Dark Universe with Euclid'. H.D. acknowledges financial support from the Spanish Ministry of Economy and Competitiveness (MINECO) under the 2014 Ramón y Cajal program MINECO RYC-2014-15686. ET acknowledges financial support from the UnivEarthS Labex program of Sorbonne Paris Cit\'{e} (ANR-10-LABX-0023 and ANR-11-IDEX-0005-02).

This paper makes use of the following ALMA data: ADS/JAO.ALMA\#2012.1.00885.S and ADS/JAO.ALMA\#2015.1.01355.S. ALMA is a partnership of ESO (representing its member states), NSF (USA) and NINS (Japan), together with NRC (Canada), MOST and ASIAA (Taiwan), and KASI (Republic of Korea), in cooperation with the Republic of Chile. The Joint ALMA Observatory is operated by ESO, AUI/NRAO and NAOJ.

This paper also makes use of JVLA program 12A-188. The National Radio Astronomy Observatory is a facility of the National Science Foundation operated under cooperative agreement by Associated Universities, Inc.

%%%%%%%%%%%%%%%%%%%%%%%%%%%%%%%%%%%%%%%%%%%%%%%%%%



\bibliographystyle{mnras}
\bibliography{EnvironmentNew.bib} % if your bibtex file is called example.bib

%%%%%%%%%%%%%%%%%%%%%%%%%%%%%%%%%%%%%%%%%%%%%%%%%%

%%%%%%%%%%%%%%%%% APPENDICES %%%%%%%%%%%%%%%%%%%%%

\appendix

\section{Image-plane vs. UV-plane flux extraction}

Flux measurement simulations were performed for both \texttt{uvfit} and GALFIT. In both cases, the 2mm continuum data were used, having first removed all real continuum sources in the uv-plane. For GALFIT, the dirty data were then transported to the image plane. Three different simulated source sizes were used - a point source (PSF), a circular Gaussian with FWHM=1", and a circular Gaussian with FWHM=2". The size of the synthesised beam in the 2mm data is 1.19"$\times$0.96", so these Gaussian sizes correspond to approximately one and two times the synthesised beam size. Additionally, these simulated sizes span the range of measured sizes for the cluster galaxies (Table~\ref{tab:sizeComp}). One thousand sources of each size were injected, one by one. Having done this, the flux of the source was then re-measured using \texttt{uvfit} or GALFIT respectively, at fixed input position and over the fixed input size. Although this is a limited scenario, we perform our simulations under these conditions and we wish to primarily quantify the difference between the flux measurements in the image plane and the uv-plane. Additionally, the majority of the flux measurement in our analysis is performed at fixed position and size.

We then inspect the average measured flux of the 1000 sources, the average measurement error returned for the 1000 sources, and the dispersion of the returned flux distribution. These results are given in Tables~\ref{tab:ImageVsUV_Uvfit} and \ref{tab:ImageVsUV_Galfit}. We find that the average returned flux is consistent with the input flux for both \texttt{uvfit} and GALFIT, although the formal flux error returned by GALFIT is underestimated. However, the dispersion of the flux distribution from GALFIT compares reasonably with the RMS noise of the input image. We find that \texttt{uvfit} performs marginally better for PSF sources than GALFIT, and that GALFIT returns a slightly smaller flux dispersion for 2" extended sources than \texttt{uvfit}.

\begin{table*}
\centering
\begin{tabular}{ccccc}
\hline
Shape & Flux$_{in}$ ($\mu$Jy) & Flux$_{ave, out}$ ($\mu$Jy) & Error$_{ave, out}$ ($\mu$Jy) & $\sigma_{flux, out}$ ($\mu$Jy) \\
\hline
PSF & 120.0 & 119.1 & 7.2 & 8.6 \\
1" Gauss. & 120.0 & 120.5 & 11.3 & 13.8 \\
2" Gauss. & 120.0 & 120.2 & 18.4 & 23.7 \\
\end{tabular}
\caption{Flux measurement simulation results for \texttt{uvfit}. The shape, input flux, average output flux, average output flux error, and 1$\sigma$ dispersion on the output fluxes are given. The RMS noise of the 2mm map is $\sim$8.5$\mu$Jy/beam. The input flux therefore corresponds to a PSF SNR of $\sim$15$\sigma$. The 1$\sigma$ dispersion of the output fluxes for the PSF closely reflects the RMS of the input map, as expected.}
\label{tab:ImageVsUV_Uvfit}
\end{table*}

\begin{table*}
\centering
\begin{tabular}{ccccc}
\hline
Shape & Flux$_{in}$ ($\mu$Jy) & Flux$_{ave, out}$ ($\mu$Jy) & Error$_{ave, out}$ ($\mu$Jy) & $\sigma_{flux, out}$ ($\mu$Jy) \\
\hline
PSF & 120.0 & 119.8 & 0.7 & 9.3 \\
1" Gauss. &120.0 & 120.5 & 1.1 & 13.7 \\
2" Gauss. & 120.0 & 120.7 & 1.1 & 21.9 \\
\end{tabular}
\caption{Flux measurement simulation results for GALFIT. The shape, input flux, average output flux, average output flux error, and 1$\sigma$ dispersion on the output fluxes are given. As above, the RMS noise of the 2mm map is $\sim$8.5$\mu$Jy/beam. The input flux therefore corresponds to a PSF SNR of $\sim$15$\sigma$. The formal errors on the fluxes reported by GALFIT  are strongly underestimated, as discussed in Section~\ref{sec:contextract}.}
\label{tab:ImageVsUV_Galfit}
\end{table*}

\section{CO[3-2] and CO[1-0] 1D Spectra}

\begin{figure*}
    \centering
    \includegraphics[width=0.8\textwidth]{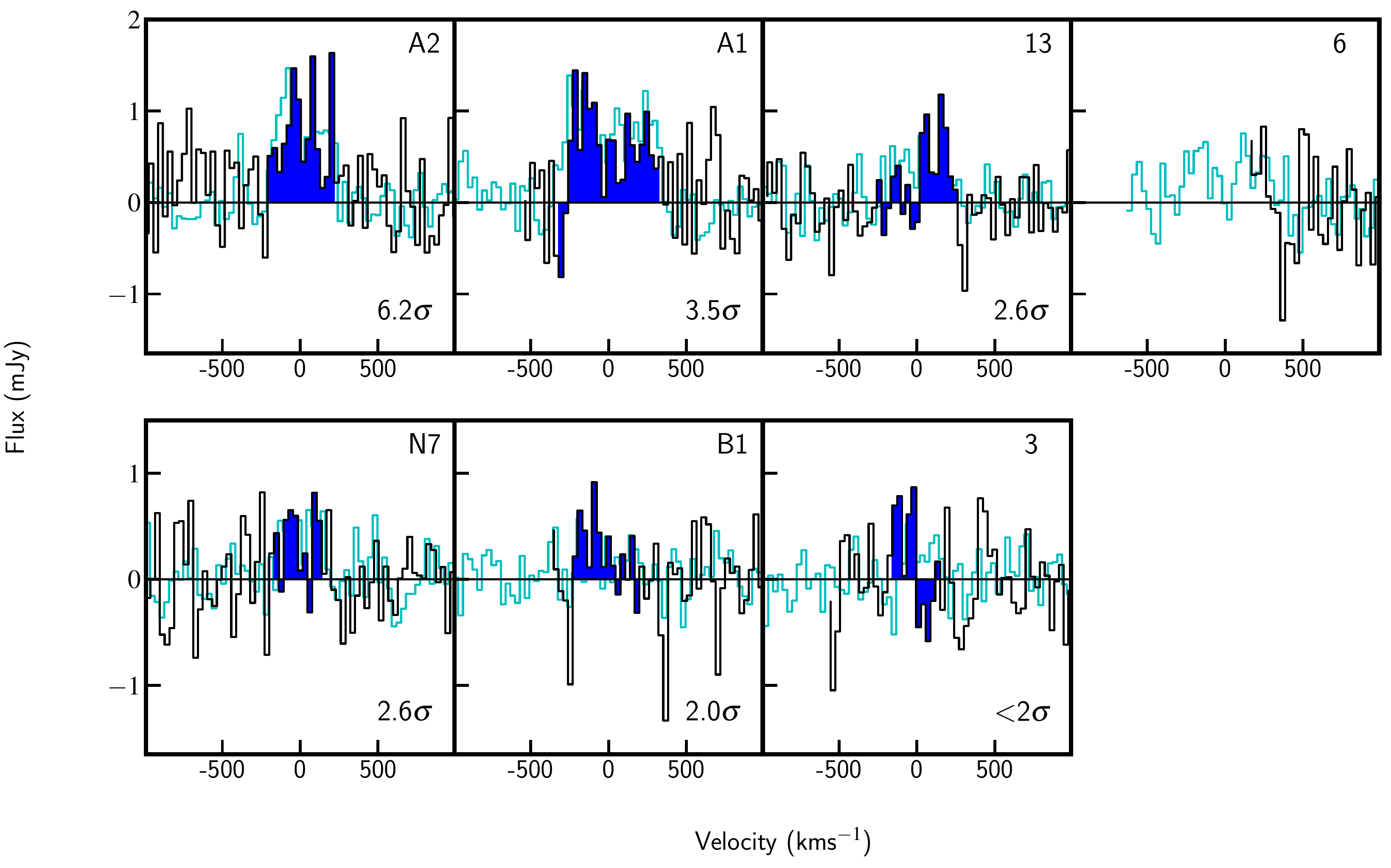}
    \caption{CO[3-2] line spectra for seven galaxies in \Cl, shown by the black line. Galaxy S7 has not been included, as the frequency coverage of the observations did not extend to the CO[3-2] position at the redshift of S7. Similarly, for galaxy 6 our band 3 data provides only partial coverage of the velocity range corresponding to the CO[4-3] line emission, shown in cyan. The blue shading indicates the velocity over which CO[3-2] emission was measured. The CO[3-2] line fluxes were extracted over this fixed velocity range, corresponding to the redshift and velocity range given by the CO[4-3] line detections. The CO[4-3] line spectra are shown in the background (cyan lines), at the same scale. The significances of the CO[3-2] line fluxes are shown are shown in the bottom-right corner of each panel, measured from the data collapsed over the linewidth shown in blue shading, over the same 2D shape as for the CO[4-3] emission of each galaxy.}
    \label{fig:CO32Lines}
\end{figure*}

\begin{figure*}
    \centering
    \includegraphics[width=0.8\textwidth]{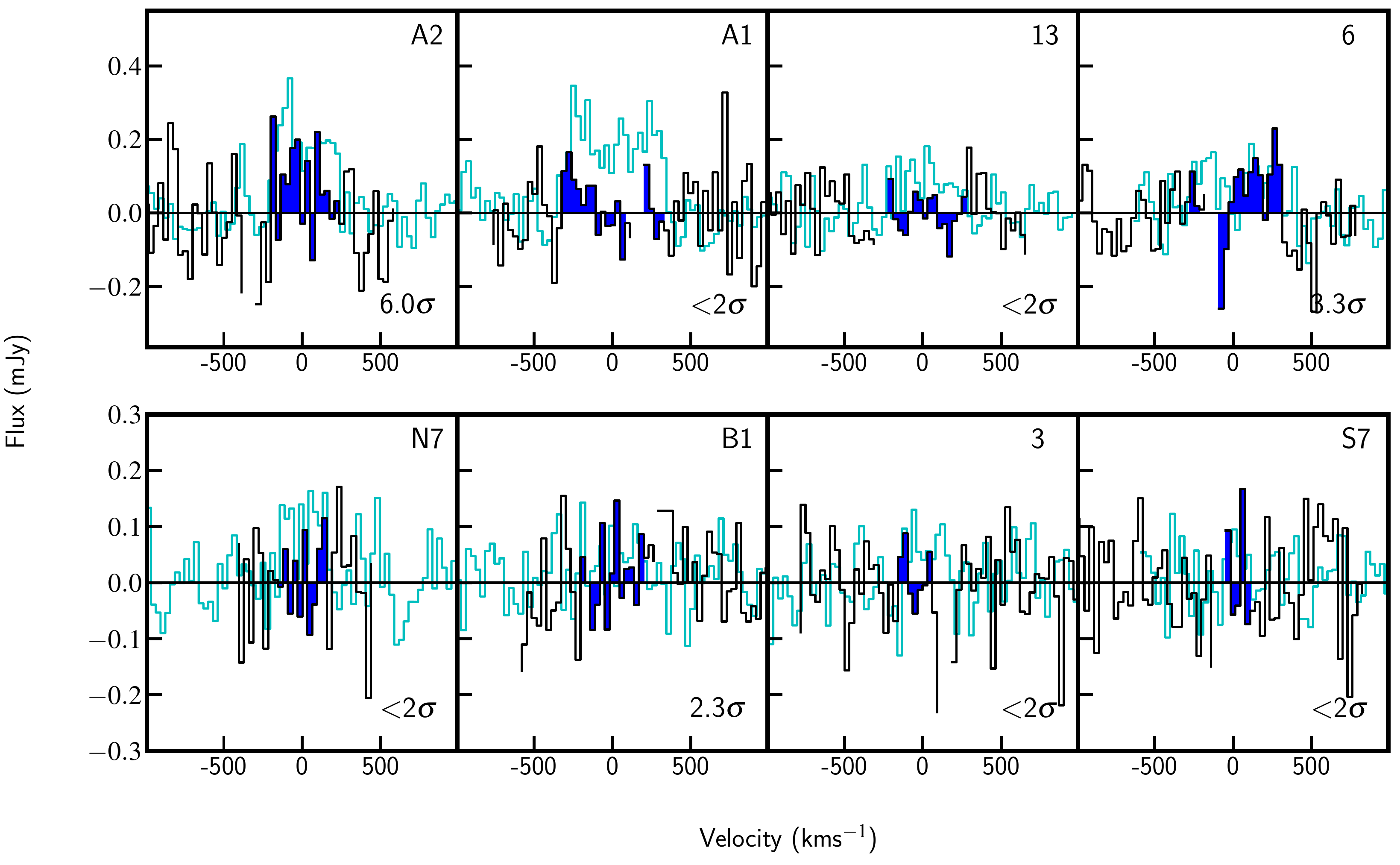}
    \caption{CO[1-0] line spectra for eight galaxies in \Cl, shown by the black line. The blue shading indicates the velocity over which CO[1-0] emission was measured. The line fluxes for the CO[1-0] line were extracted over this fixed velocity range, corresponding to the redshift and velocity range given by the CO[4-3] line detections. The CO[4-3] line spectra are shown in the background (cyan lines), with the magnitude of the flux scaled down by a factor of 4 for visualisation. The significances of the CO[1-0] line fluxes are shown are shown in the bottom-right corner of each panel, measured from the data collapsed over the linewidth indicated by the dashed lines, over the same 2D shape as for the CO[4-3] emission of each galaxy. Breaks in the CO[1-0] spectra along the velocity axis indicate the breaks between the JVLA spectral windows, between which no data is available.}
    \label{fig:CO10Lines}
\end{figure*}

The 1D spectra covering the CO[3-2] and CO[1-0] lines are shown in Figs~\ref{fig:CO32Lines} and \ref{fig:CO10Lines} respectively, for the eight galaxies in \Cl\ with CO[4-3] line detections. We did not search for CO[3-2] or CO[1-0] lines in these spectra, but instead extracted line fluxes at fixed position for each galaxy, matching the position and shape of the CO[4-3] emission. This was done in the collapsed uv-data, over a fixed velocity range corresponding to that of the CO[4-3] line detection.

\newpage

%%%%%%%%%%%%%%%%%%%%%%%%%%%%%%%%%%%%%%%%%%%%%%%%%%

% Don't change these lines
\bsp	% typesetting comment
\label{lastpage}
\end{document}